\documentstyle[12pt,epsfig,cite]{article}
\newcommand{\beq}{\begin{equation}}
\newcommand{\eeq}{\end{equation}}
\newcommand{\bea}{\begin{eqnarray}}
\newcommand{\eea}{\end{eqnarray}}

\newcommand{\ra}{\rightarrow}

\newcommand{\gsim}{\lower.7ex\hbox{$
\;\stackrel{\textstyle>}{\sim}\;$}}
\newcommand{\lsim}{\lower.7ex\hbox{$
\;\stackrel{\textstyle<}{\sim}\;$}}

\newcommand{\matel}[3]{\langle #1|#2|#3\rangle}

\newcommand{\eod}{\end{document}}

\def\op{{\bf P}}
\def\oc{{\bf C}}
\def\ot{{\bf T}}
\def\cp{{\bf CP}}
\def\cpt{{\bf CPT}}

\begin{document}
\thispagestyle{empty}
\vspace*{-22mm}

\begin{flushright}
UND-HEP-07-BIG\hspace*{.08em}03\\
hep-ph/0703132\\

\end{flushright}
\vspace*{1.3mm}

\begin{center}
{\LARGE{\bf
\cp~Violation in the SM, Quantum Subtleties and the Insights of Yogi Berra 
}}
\vspace*{19mm}

{\Large{\bf I.I.~Bigi}} \\
\vspace{7mm}

{\sl Department of Physics, University of Notre Dame du Lac}
\vspace*{-.8mm}\\
{\sl Notre Dame, IN 46556, USA}\\
{\sl email: ibigi@nd.edu}

\vspace*{10mm}

{\bf Abstract}\vspace*{-1.5mm}\\
\end{center}

\noindent
Our knowledge of flavour dynamics has undergone a `quantum jump' since just before the 
turn of the millenium: direct \cp~violation has been firmly established in $K_L \to \pi \pi$ 
decays in 1999; the first \cp~asymmetry outside $K_L$ decays has been discovered in 2001 in 
$B_d \to \psi K_S$, followed by $B_d \to \pi^+\pi^-$, $\eta^{\prime}K_S$ and $B \to K^{\pm}\pi^{\mp}$ 
establishing direct \cp~violation also in the beauty sector. Counterintuitive, yet central features 
of quantum mechanics like meson-antimeson oscillations and EPR correlations have been 
crucial in making such effects observable. 

\noindent
The CKM dynamics of the Standard Model (SM) of HEP 
allow a description of \cp~insensitive and sensitive $B$, $K$ and $D$ transitions that is 
impressively consistent even on the quantitative level. We know now that at least the lion's share 
of the observed \cp~violation is provided by the SM. Yet these novel successes do not invalidate 
the theoretical arguments for it being incomplete. We have also more direct evidence for New Physics, namely neutrino oscillations, the 
observed baryon number of the Universe, dark matter and dark energy. While the New Physics anticipated at the TeV scale is not likely to shed any light on the SM's mysteries of flavour, 
detailed and comprehensive studies of heavy flavour transitions will be essential in diagnosing salient 
features of that New Physics. Strategic principles for such studies are outlined.   
\setcounter{page}{0}
\vfill

\tableofcontents

\vspace{0.5cm}

\noindent 
{\bf Prologue}

\vspace{0.3cm}

The pronouncements of athletic coaches carry little weight in academic circles. That is not always 
justified. Take Yogi Berra for example, a well-known former catcher and baseball manager. 
He could be considered the founder of the most popular American school of philosophy, at least the 
most quoted one. He once stated: "If the world were perfect, it wouldn't be."  This can be seen as a 
memorable formulation of the principle underlying baryogenesis: if there were perfect balance between matter and antimatter in our universe, it would bear no similarity to our universe. 
On another occasion he declared: "When you come to a fork in the road, take it." I know of no more 
concise formulation of one of quantum mechanics most counter-intuitive features that underlies 
the interference pattern observed in a double-slit experiment with particle beams: even a single 
electron can pass through both slits. 

These two quotes refer to two central topics in this review: \cp~violation and its connection 
to baryogenesis on one hand and on the other the essential role played by intrinsically quantum mechanical 
effects to make \cp~violation observable. The third central topic is my judgment that none of the impressive successes of the SM invalidate the arguments for it being incomplete, that New Physics 
has to exist probably at scales as low as $\sim 1$ TeV, that whereas this nearby New Physics is unlikely to shed any light on the flavour structure of the SM, its impact on flavour transitions will 
reveal essential features of it unlikely to be obtained any other way; yet I have to add the 
caveat that this potential impact 
will probably be at most of modest strength {\em numerically} thus requiring accuracy on the experimental as well as theoretical side for proper interpretation. 

In Sect.\ref{PART1} I will provide a brief evaluation of the SM, sketch the discovery of \cp~violation 
in kaon decays and introduce the salient features of CKM dynamics leading to the `SM's Paradigm 
of Large \cp~Violation in $B$ Decays' described in Sect.\ref{PARA2000}, as it had emerged by the end of the last millenium; in 
Sect.\ref{PART2} I describe the validation of this paradigm since the turn of the millenium; 
in Sect.\ref{PART3} I outline strategies for searches for New Physics in flavour dynamics and sketch the future landscape of High Energy Physics (HEP), which in my vision should contain a 
Super-Flavour Factory, before concluding with an Epilogue.

%%%%%%%%%%%%%%%%%
\section{The SM and CKM Dynamics before 2000}
\label{PART1}
%%%%%%%%%%%

%%%%%%%%%%%%%%%%
\subsection{On the Uniqueness of the SM}
\label{SMUNIQ}
%%%%%%%%%%

A famous American Football coach once declared:"Winning is not the greatest thing -- 
it is the only thing!" This quote provides some useful criteria for sketching the status of the 
different components of the Standard Model (SM). It can be characterized by the carriers of its strong and electroweak forces that are described by {\em gauge} dynamics and the 
{\em mass matrices} for its quarks and leptons 
as follows: 
\beq 
{\rm SM}^* = SU(3)_C \times SU(2)_L \times U(1) \oplus {\rm `CKM'} (\oplus {\rm `PMNS'})  
\eeq
I have attached the asteriks to `SM' to emphasize the SM contains a very peculiar pattern of fermion 
mass parameters that is not illuminated at all by its gauge structure. 
Next I will address the status of these components. My goal is to emphasize 
the intrinsic connection of CKM dynamics, 
which is behind the observed \cp~violation, with central mysteries of the SM. 

%%%%%%%%%%%%
\subsubsection{QCD -- the `Only' Thing}
\label{QCD}
%%%%%%%%%%%%

While it is important to subject QCD again and again to quantitative tests as the theory for the strong 
interactions, one should note that these serve more as tests of our computational control over 
QCD dynamics than of QCD itself. For its features can be inferred from a few general requirements 
and basic observations. A simplified list reads as follows:  Our understanding 
of chiral symmetry as a {\em spontaneously} realized one, the measured values for 
$R=\sigma (e^+ e^- \to {\rm had.})/\sigma (e^+e^- \to \mu ^+ \mu ^-)$ (and likewise the branching ratios 
for $\pi ^0 \to \gamma \gamma $, $\tau ^- \to e^- \bar \nu _e \nu _{\tau}$ and $B \to l \nu X_c$), 
the need to deal with massless spin-one fields and the requirement of combining confinement with 
asymptotic freedom lead us quite uniquely to a local nonabelian gauge theory with three colours. A true failure of QCD would thus create a genuine paradigm 
shift, for one had to adopt an {\em intrinsically non-}local description. It should be remembered that string theory was first put forward for describing the strong interactions.

A theoretical problem arises for QCD from an unexpected quarter that is relevant 
for our context: QCD does {\em not automatically} conserve \op, \ot~ and \cp. To reflect the 
nontrivial topological structure of QCD's ground state one employs an 
{\em effective} Lagrangian containing an additional term to the usual QCD Lagrangian \cite{CPBOOK}: 
\beq 
{\cal L}_{eff} = {\cal L}_{QCD} + \theta \frac{g_S^2}{32\pi^2} G_{\mu \nu} 
\tilde G^{\mu \nu} \; ,  \; \tilde G_{\mu \nu} = \frac{i}{2} \epsilon _{\mu \nu \rho \sigma}G^{\rho \sigma} 
\label{LQCDEFF}
\eeq
Since $G_{\mu \nu} \tilde G^{\mu \nu}$ is a gauge invariant operator, its appearance in general cannot be forbidden, and what is not forbidden has to be considered allowed in a quantum field theory. It represents a total divergence, yet in QCD -- unlike in QED -- it cannot be ignored due 
to the topological structure of the ground state. 
Since under parity $\op$ and time reversal $\ot$  
\beq 
G_{\mu \nu} \tilde G^{\mu \nu} \stackrel{\op, \ot}{\Longrightarrow} - G_{\mu \nu} \tilde G^{\mu \nu} 
\; ,   
\eeq 
the last term in Eq.(\ref{LQCDEFF}) violates $\op$ as well as $\ot$. Being flavour-{\em diagonal}  
$G_{\mu \nu} \tilde G^{\mu \nu}$ generates an electric dipole moment 
(EDM) for the neutron.  From the upper bound on the latter $d_N < 0.63 \cdot 10^{-25} \; {\rm e \, cm}$  
one infers \cite{CPBOOK} $\theta < 10^{-9}$. 
Being the coefficient of a dimension-four operator $\theta$ can be renormalized to any value, even zero. Yet the modern view of renomalization is more demanding: requiring the renormalized value to be smaller than its `natural' one by {\em orders of magnitude} is frowned upon, since it requires 
{\em finetuning} between the loop corrections and the counterterms. This is what happens here. For  
purely within QCD the only intrinsically `natural' scale for $\theta$ is unity. If $\theta \sim 0.1$ 
or even $0.01$ were found, one should not be overly concerned.  Yet a bound like 
$\theta < 10^{-9}$ is viewed with great alarm as very {\em unnatural} -- unless a symmetry 
can be called upon. If any quark were massless -- most likely the $u$ quark -- chiral rotations representing symmetry transformations in that case could be employed to remove 
$\theta$ contributions. Yet a considerable phenomenological body rules against such a scenario. 

A much more attractive solution would be provided by transforming $\theta$ from a fixed parameter into the manifestation of a {\em dynamical} field -- as is done 
for gauge and fermion masses through the Higgs-Kibble mechanism, see below -- and imposing a 
Peccei-Quinn symmetry that would lead {\em naturally} to $\theta \ll {\cal O}(10^{-9})$. Alas -- this attractive solution does not come `for free': it requires the existence of axions. Those have not been observed despite great efforts to find them. 

This is a purely theoretical problem. Yet I consider the fact that it remains unresolved a significant chink 
in the SM's armour. I still have not given up hope that `victory can be snatched from the jaws of defeat': establishing a Peccei-Quinn-type solution would be a major triumph for theory.

%%%%%%%%%%%
\subsubsection{$SU(2)_L\times U(1)$ -- not even the Greatest Thing}
\label{SU(2)}
%%%%%%%%%%

The requirements of unitarity, which is nonnegotiable, and of renormalizability, which is to some degree, significantly restrict  possible theories of the electroweak interactions. There are other strong points as 
well among them: 

\noindent 
$\oplus$ Since there is a {\em single} $SU(2)_L$ group, there is a single set of gauge bosons. 
Their {\em self}-coupling controls also, how they couple to the fermion fields. As explain later in more detail, this implies the property of `weak universality'.

The generation of masses for gauge bosons and fermions is highly nontrivial, yet can be achieved 
through a feat of theoretical engineering that should be -- although rarely is -- referred to as 
Higgs-Brout-Englert-Guralnik-Hagen-Kibble mechanism. Intriguingly enough the Higgs  
doublet field can generate masses for gauge bosons and fermions alike. Those masses are controlled 
by a single vacuum expectation value 
(VEV) $\langle 0|\phi|0\rangle$ and, in the case of fermions, their Yukawa couplings -- a point we will return to.

Despite all the impressive, even  amazing successes of the SM the community is 
not happy with it for several reasons: Unlike for the strong interactions there is no uniqueness about the electroweak gauge group, 
it provides merely the minimal solution.  
With it being $SU(2)_L \times U(1)$, only partial unification has been 
achieved. 
%\footnote{A French saying describing a situation, where a decision is imposed on someone with no 
%explanation and no right of appeal.}, 
And then there is the whole issue of family replication.

%%%%%%%%%%
\subsubsection{The Family Mystery}
\label{FAMILY}
%%%%%%%%%

The twelve known quarks and leptons are arranged into three families. Those families possess identical gauge couplings and are distinguished only by their mass terms, i.e. their Yukawa couplings. 
We do not understand this family replication or why there are three families. It is not even clear 
whether the number of families represents a fundamental quantity  or is due to the more or less accidental interplay of complex forces as one encounters when analyzing the structure of nuclei. 
The only hope for a theoretical understanding we can spot on the horizon is superstring or 
M theory -- which is merely a euphemistic way of saying we hardly have a clue. 

Yet the circumstantial evidence that we miss completely a central element of Nature's `Grand Design' is even stronger in view of the strongly hierarchical pattern in the masses for up- and down-type quarks, charged leptons and neutrinos and the CKM parameters as discussed later.  In any case mass generation in particular for fermions -- including neutrinos -- and family replication constitute central 
mysteries of the SM, upon which the known gauge dynamics shed no light. It is for this reason that 
I had attached an $*$ to the term SM.

%%%%%%%%%%%%%%%
\subsection{Basics of \oc, \ot, \cp~ and \cpt}
\label{BASICSDISC}
%%%%%%%%%%

{\em Charge conjugation} exchanges particles $P$ and antiparticles $\bar P$ and thus flips the sign of all charges like electric charge, hyper-charge etc. It is described by a linear operator \oc. \cp~transformations include a parity operation as well. 

{\em Time reversal} is operationally defined as a reversal of motion 
\beq 
(\vec p, \vec l) \stackrel{\ot}{\longrightarrow} - (\vec p, \vec l)  \; , 
\eeq
which follows from $(\vec r, t) \stackrel{\ot}{\longrightarrow} (\vec r,-t)$. While the Euclidean scalar 
$\vec l_1 \cdot \vec p_2$ is invariant under the time reversal operator \ot, the triple correlations 
of (angular) momenta are not: 
\beq 
\vec v_1 \cdot (\vec v_2 \times \vec v_3)  \stackrel{\ot}{\longrightarrow} 
- \vec v_1 \cdot (\vec v_2 \times \vec v_3) \; \; {\rm with} \; \; \vec v = \vec p, \vec l \; . 
\eeq
The expectation value of such triple correlations accordingly are referred to as \ot~{\em odd moments}. 

In contrast to \op~ or \oc~ the \ot~operator is {\em anti}linear: 
\beq 
\ot (\alpha |a\rangle + \beta |b\rangle ) = \alpha ^* \ot |a \rangle + \beta ^* \ot |b \rangle
\eeq
This property of \ot~is required to have the commutation relation $[X,P] = i \hbar$ invariant 
under \ot: 
\beq 
\ot ^{-1} [X,P]\ot = - [X,P] \; \; \; {\rm and} \; \; \; 
\ot ^{-1} i\hbar \ot  = - i \hbar . 
\eeq
The anti-linearity of \ot~implies two important properties: 
\begin{itemize}
\item 
\ot~violation manifests itself through complex phases. \cpt~invariance then implies that also 
{\em \cp~violation enters through complex phases in the relevant couplings}. For \ot~or 
\cp~violation to become observable in a decay transition one thus needs the contribution 
from two different, yet coherent amplitudes.  
\item 
While a non-vanishing \op~odd moment establishes unequivocally \op~violation, this is {\em not} 
necessarily 
so for \ot~odd moments; i.e., even \ot~invariant dynamics can generate a non-vanishing 
\ot~odd moment. \ot~being antilinear comes into play when the transition amplitude is described 
{\em through second} 
(or even higher) order in the effective interaction, i.e. when final state interactions are 
included denoted symbolically by 
\beq 
\ot ^{-1} ({\cal L}_{eff}\Delta t + \frac{i}{2}({\cal L}_{eff}\Delta t)^2 + ...)\ot = 
{\cal L}_{eff}\Delta t -  \frac{i}{2}({\cal L}_{eff}\Delta t)^2 + ... \neq 
{\cal L}_{eff}\Delta t +  \frac{i}{2}({\cal L}_{eff}\Delta t)^2 + ... 
\eeq  
even for $[\ot, {\cal L}_{eff}]=0$.

\end{itemize}
%%%%%%%%%%%
\subsection{The Special Role of \cp~Invariance and its Violation}
\label{SPECIALCP}
%%%%%%%%%

While the discovery of \op~violation in the weak dynamics in 1957 caused a well documented shock 
in the community, even the theorists quickly recovered. Why then was the 
discovery of \cp~violation in 1964  through $K_L \to \pi^+\pi^-$ not viewed 
as a `deja vue all over again' in the language of Yogi Berra? I know 
of only one `heretic',  namely Okun, who in his 1963 text book \cite{OKUN63} explicitly listed the 
search for $K_L \to \pi^+\pi^-$ as a priority, i.e. one year before its discovery.

To see what moves physicists, one should {\em not} focus on what they 
say (rarely a good indicator for scientists in general), but on what they do. Point in case: 
How much this discovery shook the HEP community is best gauged by noting the efforts made to 
reconcile the observation of $K_L \to \pi^+\pi^-$ with \cp~invariance: 
\begin{itemize}
\item 
To infer \cp~violation from $K_L \to \pi \pi$  one has to invoke the superposition principle of 
quantum mechanics. One can introduce \cite{ROOS} {\em non}linear terms into the Schr\" odinger 
equation in such a way as to allow $K_L \to \pi^+\pi^-$ with \cp~invariant dynamics. While completely ad hoc, it is possible in principle. Such efforts were ruled out by further data, most decisively by 
$\Gamma (K^0(t) \to \pi^+\pi^-) \neq \Gamma (\bar K^0(t) \to \pi^+\pi^-)$. 
\item 
One can try to emulate the success of Pauli's neutrino hypothesis. An apparent violation of 
energy-momentum conservation had been observed in $\beta$ decay $n \to p e^-$, since 
the electron exhibited a {\em continuous} momentum spectrum. Pauli postulated that the reaction actually was 
\beq 
n \to p e^- \bar \nu 
\eeq  
with $\bar \nu$ a neutral and light particle that had escaped direct observation, yet led to a continuous 
spectrum for the electron. I.e., Pauli postulated a new particle -- and a most whimsical one at that -- to save a symmetry, namely the one under  translations in space and time responsible for the 
conservation of energy and momentum. Likewise it was suggested that the real reaction was 
\beq 
K_L \to \pi^+\pi^- U 
\label{UPART}
\eeq 
with $U$ a neutral and light particle with {\em odd} intrinsic \cp~parity. 
I.e., a hitherto unseen particle was introduced to save \cp~symmetry.  
This attempt at evasion was also soon rejected 
experimentally, as explained later. This represents an example of the ancient Roman saying: 

\begin{center} 
"Quod licet Jovi, non licet bovi." \\
"What is allowed [the supreme god] Jupiter, is not allowed a bull."
\end{center} 
I.e., we mere mortals cannot get away with speculations like `Jupiter' Pauli. 

\end{itemize}

There are several reasons itemized below for the physicists' reluctance to part with \cp~symmetry. 
For \op~violation being maximal (in the weak sector) -- 
{\em no right}-handed neutrinos couple to the weak interactions -- and likewise for \oc~violation, yet 
with their combined \cp~transformation describing an {\em exact} symmetry one had a natural `fall 
back' position, as described by Oscar Wilde: "... people are attracted to men with a future and women 
with a past ...". However the discovery of \cp~violation shattered this balanced picture. Furthermore even Luther's redemption of last resort  "peccate fortiter" ("sin boldly") could not be invoked, since  \cp~violation announced 
its arrival with a mere whimper: characterized by BR$(K_L \to \pi^+\pi^-) \simeq 0.0023$ or 
Im$M_{12}/M_K = 2.2 \cdot 10^{-17}$ it appears as the feeblest 
{\em observed} violation of any symmetry. \cp~symmetry as a `near-miss' is rather puzzling in view of its 
fundamental consequences listed next. 
\begin{itemize}
\item 
Let me start with an analogy from politics.  In my days as a student -- at a time long ago and a place far away -- politics was hotly debated. One of the subjects drawing out the greatest 
passions was the questions of what distinguished the `left' from the `right'. If you listened to it, you quickly found out that people almost universally defined `left' and `right' in terms of `positive' and 
`negative'. The only problem was they could not quite agree  who the good guys and the bad guys are. 

There arises a similar  conundrum  when considering decays like $\pi \to e \nu$. When saying 
that a pion decay produces a {\em left} handed charged lepton one had $\pi ^- \to e_L^- \bar \nu$ in 
mind. However $\pi^+ \to e^+_R \nu$ yields a {\em right} handed charged lepton.  
`Left' is thus defined in terms of `negative'. No matter how much \op~is violated, \cp~invariance 
imposes equal rates for these $\pi^{\pm}$ modes, and it is untrue to claim that nature makes an absolute 
distinction between `left' and `right'. The situation is analogous to the saying that `the thumb is 
left on the right hand' -- a correct, yet useless statement, since circular. 

\cp~violation is required to define `matter' vs. `antimatter', `left' vs. `right', `positive' vs. `negative' 
in a convention independent way. 

\item 
Due to the almost unavoidable \cpt~symmetry violation of \cp~implies one of \ot. 

\item 
It is one of the key ingredients in the Sakharov conditions for baryogenesis 
\cite{DOLGOV}: to obtain the 
observed baryon number of our Universe as a {\em dynamically generated} quantity rather than 
an arbitrary initial condition one needs baryon number violating transitions with \cp~violation to occur in a period, where our Universe had been out of thermal equilibrium. 

\end{itemize}

%%%%%%%%%%%%%%%%%
\subsection{On the Observability of \cp~Violation}
\label{OBCP}
%%%%%%%%%%%%

Since \cpt~symmetry confines \cp~violation to the emergence of complex phases, one needs two coherent, yet different amplitudes contribute to the same transition. The most successful 
realization of this requirement have been meson-antimeson oscillations, as described in 
Sect.\ref{CPBASICS}. The requirement can be met in a different way that can be implemented 
also for charged mesons $P$ (and baryons), when one has two amplitudes contribute with 
{\em both different weak phases and different strong phase shifts}:   
\bea 
T(P \to f) &=& e^{i\phi_{1,w}}e^{i\alpha_{1,s}}|{\cal M}_1| + 
e^{i\phi_{2,w}}e^{i\alpha_{2,s}}|{\cal M}_2|  \\  
T(\bar P \to \bar f) &=& e^{-i\phi_{1,w}}e^{i\alpha_{1,s}}|{\cal M}_1| + 
e^{-i\phi_{2,w}}e^{i\alpha_{2,s}}|{\cal M}_2|  
\eea
where I have factored out the weak and strong phases $\phi_{i,w}$ and $\alpha_{i,s}$. Then one finds for the direct \cp~asymmetry 
\beq 
\Gamma (\bar P \to \bar f) - \Gamma (P \to f) \propto 
{\rm sin}(\phi_{1,w} - \phi_{2,w})  {\rm sin}(\alpha_{1,s} - \alpha_{2,s})|{\cal M}_1||{\cal M}_2| 
\eeq
%%%%%%%%%%%%%%%%
\subsection{The Heroic Era -- \cp~Violation in $K_L$ Decays}
\label{HEROIC}
%%%%%%%%%%%%% 

%%%%%%%%%%%%%
\subsubsection{Basic Phenomenology}
\label{KPHEN}
%%%%%%%%%%%

The discussion here will be given in terms of strangeness $S$, yet can be generalized to any 
other flavour quantum number $F$ like beauty, charm, etc. 

Weak dynamics can drive $\Delta S = 1 \& 2$ transitions, i.e. decays and oscillations. 
While the underlying theory has to account  for both, it is useful to differentiate between them on the 
phenomenological level. The interplay between $\Delta S = 1\& 2$ affects also \cp~violation and how it 
can manifest itself. Consider $K_L \to \pi \pi $: while $\Delta S =2$ dynamics transform the flavour 
eigenstates $K^0$ and $\bar K^0$ into mass eigenstates $K_L$ and $K_S$, $\Delta S=1$ forces produce the decays into pions. 
\beq 
[K^0 \stackrel{\Delta S =2}{\longleftrightarrow} \bar K^0] \Rightarrow 
K_L \stackrel{\Delta S=1}{\longrightarrow} \pi \pi 
\eeq
Both of these reactions can exhibit \cp~violation, which is usually 
expressed as follows: 
\bea 
\nonumber 
\eta_{+-[00]} &\equiv& \frac{T(K_L \to \pi^+\pi^- [\pi^0\pi^0])}{T(K_S \to \pi^+\pi^- [\pi^0\pi^0])} \\
\eta_{+-} &\equiv& \epsilon_K + \epsilon^{\prime}\; , \; \; \; 
\eta_{00} \equiv \epsilon_K - 2 \epsilon^{\prime} 
\eea
Both $\eta_{+-}, \eta_{00} \neq 0$ signal \cp~violation; $\epsilon_K$ is common to both 
observables and reflects the \cp~properties of the state mixing that drives oscillations, 
i.e. in $\Delta S =2$ dynamics; 
$\epsilon^{\prime}$ on the other hand differentiates between the two final states and parametrizes 
\cp~violation in $\Delta S =1$ dynamics. With an obvious lack in Shakespearean flourish 
$\epsilon_K \neq 0$ is referred to as `indirect' or `superweak' \cp~violation and 
$\epsilon^{\prime}\neq 0$ as `direct' \cp~violation. As long as \cp~violation is seen only through a 
single mode of a neutral meson -- in this case {\em either} $K_L \to \pi^+\pi^-$ {\em or} 
$K_L \to \pi^0\pi^0$ -- the  distinction between direct and indirect \cp~violation is somewhat 
arbitrary, as explained later for $B_d$ decays.

Five types of \cp~violating observables have emerged reflecting the fact that $K_L$ is not 
an exact mass eigenstate and involving $K^0 - \bar K^0$ oscillations in one way or another:  
\begin{itemize}
\item 
{\em Existence} of a transition: $K_L \to \pi^+\pi^-,\, \pi^0\pi^0$; 
\item 
An {\em asymmetry} due to the {\em initial} state: $K^0 \to \pi^+\pi^-$ vs. $\bar K^0 \to \pi^+\pi^-$; 
\item 
An {\em asymmetry} in the {\em final} state: $K_L \to l^+\nu \pi^-$ vs. $K_L \to l^- \bar \nu \pi^+$, 
$K_L \to \pi^+\pi^-$ vs. $K_L \to \pi^0 \pi^0$; 
\item 
A {\em microscopic} \ot~asymmetry: rate$(K^0 \to \bar K^0)$ $\neq$ rate$(\bar K^0 \to K^0)$; 
\item 
A \ot~{\em odd correlation} in the final state: $K_L \to \pi^+\pi^- e^+e^-$. 

\end{itemize}

\noindent {\bf (i)} 
Using today's numbers \cite{PDG06} 
\beq 
{\rm BR}(K_L \to \pi^+\pi^-) = (1.976  \pm 0.008 ) \cdot 10^{-3}
\label{BRKPIPI}
\eeq
one derives 
\beq 
|\eta_{+-}| = (2.236 \pm 0.007)\cdot 10^{-3} \; ; 
\eeq 
It allows to describe also the asymmetry in semileptonic $K_L$ decays 
\beq 
\delta _l  \equiv  
\frac{\Gamma (K_L \to l^+\nu \pi^-) - \Gamma (K_L \to l^- \bar \nu \pi^+)}
{\Gamma (K_L \to l^+\nu \pi^-) + \Gamma (K_L \to l^- \bar \nu \pi^+)} = (3.32 \pm 0.06) \cdot 10^{-3} 
\label{CHARDEF} \; , 
\eeq
since 
\beq 
\left. \delta _l \right| _{K_L \to 2\pi}   
\simeq 2 {\rm Re}\epsilon = (3.16 \pm 0.01) \cdot 10^{-3} 
\label{CHARDEF2}
\eeq

\noindent {\bf (ii)} 
\cpt~invariance -- an (almost) inescapable property of relativistic local quantum field  theories -- 
tells us that for every violation of \cp~symmetry there has to be a commensurate one for 
\ot~invariance. Verifying this statement experimentally is far from straightforward though. For in a decay process 
$A \to B +C$ practical considerations prevent one from creating 
the time reversed sequence $B+C \to A$. 

Meson-antimeson oscillations (and likewise neutrino oscillations) provide unique opportunities to probe 
\ot~violations. For one can compare directly the rates for $K^0 \Rightarrow \bar K^0$ and 
$\bar K^0 \Rightarrow K^0$, which is referred to as `Kabir test' \cite{PASHA}, listed as the fourth item above. For that purpose one has to determine the flavour of the final state 
-- $K^0$ or $\bar K^0$ -- as well as 
tag the flavour of the initial one. Semileptonic channels can achieve the former through the 
SM $\Delta S = \Delta Q$ selection rule. For the latter one can rely on associated production in, 
say, proton-antiproton annihilation: $p \bar p \to K^+\bar K^0 \pi^-$ vs. 
$p \bar p \to K^- K^0 \pi^+$.  I.e., one compares the sequences 
$p\bar p \to K^+\bar K^0 \pi^- \to K^+(l^+\nu \pi^-)\pi^-$ and 
$p\bar p \to K^-K^0 \pi^+ \to K^-(l^-\bar \nu \pi^-)\pi^-$. 
Using this technique the CPLEAR collaboration found \cite{CPLEARPAP}: 
\beq 
A_T = \frac{{\rm rate}(\bar K^0 \to K^0) - {\rm rate}(K^0 \to \bar K^0)}
{{\rm rate}(\bar K^0 \to K^0)+ {\rm rate}(K^0 \to \bar K^0)} = (6.6 \pm 1.6) \cdot 10^{-3} 
\eeq
in full agreement with what is inferred from BR$(K_L \to \pi^+\pi^-)$: 
\beq 
\left. A_T\right| _{K_L \to 2\pi}  \simeq 4 {\rm Re}\epsilon = (6.32 \pm 0.02) \cdot 10^{-3}
\eeq

\noindent {\bf (iii)} 
The fifth item in the list above has a novel aspect to it: it reflects a \cp~asymmetry in a 
final state distribution. Consider $K_L \to \pi^+\pi^- e^+e^-$ and define 
$\phi$ as the angle between the planes 
spanned by 
the two pions and the two leptons in the $K_L$ 
restframe:  
\beq    
\phi \equiv \angle ( \vec n_l, \vec n_{\pi})\; , \;   
\vec n_l = \vec p_{e ^+}\times \vec p_{e ^-}/
|\vec p_{e ^+}\times \vec p_{e ^-}| \; , \;  
\vec n_{\pi} = \vec p_{\pi ^+}\times \vec p_{\pi ^-}/ 
|\vec p_{\pi ^+}\times \vec p_{\pi ^-}| \; . 
\label{PHISEHGAL}
\eeq    
One analyzes 
the decay rate as a function of $\phi$: 
\beq 
\frac{d\Gamma}{d\phi} = \Gamma _1 {\rm cos}^2\phi + 
\Gamma _2 {\rm sin}^2\phi + 
\Gamma _3 {\rm cos}\phi \, {\rm sin} \phi 
\eeq 
Since  
\beq 
{\rm cos}\phi \, {\rm sin} \phi = 
(\vec n_l \times \vec n_{\pi}) \cdot 
(\vec p_{\pi ^+} + \vec p_{\pi ^-}) 
(\vec n_l \cdot \vec n_{\pi})/
|\vec p_{\pi ^+} + \vec p_{\pi ^-}| 
\eeq
one notes that 
\beq 
{\rm cos}\phi \, {\rm sin} \phi \; \; \; 
\stackrel{{\bf T},{\bf CP}}{\longrightarrow} \; \; \; 
- \; {\rm cos}\phi \, {\rm sin} \phi 
\eeq    
under both \ot~ and \cp~transformations; i.e. the observable  
$\Gamma _3$ represents a \ot- and \cp-odd correlation. 
It can be projected out by comparing the $\phi$ 
distribution integrated over two quadrants: 
\beq 
A = 
\frac{\int _0^{\pi/2} d\phi \frac{d\Gamma}{d\phi} - 
\int _{\pi /2}^{\pi} d\phi \frac{d\Gamma}{d\phi}}
{\int _0^{\pi} d\phi \frac{d\Gamma}{d\phi}} = 
\frac{2\Gamma _3}{\pi (\Gamma _1 + \Gamma _2)} 
\eeq
It was first measured by KTEV and then confirmed by NA48 \cite{PDG06}:  
\beq 
A = (13.7 \pm 1.5)\% \, .
\label{KTEVSEHGAL2}
\eeq 
$A\neq 0$ is induced by $\epsilon_K$, the \cp~violation in the $K^0 - \bar K^0$ mass matrix, 
leading to the prediction \cite{SEGHALKL}
\beq 
A = (14.3 \pm 1.3)\% \, .
\eeq
The observed value for the \ot~odd moment $A$ is fully consistent with \ot~violation. Yet 
$A\neq 0$ {\em by itself} does not establish 
\ot~violation \cite{BSTODD}. 

It is actually easy to see how this sizable forward-backward asymmetry is generated from the tiny quantity 
$|\eta_{+-}| \simeq 0.0023$. For 
$K_L \to \pi ^+ \pi ^- e^+ e^-$ is driven by the two sub-processes 
\bea 
K_L &\stackrel{\not {\cp} \&\Delta S =1}{\longrightarrow} \pi^+\pi^- 
\stackrel{E1}{\longrightarrow} \pi^+\pi^- \gamma ^* \to \pi ^+ \pi ^- e^+ e^- 
\\
K_L &\stackrel{M1\& \Delta S =1}{\longrightarrow} \pi^+\pi^- \gamma ^* \to \pi ^+ \pi ^- e^+ e^- \; , 
\eea
where the first reaction is suppressed, since it requires \cp~violation in  
$K_L \to 2\pi$, and the second one, since it involves an $M1$ transition.  Those two a priori very 
different suppression mechanisms happen to yield comparable amplitudes, which thus generate 
sizable interference. The price one pays is the small branching ratio, namely 
BR$(K_L \ra \pi ^+ \pi ^- e^+ e^-) = 
(3.32 \pm 0.14 \pm 0.28 ) \cdot 10^{-7}$.  I will revisit the issue of \cp~violation in final state distributions. 

\begin{figure}[t]
\vspace{7.0cm}
\includegraphics{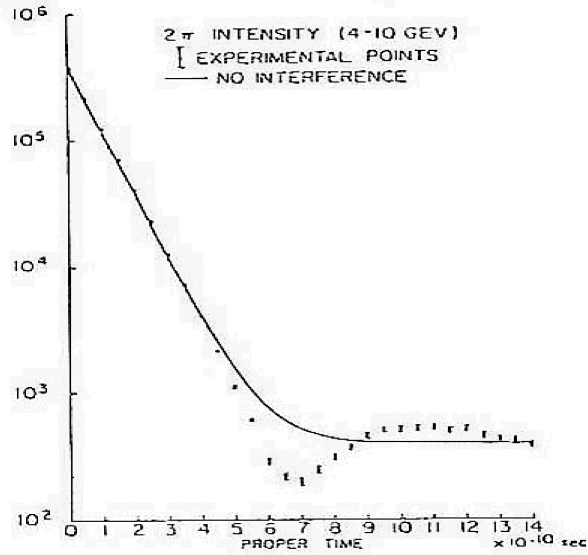}
\caption{ $K^0(t) \to \pi^+\pi^-$ with $K_S-K_L$ interference.  \label{CPINTfig} }
\end{figure}

\noindent 
{\bf (iv)} The fact that $K^0 - \bar K^0$ oscillations are involved in an essential way can most directly be established by analyzing $K^0 (t) \to \pi^+\pi^-$ -- i.e. the decay rate evolution as a function of the (proper) time of decay $t$ of a beam that initially 
contained only $K^0$ mesons, see Fig.\ref{CPINTfig}. Comparing it with its \cp~conjugate 
$\bar K^0 (t) \to \pi^+\pi^-$, as listed as the second item in the 
list above,  one finds clear $t$ {\em dependent} asymmetries in both the pure $K_L$ as well as the $K_S-K_L$ interference domains, though not in the pure $K_S$ domain. Studying just 
$K^0 (t) \to \pi^+\pi^-$ is actually sufficient to establish \cp~violation due to the following 

\begin{center}
{\bf Theorem:}
\end{center}

\noindent If one finds that the evolution of the decays of an arbitrary linear combination of neutral 
mesons into a \cp~{\em eigenstate} as a function of (proper) time of decay cannot be described by a {\em single exponential}, then \cp~invariance must be broken. Or formulated more concisely for the case at hand: 
\beq 
\frac{d}{dt}e^{\Gamma t}\Gamma (K^{neut}(t) \to \pi^+\pi^-) \neq \; 0 \; \; {\rm for\; all 
\; real}\; \Gamma 
\Longrightarrow \; \; \cp~{\rm violation!}
\label{EQTH}
\eeq
The proof is elementary: With \cp~being conserved, mass eigenstates have to be 
either even or odd \cp~eigenstates as well and can decay only into final states of the same \cp~parity. 
Their decay rate evolution thus has to be given by a single exponential in time; q.e.d.

The fact that the curve in Fig.\ref{CPINTfig} shown an interference region between the practically 
pure $K_S$ and $K_L$ regimes provided the conclusive evidence against one of the aforementioned 
attempts to maintain \cp~symmetry by postulating that $K_L \to \pi^+\pi^- U$ was occurring, 
see Eq.(\ref{UPART}): for there 
can be no interference between $K_S \to \pi^+\pi^-$ and $K_L \to \pi^+\pi^- U$.

%%%%%%%%%%%%
\subsubsection{Completion of the Heroic Era: Direct \cp~Violation}
\label{EPSPRIME}
%%%%%%%%%%%

In the decades after 1964 dedicated searches for {\em direct} 
\cp~violation were undertaken. Measurements launched in the 1980's 
yielded intriguing, though not conclusive evidence:  
\beq 
\frac{\epsilon ^{\prime}}{\epsilon_K} = 
\left\{ 
\begin{array}{ll} 
(2.30 \pm 0.65)\cdot 10^{-3} & {\rm NA31} \\
(0.74 \pm 0.59)\cdot 10^{-3} & {\rm E731}
\end{array}
\right.
\eeq
{\em Direct} CP violation 
has been unequivocally established in 1999. The present world average dominated 
by the data from NA48 and KTeV reads as follows \cite{SOZZI}: 
\beq 
\langle \epsilon ^{\prime}/\epsilon _K \rangle = 
(1.63 \pm 0.22) \cdot 10^{-3} 
\eeq 
Quoting the result in this way does not do justice to the experimental 
achievement, since $\epsilon _K$ is a very small number itself. 
The sensitivity achieved becomes more obvious when quoted in terms of actual 
widths \cite{SOZZI}:
\beq 
\frac{\Gamma (K^0 \to \pi ^+ \pi ^-) - 
\Gamma (\bar K^0 \to \pi ^+ \pi ^-)}
{\Gamma (K^0 \to \pi ^+ \pi ^-) + 
\Gamma (\bar K^0 \to \pi ^+ \pi ^-)} = 
(5.04 \pm 0.82) \cdot 10^{-6} \; !
\label{DIRECTCPV}
\eeq
This represents a discovery of the very first rank 
\footnote{As a consequence of Eq.(\ref{DIRECTCPV}) I am not impressed by \cpt~tests falling short 
of the $10^{-6}$ level.}. Its significance does not depend on whether the  
SM can reproduce it or not -- which is the most concise confirmation of 
how important it is. 
The HEP community can take pride 
in this achievement; the tale behind it is a most fascinating one about imagination and perseverance. The two groups and their predecessors deserve our respect; they 
have certainly earned my admiration.

%%%%%%%%%%%%%%
\subsection{CKM Dynamics -- an `Accidental Miracle'}
\label{CKMDYN}
%%%%%%%%%%%%

The  existence of three quark-lepton families that differ only in their mass related parameters -- 
and within the SM thus only in their Yukawa couplings -- is one of the profound puzzles about 
the SM. The latter's Yukawa sector is indeed its most unsatisfactory feature. Yet  
this three family structure is an observed fact, and it gives rise to a very rich phenomenology in 
weak dynamics based on a huge body of data -- including \cp~violation -- that so far is fully consistent with the SM's predictions.

\cp~violation was discovered in 1964 through the observation of $K_L \to \pi^+\pi^-$, yet it was not realized for a number of years that dynamics known at {\em that} time could {\em not} generate it. 
We should not be too harsh on our predecessors for that oversight: as long as one did not have a 
renormalizable theory for the weak interactions and thus had to worry about {\em infinities} 
in the calculated rates, one can be excused for ignoring a seemingly marginal rate with a branching 
ratio of $2\cdot 10^{-3}$. Yet even after the emergence of the renormalizable 
Glashow-Salam-Weinberg model its {\em phenomenological} incompleteness was not recognized right 
away. There is a short remark by Mohapatra in a 1972 paper invoking the need for right-handed 
currents to induce \cp~violation. 

It was the 1973 paper by Kobayashi and Maskawa \cite{KM} that fully stated the inability of even a 
two-family SM to produce \cp~violation and that explained what had to be added to it: right-handed 
charged currents, extra Higgs doublets -- or (at least) a third quark family. Of the three options 
Kobayashi and Maskawa listed, their name has been attached only to the last one as the CKM 
description. They were helped by the `genius loci' of Nagoya University: 
\begin{itemize}
\item 
Since it was the home of the Sakata school and the Sakata model of elementary particles quarks 
were viewed as physical degrees of freedom from the start. 
\item 
It was also the home of Prof. Niu who in 1971 had observed 
\cite{NIU} a candidate for a charm decay in emulsion 
exposed to cosmic rays and actually recognized it as such. The existence of charm, its association 
with strangeness and thus of two complete quark families were thus taken for granted at Nagoya. 

\end{itemize}
Their argument went as follows. The six quark flavours of the SM are arranged in three up-type and three down-type quarks fields 
that can be written as vectors $U^F= (u,c,t)^F$ and $D^F=(d,s,b)^F$, respectively,  
in terms of the {\em flavour} eigenstates denoted 
by the superscript $F$. One can form two $3\times 3$ mass matrices  
\beq 
{\cal L}_M \propto \bar U_L^F{\cal M}_U U_R^F + \bar D_L^F{\cal M}_D D_R^F  \; . 
\label{MASSLAG}
\eeq
There is no a priori reason 
why the matrices ${\cal M}_{U/D}$ should be diagonal. Applying bi-unitary rotations 
${\cal J}_{U/D,L}$ will 
allow to diagonalize them
\beq 
{\cal M}_{U/D}^{\rm diag} =     {\cal J}_{U/D,L} {\cal M}_{U,D}{\cal J}_{U/D,R}^{\dagger}
\eeq
and obtain the {\em mass} eigenstates of the quark fields: 
\beq 
U^m_{L/R} = {\cal J}_{U,L/R} U_{L/R}^F \; , \; \; D^m_{L/R} = {\cal J}_{D,L/R} D_{L/R}^F
\eeq
I.e., the flavour eigenstates `mix' to form the  mass eigenstates. 
The eigenvalues of ${\cal M}_{U/D}$ represent the masses of the quark fields. The measured 
values exhibit a very peculiar hierarchical pattern for up- and down-type quarks, charged and neutral leptons that hardly appears to be accidental. 

There is much more to it. The neutral current coupling keeps its form when expressed in terms of the mass eigenstates: 
\beq 
{\cal L}_{NC}^{U[D]} \propto \bar g_Z \bar U^F[\bar D^F] \gamma _{\mu} U^F[D^F] Z^{\mu} \; \; \; 
\Longrightarrow  \; \; \; 
{\cal L}_{NC}^{U[D]} \propto \bar g_Z \bar U^m[\bar D^m] \gamma _{\mu} U^m[D^m] Z^{\mu} \; ; 
\label{LAGNC2} 
\eeq 
i.e., there are {\em no tree-level} flavour changing neutral currents. This important property is 
referred to 
as the `generalized' GIM mechanism \cite{GIM}. 

%Since meson-antimeson oscillations require dynamics that changes the flavour quantum number 
%by two units, but not the electric charge, one has to turn to weak 
%{\em charged} currents. Iterating them to second order leads to {\em effective} 
%flavour changing neutral currents. 
The charged currents do change their form when going from  
flavour to mass eigenstates; 
\beq 
{\cal L}_{CC} \propto \bar g_W \bar U^F_L \gamma _{\mu} D^F W^{\mu} = 
\bar g_W \bar U^m_L \gamma _{\mu} V_{CKM}D^mW^{\mu} \; \; , \; \; \; 
V_{CKM} = {\cal J}_{U,L}{\cal J}_{D,L}^{\dagger}  
\label{LAGCC}
\eeq
While the matrix $V_{CKM}$ has to be unitary within the SM, there is no reason why it should be 
the identity matrix or even diagonal \footnote{Even if some speculative dynamics were to enforce an alignment 
between the $U$ and $D$ quark fields at some high scales causing their mass matrices 
to get diagonalized by the same bi-unitary transformation, this alignment would probably 
get upset by renormalization down to the electroweak scales.}.  
It means the charged 
current couplings of the mass eigenstates will be modified in an observable way. In which way and by how much this happens requires further analysis since the phases of fermion fields are not necessarily 
observables. An $N\times N$ unitary matrix contains $N^2$ independent real parameters. Since the 
phases of quark fields like other fermion fields can be rotated freely, $2N-1$ phases can be removed 
from ${\cal L}_{CC}$ reducing the number of {\em independent physical} parameters to 
$(N-1)^2$. An $N\times N$ {\em orthogonal} matrix has $N(N-2)/2$ angles; thus we can conclude 
that an $N\times N$ {\em unitary} matrix contains $(N-1)(N-2)/2$ {\em physical phases} in addition. 
Accordingly for $N=2$ families we have only the Cabibbo angle and no phases, while for $N=3$ 
we obtain three angles and one irreducible phase. I.e., a three family ansatz can support \cp~violation 
with a single source -- the `CKM phase'. 

For three families the unitarity of the CKM matrix 
\beq 
{\bf V}_{CKM} = 
\left(  
\begin{array}{ccc} 
V(ud) & V(us) & V(ub) \\
V(cd) & V(cs) & V(cb) \\
V(td) & V(ts) & V(tb) 
\end{array} 
\right) 
\eeq
yields three {\em universality} relations 
\beq 
\sum_{j=d,s,b}|V(ij)|^2 = 1 \; , \; \; i=u,c,t
\label{UNIV}
\eeq
as well as six {\em orthogonality} conditions
\beq 
\sum_{j=u,c,t} V^*(ji)V(jk) = 0 \; , \; \; i\neq k = d,s,b \; . 
\label{ORTHO}
\eeq
Eqs.(\ref{ORTHO}) represent triangle relations in the complex plane, a point I will repeatedly return to. Changing the phase conventions 
for the quark fields will change the orientations of these triangles in the complex plane, but 
{\em not} their internal angles. Those represent the {\em relative} phases of the elements 
of $V_{CKM}$, which in turn can give rise to observable \cp~asymmetries. 

This graphic interpretation also makes it transparent, why the charged currents cannot generate 
\cp~violation with two families. In that case the orthogonality relations of the corresponding $2\times 2$ matrix are trivial stating that two products of a priori complex matrix elements had to add to zero, i.e. cannot exhibit a {\em nontrivial relative} phase. 

For the three families of the SM there are six triangles. They can and do vary greatly in their shapes, 
yet have to possess the same area -- a consequence of there being just a single 
CKM phase, as stated above. PDG suggests the following parametrization:  
\beq 
{\bf V}_{CKM} =  
\left( 
\begin{array}{ccc} 
c_{12}c_{13} & s_{12}c_{13} & s_{13}e^{-i \delta _{13}}  \\
- s_{12}c_{23} - c_{12}s_{23}s_{13}e^{i \delta _{13}} &
c_{12}c_{23} - s_{12}s_{23}s_{13}e^{i \delta _{13}} & 
c_{13}s_{23} \\
s_{12}s_{23} - c_{12}c_{23}s_{13}e^{i \delta _{13}} &
- c_{12}s_{23} - s_{12}c_{23}s_{13}e^{i \delta _{13}} &
c_{13}c_{23} 
\end{array}
\right) 
\label{PDGKM} 
\eeq 
where 
\beq 
c_{ij} \equiv {\rm cos} \theta _{ij} \; \; , \; \;  
s_{ij} \equiv {\rm sin} \theta _{ij}
\eeq  
with $i,j = 1,2,3$ label the families. 

This is a completely general, yet not unique representation: a 
different set of 
Euler angles could be chosen; the phases can be shifted around 
among the matrix elements 
by using a different phase convention.

The CKM implementation of \cp~violation depends on the form of the quark mass matrices 
${\cal M}_{U,D}$, not so much on how those are generated. Nevertheless something can be inferred 
about the latter: within the SM all fermion masses are driven by a {\em single} vacuum expectation 
value of a neutral Higgs field (VEV); to obtain an 
irreducible relative phase between different quark couplings thus requires such a phase in 
quark Yukawa couplings; this means that in the SM \cp~violation arises in dimension-{\em four} 
couplings, i.e., is `hard' in the language of quantum field theory. 

%%%%%%%%%%%%
\subsection{`Maximal' \cp~Violation?}
\label{MAXCPV}
%%%%%%%%

As already mentioned charged current couplings with their $V-A$ structure break parity and 
charge conjugation {\em maximally}. Since due to \cpt~invariance \cp~violation is expressed 
through couplings with complex phases, one might say that maximal \cp~violation 
is characterized by complex phases of $90^o$. However this would be fallacious: for by changing 
the phase {\em convention} for the quark fields one can change the phase of a given 
CKM matrix element and even rotate it away; it will of course re-appear in other matrix elements. 
For example $|s\rangle \to e^{i\delta _s}|s\rangle$ leads to $V_{qs} \to  e^{i\delta _s}V_{qs}$ 
with $q=u,c,t$. In that sense the CKM phase is like the `Scarlet Pimpernel': "Sometimes here, 
sometimes there, sometimes everywhere." 

One can actually illustrate with a general argument, why there can be no straightforward definition for maximal \cp~violation. Consider neutrinos: Maximal \cp~violation means there are 
$\nu _L$ and $\bar \nu _R$, yet no $\nu _R$ or $\bar \nu _L$  
\footnote{To be more precise: $\nu_L$ and $\bar \nu _R$ couple to weak gauge bosons, 
$\nu _R$ or $\bar \nu _L$ do not.}. Likewise there are $\nu_L$ and $\bar \nu_R$, but 
not $\bar \nu _L$ or $\nu _R$. One might then suggest that maximal \cp~violation means 
that $\nu_L$ exists, but $\bar \nu _R$ does not. Alas -- \cpt~invariance already enforces the existence 
of both. 

Similarly -- and maybe more obviously -- it is not clear what maximal \ot~violation would mean although 
some formulations have entered daily language like the `no future generation', the `woman 
without a past' or the `man without a future'. 

%%%%%%%%%%%%%%%
\section{The SM's Paradigm of Large \cp~Violation in $B$ Decays before 2000}
\label{PARA2000}
%%%%%%%%%

%%%%%%%%%%%%
\subsection{Basics}
\label{CPBASICS}
%%%%%%%%%%

Since \cp~violation enters the dynamics through complex couplings, one needs two different 
amplitudes contribute coherently to a reaction for an observable \cp~asymmetry to emerge. In 1979 it was pointed out that 
$B^0 - \bar B^0$ oscillations are well suited to satisfy this requirement for final states $f$  
that can be fed both by $B^0$ and $\bar B^0$ decays, in particular since those oscillation 
rates were expected to be sizable \cite{CARTER}: 
\beq  
B^0 \Rightarrow \bar B^0 \to f \leftarrow B^0  \hspace{1cm} vs. \hspace{1cm} 
\bar B^0 \Rightarrow B^0 \to \bar f \leftarrow \bar B^0
\label{BASICIDEA}
\eeq
In 1980 it was predicted \cite{BS80} that in particular $B_d \to \psi K_S$ should exhibit a 
\cp~asymmetry larger by two orders of magnitude than the corresponding one in  
$K^0 \to 2\pi$ vs. $\bar K^0 \to 2 \pi$, {\em if} CKM theory provides the main driver of $K_L \to \pi^+\pi^-$;  
even values close to 100 \% were suggested as conceivable. The analogous mode 
$B_s \to \psi \phi$ should however show an asymmetry not exceeding the few percent level. 

It was also suggested that in rare modes like $\bar B_d \to K^- \pi^+$ sizable {\em direct} \cp~violation 
could emerge due to intervention of `Penguin' operators \cite{SONI}. 

We now know that these predictions were rather prescient. It should be noted that at the time 
of these predictions very little was known about $B$ mesons. 
While their existence had been inferred from the 
discovery of the $\Upsilon (1S - 4S)$ family at FNAL in 1977ff,  none of their exclusive decays had been 
identified, and their lifetime were unknown as were a forteriori their oscillation rates. While 
very little was thus known about the values of the contributing CKM parameters the 
relevant formalism for \cp~asymmetries involving $B^0 - \bar B^0$ oscillations was already fully given 
inspired by what we had learnt from the $K^0 - \bar K^0$ complex. 

To describe oscillations in the presence of \cp~violation one applies the Weisskopf-Wigner 
approximation \cite{WIGNERAPPROX} and turns to solving a nonrelativistic 
Schr\" odinger equation, which I formulate for the general case of a pair of neutral 
mesons $P^0$ and $\bar P^0$ with flavour quantum number $F$; it can denote 
a $K^0$, $D^0$ or $B^0$:  
\beq 
i\frac{d}{dt} \left( 
\begin{array}{ll}
P^0 \\
\bar P^0
\end{array}  
\right)  = \left( 
\begin{array}{ll}
M_{11} - \frac{i}{2} \Gamma _{11} & 
M_{12} - \frac{i}{2} \Gamma _{12} \\ 
M^*_{12} - \frac{i}{2} \Gamma ^*_{12} & 
M_{22} - \frac{i}{2} \Gamma _{22} 
\end{array}
\right) 
\left( 
\begin{array}{ll}
P^0 \\
\bar P^0
\end{array}  
\right) 
\label{SCHROED} 
\eeq
\cpt~ invariance imposes 
\beq 
M_{11}= M_{22} \; \; , \; \; \Gamma _{11} = \Gamma _{22} \; . 
\label{CPTMASS}
\eeq  

The subsequent discussion might strike the reader as overly 
technical, yet I hope (s)he will bear with me since 
these remarks will lay important groundwork for a proper 
understanding of \cp~asymmetries in $B$ decays. 

The mass eigenstates obtained through diagonalising this matrix 
are given by (for details see \cite{CPBOOK}) 
\beq 
 |P_{A[B]}\rangle = 
\frac{1}{\sqrt{|p|^2 + |q|^2}} \left( p |P^0 \rangle + [-] 
q |\bar P^0\rangle \right)
\label{P1P2}
\eeq 
with eigenvalues 
\beq 
M_{A[B]} - \frac{i}{2}\Gamma _{A[B]} 
= M_{11} - \frac{i}{2} \Gamma _{11} 
+[-]\frac{q}{p}\left( M_{12} - \frac{i}{2} \Gamma _{12}\right) 
\label{EV} 
\eeq 
as long as 
\beq 
\left( \frac{q}{p}\right) ^2 = 
\frac{M_{12}^* - 
\frac{i}{2} \Gamma ^*_{12}}
{M_{12} - 
\frac{i}{2} \Gamma _{12}} 
\label{Q/PSQ} 
\eeq 
holds. I am using letter subscripts 
$A$ and $B$ for labeling the 
mass eigenstates rather than numbers $1$ and $2$ 
as it is usually done. For I want to 
avoid confusing them with the matrix indices 
$1,2$ in $M_{ij} - \frac{i}{2}\Gamma _{ij}$. 
 
Eqs.(\ref{EV}) yield for the differences in mass and width 
\bea 
\Delta M &\equiv& M_B - M_A = 
-2 {\rm Re} \left[ \frac{q}{p}(M_{12} - 
\frac{i}{2}\Gamma _{12})\right]  \\
\Delta \Gamma  &\equiv& \Gamma _A - 
\Gamma _B = 
-2 {\rm Im}\left[ \frac{q}{p}(M_{12} - 
\frac{i}{2}\Gamma _{12})\right] 
\label{DELTAEV} 
\eea
Note that the subscripts $A$, $B$ have been swapped in 
going from $\Delta M$ to $\Delta \Gamma$. This is 
done to have both quantities {\em positive} 
for kaons. 

In expressing the mass eigenstates $P_A$ and $P_B$ 
explicitely in terms of the flavour eigenstates -- 
Eq.(\ref{P1P2}) -- one needs $\frac{q}{p}$. There 
are two solutions to Eq.(\ref{Q/PSQ}):  
\beq 
\frac{q}{p} = \pm 
\sqrt{\frac{M_{12}^* - \frac{i}{2} \Gamma _{12}^*}
{M_{12} - \frac{i}{2} \Gamma _{12}}}
\label{Q/P} 
\eeq
There is actually a more general ambiguity than this 
binary one. For antiparticles
are defined up to a phase only: 
\beq 
{\bf CP} |P^0 \rangle = 
\eta | \bar P^0 \rangle  \; \; \; {\rm with} 
\; \; |\eta| =1
\eeq 
Adopting a different phase convention will change 
the phase for $M_{12} - \frac{i}{2} \Gamma _{12}$ 
as well as 
for $q/p$: 
\beq 
|\bar P^0 \rangle \ra e^{i\xi}|\bar P^0 \rangle \; 
\Longrightarrow \; 
(M_{12}, \Gamma _{12}) \ra e^{i\xi} 
(M_{12}, \Gamma _{12}) \; 
\& \; 
\frac{q}{p} \ra e^{-i\xi} \frac{q}{p} \; , 
\eeq
yet leave $(q/p) (M_{12} - \frac{i}{2} \Gamma _{12})$ 
invariant -- as it has to be since the eigenvalues, 
which are observables, depend on this combination, see 
Eq.(\ref{EV}). Also $\left| \frac{q}{p}\right|$ is an 
observable; its {\em deviation} from unity is one 
measure of \cp~violation in $\Delta F =2$ dynamics.

By {\em convention} most authors pick the 
{\em positive} sign in Eq.(\ref{Q/P}) 
\beq 
\frac{q}{p} = +  
\sqrt{\frac{M_{12}^* - \frac{i}{2} \Gamma _{12}^*}
{M_{12} - \frac{i}{2} \Gamma _{12}}} \; . 
\label{QPPOS} 
\eeq 
Up to this point the two states 
$|P_{A,B}\rangle$ are merely 
{\em labelled} by their subscripts. 
Indeed $|P_A\rangle$ and $|P_B\rangle$ switch places 
when selecting the minus rather than the plus sign in 
Eq.(\ref{Q/P}). 

One can define the labels $A$ and $B$ such that $\Delta M \equiv M_B - M_A > 0$  
is satisfied. Once this {\em convention} 
has been adopted, it becomes a sensible question 
whether $ \Gamma _B > \Gamma _A$ or 
$\Gamma _B < \Gamma _A$  
holds, i.e. whether the heavier state is shorter or 
longer lived. 

One can write the general 
mass eigenstates in terms of the 
\cp~eigenstates as well: 
\bea 
|P_A \rangle &=& \frac{1}{\sqrt{1+|\bar \epsilon |^2}} 
\left( |P_+ \rangle + \bar \epsilon |P_-\rangle \rangle 
\right)  
\; \; \; , \; CP|P_{\pm}\rangle = \pm |P_{\pm}\rangle \\  
|P_B \rangle &=& \frac{1}{\sqrt{1+|\bar \epsilon |^2}} 
\left( |P_- \rangle + \bar \epsilon |P_+\rangle \rangle 
\right) 
\; ; 
\eea
$\bar \epsilon = 0$ means that the mass and 
\cp~eigenstates coincide, i.e. \cp~is conserved in 
$\Delta F=2$ dynamics driving $P^0 - \bar P^0$ 
oscillations. With the phase between the 
orthogonal states $|P_+\rangle$ and 
$|P_-\rangle$ arbitrary, the phase of 
$\bar \epsilon$ can be changed at will and is not an 
observable; $\bar \epsilon$ can be expressed in terms of 
$\frac{q}{p}$, yet in a way that depends on the 
convention for the phase of antiparticles. For 
${\bf CP}|P^0\rangle = |\bar P^0\rangle$ one has 
\beq 
\bar \epsilon = 
\frac{1 - \frac{q}{p}}{1 +  \frac{q}{p}} 
\eeq  

Later I will describe how to evaluate $M_{12}$ and thus 
also $\Delta M$ within a given 
theory for the $P^0-\bar P^0$ complex. The examples just listed 
illustrate that some care has to be applied in interpreting 
such results. For expressing mass eigenstates explicitely 
in terms of flavour eigenstates involves some conventions. 
Once adopted we have to stick with a convention; yet our 
original choice cannot influence observables.  

Decay rates for \cp~conjugate channels can be expressed as follows: 
\beq   
{\rm rate} (B^0(t)[\bar B^0] \to f[\bar f]) = e^{-\Gamma _Bt}G_f(t)[\bar G_{\bar f}]  
\label{DECGEN}
\eeq 
where \cpt~invariance has been invoked to assign the same lifetime 
$\Gamma _B^{-1}$ to $B$ and $\bar B$ hadrons. Obviously if 
\beq
\frac{G_f(t)}{\bar G_{\bar f}(t)} \neq 1 
\eeq 
is observed, \cp~violation has been found. Yet one should 
keep in mind that this can manifest itself in two (or three) 
qualitatively different ways: 
\begin{enumerate} 
\item 
\beq 
\frac{G_f(t)}{\bar G_{\bar f}(t)} \neq 1 
\; \; {\rm with} \; \; 
\frac{d}{dt}\frac{G_f(t)}{\bar G_{\bar f}(t)} =0 \; ; 
\label{DIRECTCP1}
\eeq   
i.e., the {\em asymmetry} is the same for all times of decay. This 
is true for {\em direct} \cp~violation; yet, as explained later, it also 
holds for \cp~violation {\em in} the oscillations.  
\item 
\beq 
\frac{G_f(t)}{\bar G_{\bar f}(t)} \neq 1 
\; \; {\rm with} \; \; 
\frac{d}{dt}\frac{G_f(t)}{\bar G_{\bar f}(t)} \neq 0 \; ; 
\label{DIRECTCP2}
\eeq   
here the asymmetry varies as a function of the time of decay. 
This can be referred to as \cp~violation {\em involving} oscillations. 
\end{enumerate} 

A straightforward application of quantum mechanics with its linear superposition principle yields   
\cite{CPBOOK} for $\Delta \Gamma = 0$, which holds for $B^{\pm}$ and 
$\Lambda_b$ exactly and for $B_d$ to a good approximation 
\footnote{Later I will address the scenario with $B_s$, where $\Delta \Gamma$ presumably reaches 
a measurable level.}:
\beq 
\begin{array}{l}
G_f(t) = |T_f|^2 
\left[ 
\left( 1 + \left| \frac{q}{p}\right| ^2|\bar \rho _f|^2 \right) + 
\left( 1 - \left| \frac{q}{p}\right| ^2|\bar \rho _f|^2 \right) 
{\rm cos}\Delta m_Bt 
- 2 ({\rm sin}\Delta m_Bt) {\rm Im}\frac{q}{p} \bar \rho _f 
\right]  \\ 
\bar G_{\bar f}(t) = |\bar T_{\bar f}|^2 
\left[ 
\left( 1 + \left| \frac{p}{q}\right| ^2|\rho _{\bar f}|^2 \right) + 
\left( 1 - \left| \frac{p}{q}\right| ^2|\rho _{\bar f}|^2 \right) 
{\rm cos}\Delta m_Bt 
- 2 ({\rm sin}\Delta m_Bt) {\rm Im}\frac{p}{q} \rho _{\bar f}  
\right]  
\end{array}
\label{GGBAR}
\eeq 
The amplitudes for the instantaneous $\Delta B=1$ 
transition into a 
final state $f$ are denoted by 
$T_f = T(B \ra f)$ and $\bar T_f = T(\bar B \ra f)$ and  
\beq 
\bar \rho _f = \frac{\bar T_f}{T_f} \; \; , 
\rho _{\bar f} = \frac{T_{\bar f}}{\bar T_{\bar f}} 
\eeq

Staring at the general expression is not always very illuminating; 
let us therefore consider three limiting cases: 
\begin{itemize}
\item
$\Delta m_B = 0$, i.e. {\em no} $B^0- \bar B^0$ oscillations: 
\beq 
G_f(t) = 2|T_f|^2 \; \; , \; \; 
\bar G_{\bar f}(t) = 2|\bar T_{\bar f}|^2 
\leadsto \frac{\bar G_{\bar f}(t)}{G_{ f}(t)} = 
\left|
\frac{\bar T_{\bar f}}{T_{ f}}
\right|^2 \; \; , \frac{d}{dt}G_f (t) \equiv 0 \equiv 
\frac{d}{dt}\bar G_{\bar f} (t) 
\eeq 
This is explicitely what was referred to above as {\em direct} 
\cp~violation. 
\item 
$\Delta m_B \neq  0$   
and $f$ a flavour-{\em specific} final state with {\em no} 
direct \cp~violation; i.e., 
$T_{f} = 0 = \bar T_{\bar f}$ and $\bar T_f = T_{\bar f}$   
\footnote{For a flavour-specific mode one has in general 
$T_f \cdot T_{\bar f} =0$; the more intriguing case arises  
when one considers a transition that requires oscillations 
to take place.}: 
\beq 
\begin{array} {c} 
G_f (t) = \left| \frac{q}{p}\right| ^2 |\bar T_f|^2 
(1 - {\rm cos}\Delta m_Bt )\; \; , \; \; 
\bar G_{\bar f} (t) = \left| \frac{p}{q}\right| ^2 |T_{\bar f}|^2 
(1 - {\rm cos}\Delta m_Bt) \\ 
\leadsto 
\frac{\bar G_{\bar f}(t)}{G_{ f}(t)} = 
\left| \frac{q}{p}\right| ^4 
\; \; , \; \; \frac{d}{dt} \frac{\bar G_{\bar f}(t)}{G_{ f}(t)} 
\equiv 0  
\; \; , \; \; \frac{d}{dt} \bar G_{\bar f}(t) \neq 0 \neq 
\frac{d}{dt} G_ f(t)
\end{array} 
\eeq 
This constitutes \cp~violation {\em in the 
oscillations}. For the \cp~conserving decay into the 
flavour-specific 
final state is used merely to track the flavour identity of the 
decaying meson. This situation can therefore be denoted also 
in the following way: 
\beq 
\frac{{\rm Prob}(B^0 \Rightarrow  \bar B^0; t) - 
{\rm Prob}(\bar B^0 \Rightarrow  B^0; t)}
{{\rm Prob}(B^0 \Rightarrow \bar B^0; t) + 
{\rm Prob}(\bar B^0 \Rightarrow  B^0; t)} = 
\frac{|q/p|^2 - |p/q|^2}{|q/p|^2 + |p/q|^2} = 
\frac{1- |p/q|^4}{1+ |p/q|^4} 
\eeq 

\item 
$\Delta m_B \neq  0$ with $f$ now being a  
flavour-{\em non}specific final state -- a final state 
{\em common} 
to $B^0$ and $\bar B^0$ decays -- of a special nature, namely 
a \cp~eigenstate -- $|\bar f\rangle = {\bf CP}|f\rangle = 
\pm |f\rangle $ -- {\em without} direct \cp~violation --  
$|\bar \rho _f| = 1 = |\rho _{\bar f}| $: 
\beq 
\begin{array} {c} 
G_f(t) = 2 |T_f|^2 
\left[ 1 - ({\rm sin}\Delta m_Bt) \cdot 
{\rm Im} \frac{q}{p} \bar \rho _f 
\right] \\  
\bar G_f(t) = 2 |T_f|^2 
\left[ 1 + ({\rm sin}\Delta m_Bt )\cdot 
{\rm Im} \frac{q}{p} \bar \rho _f 
\right] \\ 
\leadsto 
\frac{d}{dt} \frac{\bar G_f(t) }{G_f(t)} \neq 0\\
\frac{\bar G_f(t) - G_f(t)}{\bar G_f(t) + G_f(t)} = 
({\rm sin}\Delta m_Bt )\cdot {\rm Im} \frac{q}{p} \bar \rho _f 
\end{array} 
\label{GGBAR}
\eeq 
is the concrete realization of what was called \cp~violation 
{\em involving oscillations}. 

\noindent 
For $f$ still denoting a \cp~eigenstate, yet with $|\bar \rho_f| \neq 1$ one has the more complex 
asymmetry expression
\beq 
\frac{\bar G_f(t) - G_f(t)}{\bar G_f(t) + G_f(t)} = S_f\cdot  ({\rm sin}\Delta m_Bt ) - 
C_f \cdot ({\rm cos}\Delta m_Bt )
\label{CSASYM}
\eeq
with 
\beq 
S_f = \frac{2{\rm Im}\frac{q}{p}\bar \rho_{\pi^+\pi^-}}{1+\left|\frac{q}{p}\bar \rho_{\pi^+\pi^-}\right| ^2} \; , \; \; 
C_f = \frac{1 - \left|\frac{q}{p}\bar \rho_{\pi^+\pi^-}\right| ^2}
{1+\left|\frac{q}{p}\bar \rho_{\pi^+\pi^-}\right| ^2} 
\eeq
\end{itemize} 

An obvious, yet still useful criterion for \cp~observables is that they must be 
`re-phasing' invariant under $|\bar B^0\rangle \to e^{-i\xi}|\bar B^0\rangle$. The expressions 
above show there are three 
classes of such observables:
\begin{itemize}
\item 
An asymmetry in the {\em instantaneous} transition amplitudes for \cp~conjugate modes:
\beq 
|T(B \to f )| \neq |T(\bar B \to \bar f)| \hspace{1cm} 
\Longleftrightarrow \hspace{1cm} \Delta B =1 \; .  
\label{class1}
\eeq
It reflects pure $\Delta B = 1$ dynamics and thus amounts 
to {\em direct} \cp~violation. 
Those modes are most likely to be nonleptonic; in the SM they practically have to be. 

\item 
\cp~violation {\em in} $B^0 - \bar B^0$ oscillations: 
\beq 
|q| \neq |p| \hspace{1cm} 
\Longleftrightarrow \hspace{1cm} \Delta B =2 \; .  
\label{class2}
\eeq 
It requires \cp~violation {\em in} $\Delta B =2$ dynamics. 
The theoretically cleanest modes here are semileptonic ones due to the SM 
$\Delta Q = \Delta B$ selection rule.  

\item 
\cp~asymmetries {\em involving} oscillations
\footnote{This condition is formulated for the simplest case of $f$ being a \cp~eigenstate.}:  
\beq 
{\rm Im}\frac{q}{p}\bar \rho (f) \neq 0 \; ,\; \;  \bar \rho (f) = \frac{T(\bar B \to f)}{T(B\to f)}\hspace{1cm} 
\Longleftrightarrow \hspace{1cm} \Delta B =1 \& 2 \; . 
\label{class3}
\eeq
Such an effect requires the interplay of $\Delta B=1 \& 2$ forces. 

While $C_f \neq 0$ unequivocally signals {\em direct} \cp~violation in Eq.(\ref{CSASYM}), 
the interpretation of 
$S_f \neq 0$ is more complex. (i) As long as one has measured $S_f$ only in a single mode, the distinction 
between {\em direct} and {\em indirect} \cp~violation -- i.e. \cp~violation in $\Delta B =1$ and 
$\Delta B =2$ dynamics -- is convention dependent, since a change in phase for $\bar B^0$ -- 
$|\bar B^0 \rangle \to e^{-i\xi} |\bar B^0 \rangle$ -- leads to $\bar \rho_f \to e^{-i\xi}\bar \rho_f$ and 
$(q/p) \to e^{i\xi} (q/p)$, i.e. can shift any phase from $(q/p)$ to $\bar \rho_f$ and back while 
leaving $(q/p)\bar \rho_f$ invariant. However once $S_f$ has  been measured for two different 
final states $f$, then the distinction becomes truly meaningful independent of theory: 
$S_{f_1} \neq S_{f_2}$ implies $(q/p)\bar \rho_{f_1} \neq (q/p)\bar \rho_{f_2}$ and thus 
$\bar \rho_{f_1} \neq \bar \rho_{f_2}$, i.e. \cp~violation in the $\Delta B =1$ sector. One should note 
that this {\em direct} \cp~violation might not generate a $C_f$ term, see Sect.\ref{OBCP}. For 
$\bar \rho_{f_1} = e^{i\phi_{1,w}}$ and $\bar \rho_{f_2} = e^{i\phi_{2,w}}$  causing 
$S_{f_1} \neq S_{f_2}$ 
would both lead to $C_{f_1} = 0 = C_{f_2}$.

\end{itemize}
In summary: to observe such \cp~asymmetries in $B^0$ decays, one hopes 
for three conditions to be satisfied to a sufficient degree: 
\begin{itemize}
\item 
With the asymmetry parameter given by Im$\frac{q}{p}\bar \rho _f$, one needs weak complex phases 
to enter through $\Delta B = 2$ and/or $\Delta B =1$ dynamics -- $\frac{q}{p}$ and/or 
$\bar \rho_f$, respectively. 
\item 
The coefficient of the asymmetry parameter -- sin$\Delta M_B t$ -- reflects  the presence of oscillations needed to provide the second amplitude. With $\Delta M_B$ denoting the oscillation rate one 
wishes for $\Delta M_B \simeq \Gamma _B$ as the optimal scenario: while $\Delta M_B \ll \Gamma _B$ 
would suppress the signal, $\Delta M_B \gg \Gamma _B$ would provide greater experimental challenges in resolving the signal. This can be read off from the time-integrated asymmetry obtained 
from Eqs.(\ref{DECGEN},\ref{GGBAR}), which carries the factor $x/(1+ x^2)$ with $x=\Delta M_B/\Gamma_B$: it is maximal for 
$x=1$, vanishes like $x$ for $x\to 0$ and like $1/x$ for $x \to \infty$. 
\item 
While time integrated \cp~asymmetries can be extracted (from data well above threshold), being able to measure the peculiar time dependance predicted adds greatly to the experimental sensitivity and 
provides an excellent tool to reject background. I.e., one wishes not only $x \sim 1$, but also 
$\Gamma_B$ and thus $\Delta M_B$ sufficiently small so that decay and oscillation times can be resolved experimentally. 

Nature provided us with a generous gift in this context. Neutral charm mesons had been observed with lifetimes around $4\cdot 10^{-13}$ sec in the late 1970's. If the three times heavier $B$ mesons had a lifetime 
around $10^{-12}$ sec, then one could rely on the microvertex detectors developed for charm lifetime 
measurements to track the decay rate evolution of $B$ mesons as well.

\end{itemize}

%%%%%%%%%%%%%%
\subsection{The First Central Pillar of the Paradigm: Long Lifetimes}
\label{BLIFE}
%%%%%%%%%%%%

Beauty, the existence of which had been telegraphed by the discovery of the $\tau$ as the third 
charged lepton 
was indeed observed exhibiting a surprising feature: starting in the early 1980's its lifetime was 
found to be about $10^{-12}$ sec.  This was considered `long'. For one can get an estimate 
for $\tau (B)$ by relating it to the muon lifetime: 
\beq 
\tau (B) \simeq \tau _b \sim \tau (\mu ) \left( \frac{m(\mu )}{m(b)}  \right) ^5 \frac{1}{9} 
\frac{1}{|V(cb)|^2} \simeq 3 \cdot 10^{-14} \left| \frac{{\rm sin}\theta_C}{V(cb)}  \right|^2 \; {\rm sec}
\eeq
One had expected $|V(cb)|$ to be suppressed, since it represents an out-of-family coupling. 
Yet  one had assumed without deeper reflection that $|V(cb)| \sim {\rm sin}\theta_C$ -- what 
else could it be? The measured value for $\tau (B)$ however pointed to  
$|V(cb)| \sim |{\rm sin}\theta_C|^2$.  By the end of the millenium one had obtained a rather accurate value: 
$\tau (B_d) = (1.55\pm 0.04) \cdot 10^{-12}$ s. Now the data have become even more precise: 
\beq 
\tau (B_d) = (1.530 \pm 0.009) \cdot 10^{-12} \; {\rm s}\; , \; \; 
\tau (B^{\pm}) /\tau (B_d) = 1.071 \pm 0.009 
\eeq
The lifetime {\em ratio}, which reflects the impact of hadronization, had been predicted 
\cite{MIRAGE} successfully 
well before data of the required accuracy had been available.

%%%%%%%%%%%%%%%
\subsection{The Second Central Pillar: $B^0 -\bar B^0$ Oscillations}
\label{SECONDPILLAR}
%%%%%%%%%%

Many lessons learnt from $K^0 - \bar K^0$ oscillations have been applied profitably 
to $B^0 - \bar B^0$ oscillations. 
The generalized mass matrix introduced in Sect.\ref{CPBASICS} is generated from 
$\Delta B =2$ dynamics:  
\beq 
{\cal M}_{12} = M_{12} + \frac{i}{2}{ \Gamma}_{12} = 
\matel{B^0}{{\cal L}_{eff}(\Delta B =2)}{\bar B^0}
\eeq
In the SM ${\cal L}_{eff} (\Delta B=2)$ is produced by iterating 
two $\Delta B=1$ operators: 
\beq 
{\cal L}_{eff} (\Delta B=2) = {\cal L}(\Delta B=1) \otimes 
{\cal L}(\Delta B=1) 
\eeq  
This leads to the well known 
quark box diagrams, which generate a {\em local} [{\em short-distance}] $\Delta B=2$ operator 
for $M_{12}$ [$\Gamma_{12}$]. 
The contributions that do 
{\em not} depend on the mass of the internal quarks cancel against 
each other due to the GIM mechanism, which leads to highly convergent diagrams. The situation is actually 
simpler and under better theoretical control than for the $\Delta S=2$ case: 
(i) The matrix $M$ is dominated by a single contribution, namely from 
$t\bar t$ internal quarks. (ii) The matrix $\Gamma$ receives its leading contribution 
from $c \bar c$ internal quarks, and it is a reasonable ansatz (albeit not a guaranteed one) that also 
$\Delta \Gamma$ 
can be evaluated using the quark box diagram. 

The overall leading contribution is obtained by evaluating 
the quark box diagram with internal $W$ and top quark lines corresponds to integrating those 
heavy degrees of freedom out in a straightforward way leading to: 
\beq   
{\cal L}_{eff}^{box}(\Delta B=2, \mu ) \simeq 
\left( \frac{G_F}{4\pi }\right) ^2 M_W^2\cdot     
\xi _t^2 E(x_t) \eta _{tt} 
\left( \bar q \gamma _{\mu}(1- \gamma _5) b\right) ^2 
+ h.c. 
\label{LAGDELTAS2}
\eeq  
with $q = d,s$, $\xi _i = V(is)V^*(id) \; , \; \; i=c,t$, $\eta _{tt} \simeq 0.57 \pm 0.01$ denoting the 
QCD radiative correction, $x_t = m_t^2/M_W^2$ and $E(x_t)$ reflecting the box loop 
with internal $W$ and top quarks:  
\beq 
E(x_t) = x_t 
\left(   
\frac{1}{4} + \frac{9}{4(1- x_t)} - \frac{3}{2(1- x_t)^2} 
\right) 
- \frac{3}{2} \left( \frac{x_t}{1-x_t}\right) ^3 
{\rm log} x_t  \; . 
\label{TOPBOX}
\eeq 
Evaluating $\Delta M_B$ as a function of the top quark mass one obtains from 
Eq.(\ref{TOPBOX}) 
\beq 
\Delta M_B \propto \left( \frac{m_t}{M_W}\right) ^2     \; \; {\rm for} \; \; m_t \gg M_W
\eeq
The factor on the right hand side reflects the familiar GIM suppression for $m_t \ll M_W$. 
Yet for $m_t \gg M_W$ it represents a huge enhancement that would seem to contradict the 
expected decoupling. For it would mean that the low energy observable $\Delta M_B$ is 
controlled more and more by a field, namely that for the top quark, at asymptotically  high scales. 
The resolution of this seeming paradox arises in an intriguing way: the massive $W$ boson has   
`ancestors', namely the original massless gauge boson forming the transverse components -- 
and the charged scalar component of the Higgs doublet field introduced to drive electroweak 
symmetry breaking, which is re-incarnated as the longitudinal $W$ component. The latter, for which 
there is no decoupling theorem, generates the $(m_t/M_W)^2$ contribution.  

There are two types of $B^0$ mesons that can oscillate, namely $B_d$ and $B_s$ mesons, 
with the reliable SM predictions 
\beq 
\Delta M_{B_d} \ll \Delta M_{B_s} \;  , \; \; \Delta \Gamma_{B_d} \ll \Delta \Gamma_{B_s} \; , 
\; \; \Delta \Gamma_B \ll \Delta M_B
\label{DELTABDBS}
\eeq
and accordingly (since $\Gamma_{B_d} \simeq \Gamma_{B_s}$) 
\beq 
x_d \equiv \frac{\Delta M_{B_d}}{\Gamma_{B_d}} \ll x_s \equiv \frac{\Delta M_{B_s}}{\Gamma_{B_s}}
\label{XBDBS}
\eeq
Due to the SM selection rule $\Delta B = \Delta Q$ $x$ can be extracted from the ratio of `wrong-sign' 
vs. `right-sign' leptons in semileptonic $B^0$ decays: 
\beq 
r_B \equiv \frac{\Gamma (\bar B^0 \to l^+\nu X_c^-)}{\Gamma (\bar B^0 \to l^-\nu X_c^+)}= 
\frac{x^2}{2+ x^2} \; , \; \; 
\chi_B \equiv \frac{\Gamma (\bar B^0 \to l^+\nu X_c^-)}{\Gamma (\bar B^0 \to l^{\mp}\nu X_c^{\pm})}= 
\frac{r_B}{1+ r_B} 
\label{RCHI}
\eeq

%%%%%%%%%%%
\subsubsection{The Discovery of $B_d - \bar B_d$ Oscillations}
\label{BDDISC}
%%%%%%%%%%%

When ARGUS discovered $B_d - \bar B_d$ oscillations in 1986, it caused quite a stir, for their 
observation of $x\sim 0.7$ was much larger than the quantitative theoretical predictions given before. 
Yet in all fairness one should understand the main 
reason behind this underestimate:  
$\Delta M_{B_d}$ is very sensitive to the value of the top quark mass $m_t$. In the early 1980's there 
had been an experimental claim by UA1 to have discovered top quarks in 
$p\bar p$ collisions with $m_t = 40 \pm 10$ GeV. To their later chagrin theorists by and large had 
accepted this claim. Yet after the ARGUS discovery theorists quickly concluded that top quarks had to 
be much heavier than previously thought, namely $m_t > 100$ GeV \cite{CPBOOK}; 
this was the first indirect 
evidence for top quarks being `super-heavy' before they were discovered in 
$p \bar p$ collisions at Fermlab. Present data tell us \cite{PDG06}   
\bea 
x_d &\equiv& \frac{\Delta M_{B_d}}{\Gamma _{B_d}} = 0.776 \pm 0.008 \\
m_t &=& 174.2 \pm 3.3 \; {\rm GeV} 
\label{DELTABDEXP}
\eea
The measured values of $\Delta M_{B_d}$ and $m_t$ 
are completely compatible with the SM. 

The important consequence for the future was that the observed value for $x_d$ is close to 
optimal for studying \cp~violation, as explained before. This gave great impetus to the plans for 
building a $B$ factory. 

%%%%%%%%%%%
\subsubsection{The `hot' news: $B_s - \bar B_s$ oscillations}
\label{HOT}
%%%%%%%%

Nature has actually provided us with an `encore'. 
As explained above, within the SM one predicts 
$\Delta M(B_s) \gg \Delta M(B_d)$, i.e. that $B_s$ mesons oscillate much faster than 
$B_d$ mesons. Those rapid oscillations have been resolved now \cite{D0BSPUB, CDFBSPUB}: 
\bea 
\Delta M_{B_s} &=& 
\left\{   
\begin{array}{ll}
\left(19 \pm 2 \right)  {\rm ps}^{-1} = \left( 1.25 \pm 0.13\right) \cdot 10^{-2}\,   
{\rm eV}& {\rm D0}  \\
\left(17.77 \pm 0.12 \right) {\rm ps}^{-1} = \left( 1.17 \pm 0.01 \right) \cdot 10^{-2} \, 
{\rm eV}& {\rm CDF} \\ 
\end{array} 
\right. \\
x_s &=& \frac{\Delta M_{B_s}}{\Gamma _{B_s}} \simeq 25
\label{XSDATA} 
\eea 
to be compared with the theoretical prediction: 
\beq 
\Delta M_{B_s} = \left(18.3 ^{+6.5}_{-1.5}\right) \, {\rm ps}^{-1} = 
\left( 1.20 ^{+0.43} _{-0.10} \right) \cdot 10^{-2}\,   
{\rm eV}\; \;  {\rm CKM\; fit} 
\eeq 
This finding represents another triumph of CKM theory even more impressive 
than a mere comparison of the observed and predicted values of $\Delta M_{B_s}$, as explained later. 

There is also marginal evidence for $\Delta \Gamma _{B_s} \neq 0$ \cite{HFAG}  
\beq 
\frac{\Delta \Gamma _{B_s}}{\Gamma _{B_s}} = 0.31 \pm 0.13 \; . 
\eeq

%%%%%%%%%%%%%
\subsection{Large \cp~Asymmetries in $B$ Decays Without 
 `Plausible Deniability'}
\label{PLAUS}
%%%%%%%%

With both the lifetime and oscillation rate falling into the aforementioned practically 
optimal range and thus satisfying two of the conditions listed at the end of Sect.\ref{CPBASICS}, one 
turns to the third one concerning large weak phases.  

The above mentioned observation of a long $B$ lifetime pointed to 
$|V(cb)| \sim {\cal O}(\lambda ^2)$ with $\lambda = {\rm sin}\theta_C$. Together with the 
observation $|V(ub)| \ll |V(cb)|$ and coupled with the assumption 
of three-family unitarity this allows to expand the CKM matrix in powers of $\lambda$, which 
yields the following most intriguing result through order $\lambda ^5$, as first recognized by Wolfenstein: 
\beq 
{\bf V}_{CKM} = 
\left( 
\begin{array}{ccc} 
1 - \frac{1}{2} \lambda ^2 & \lambda & 
A \lambda ^3 (\rho - i \eta + \frac{i}{2} \eta \lambda ^2) \\
- \lambda & 1 - \frac{1}{2} \lambda ^2 - i \eta A^2 \lambda ^4 & 
A\lambda ^2 (1 + i\eta \lambda ^2 ) \\ 
A \lambda ^3 (1 - \rho - i \eta ) \\
& - A\lambda ^2 & 1 
\end{array}
\right) 
\label{WOLFKM}
\eeq  
The three Euler angles and one complex phase of the representation given in Eq.(\ref{PDGKM}) 
is taken over by the four real quantities $\lambda$, $A$, $\rho$ and $\eta$; 
$\lambda$ is the expansion parameter with $\lambda \ll 1$, whereas $A$, $\rho$ and $\eta$ 
are a priori of order unity. I.e., the `long' 
lifetime of beauty hadrons of around 1 psec together with beauty's affinity to transform itself into charm 
and the assumption of only three quark families tell us 
that the CKM matrix exhibits a very peculiar hierarchical pattern in powers of $\lambda$: 
\beq 
V_{CKM} = 
\left( 
\begin{array}{ccc} 
1 & {\cal O}(\lambda ) & {\cal O}(\lambda ^3) \\ 
{\cal O}(\lambda ) & 1 & {\cal O}(\lambda ^2) \\ 
{\cal O}(\lambda ^3) & {\cal O}(\lambda ^2) & 1 
\end {array} 
\right) 
\; \; \; , \; \; \; \lambda = {\rm sin}\theta _C 
\eeq 
We know within the SM this matrix has to be unitary. Yet in addition it is almost 
the identity matrix, almost symmetric and the moduli of its elements shrink with the distance from the diagonal. 
It has to contain a message from nature -- albeit in a highly encoded form. 

%%%%%%%%%%%%
\begin{figure}[t]
\vspace{6.0cm}
\includegraphics{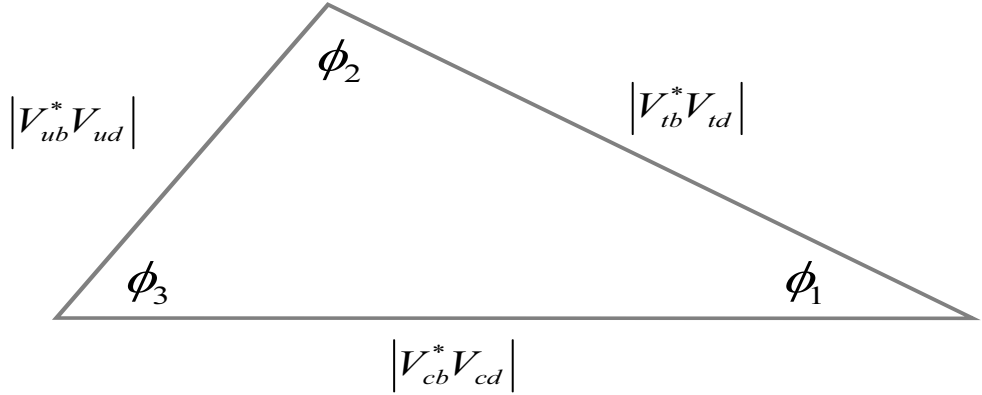}
\caption{The CKM Unitarity Triangle}
\label{CKMTRIANGLENOT1} 
\end{figure}
%%%%%%%%%%%%%%%
Among the six unitarity triangles is a special one, shown in Fig.\ref{CKMTRIANGLENOT1}. 
The sides of this triangle are 
given by $\lambda \cdot V(cb)$, $V(ub)$ and 
$V^*(td)$. Therefore their lengths are of the same order, namely $\lambda ^3$ and their 
angles thus naturally large,  i.e. $\sim$ several $\times$ $10$ degrees. The sides control the rates 
for CKM favoured and disfavoured 
$B_{u,d}$ decays and $B_d - \bar B_d$ oscillations, and the angles their \cp~asymmetries.

Some comments on notation might be useful. The BABAR collaboration 
and its followers refer to the three angles of the CKM unitarity triangle as 
$\alpha$, $\beta$ and $\gamma$; the BELLE collaboration instead has adopted the 
notation $\phi_1$, $\phi_2$ and $\phi_3$. While it poses no problem to be 
conversant in both languages, the latter has not only historical priority on its side 
\cite{BJSANDA}, but is 
also more rational. For the angles $\phi_i$ in the 
`$bd$' triangle of Fig.(\ref{CKMTRIANGLENOT1}) are always opposite the side defined by 
$V^*(id)V(ib)$.   Furthermore this classification scheme can readily be generalized to all six 
unitarity triangles;  those triangles can be labeled by $kl$ with 
$k \neq l= d,s,b$ or $k\neq l = u,c,t$. Its 18 angles 
can then be unambiguously denoted by $\phi_i^{kl}$: it is the angle in triangle $kl$ opposite 
the side $V^*(ik)V(il)$ or $V^*(ki)V(li)$, respectively. {\em Therefore I view the notation 
$\phi_i^{(kl)}$ as the only truly Cartesian one.}  

The discovery of $B_d - \bar B_d$ oscillations {\em defined} the 
`CKM Paradigm of Large \cp~Violation in $B$ Decays' that had been anticipated in 1980: 
\begin{itemize}
\item 
A host of nonleptonic $B$ channels has to exhibit sizable \cp~asymmetries. 
\item 
For $B_d$ decays to flavour-nonspecific final states (like \cp~eigenstates) the \cp~asymmetries 
depend on the time of decay in a very characteristic manner; their size should typically be measured 
in units of 10\% rather than 0.1\%. 
\item 
{\em There is no plausible deniability for the CKM description, if such asymmetries are 
not found.}  
\item 
For $m_t \geq 150$ GeV the SM prediction for $\epsilon_K$ is dominated by the top quark 
contribution like $\Delta M_{B_d}$. It thus drops out from their ratio, and sin$2\phi_1$ can be 
predicted within the SM irrespective of the (superheavy) top quark mass. In the early 1990's, i.e., 
{\em before} the direct discovery of top quarks, it was predicted \cite{BEFORETOP} 
\beq 
\frac{\epsilon_K}{\Delta M_{B_d}} \propto {\rm sin}2\phi_1 \sim 0.6 - 0.7
\label{BEFORE}
\eeq  
with values for $B_Bf_B^2$ inserted as now estimated by Lattice QCD. 
\item 
The \cp~asymmetry in the Cabibbo favoured channels $B_s \to \psi \phi/\psi \eta$ is Cabibbo suppressed, i.e. below 4\%, for reasons very specific to CKM theory, as pointed out already in 
1980 \cite{BS80}. 

\end{itemize}

%%%%%%%
\subsection{CKM Theory at the End of the 2nd Millenium}
%%%%%%%

\begin{figure}[t]
\vspace{7.0cm}
\includegraphics{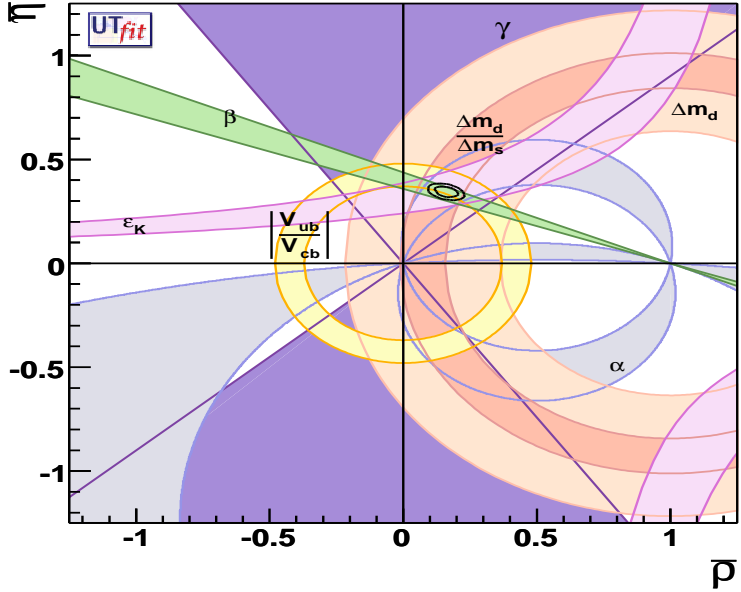}
\caption{The CKM Unitarity Triangle fit.}
\label{CKMTRIANGLEFIT1} 
\end{figure}

It is indeed true that large fractions of the observed values for $\Delta M_K$, $\epsilon_K$ and 
$\Delta M_B$ and even most of $\epsilon^{\prime}$ could be due to New Physics given 
the limitations in our theoretical control over hadronic matrix elements. Equivalently constraints from 
these and other data translate into `broad' bands in plots of the unitarity triangle, see 
Fig.\ref{CKMTRIANGLEFIT1}. 
%%%%%%%%%%%%%
%\begin{figure}[ht]
%\begin{center}
%\epsfig{%bbllx=0.5cm,bblly=16cm,bburx=20cm,bbury=23cm,
%height=5truecm, width=10truecm,
 %       figure=CKM_fit_today.eps}%\vskip1cm 
%}
%\end{center}
%\end{figure}
%%%%%%%%%%%%
%%%%%%%%%%%%%%%

The problem with this statement is that it is not even wrong -- it misses the real point. Observables like 
$\Gamma (B \to l \nu X_{c,u})$, $\Gamma (K \to l \nu \pi)$, $\Delta M_K$, $\Delta M_B$, 
$\epsilon _K$ and sin$2\phi_1$ etc. represent very different dynamical regimes that proceed on 
time scales hat span several orders of magnitude.  The very fact that CKM theory 
can accommodate such diverse observables always within a factor two or better and 
relate them in such a manner that its parameters can be plotted as meaningful constraints on a 
triangle is highly nontrivial and -- in my view -- must reflect some underlying, yet unknown 
dynamical layer. Furthermore the CKM parameters exhibit an unusual hierarchical pattern -- 
$|V(ud)| \sim |V(cs)| \sim |V(tb)| \sim 1$, $|V(us)| \simeq |V(cd)| \simeq \lambda$, 
$|V(cb)| \sim |V(ts)| \sim {\cal O}(\lambda ^2)$, $|V(ub)| \sim |V(td)| \sim {\cal O}(\lambda ^3)$ -- 
as do the quark masses culminating in $m_t \simeq 175$ GeV. Picking such values for these parameters would have been seen as frivolous at best -- had they not been forced upon us by 
(independent) data. Thus I view it already as a big success for CKM theory that the experimental constraints on its parameters as they existed in 2000 can be represented through triangle plots in a meaningful way. 

\begin{center}
%%%%%%%%%%%%%%
{\bf Interlude: Singing the Praise of Hadronization}
%%%%%%%%%
\end{center}

\noindent 
Hadronization and nonperturbative dynamics in general are usually viewed as unwelcome 
complication, if not outright nuisances. A case in point was already mentioned: while 
I view the CKM predictions for $\Delta M_K$, $\Delta M_B$, $\epsilon_K$ to be in 
remarkable agreement with the data, significant contributions from New Physics could 
be hiding there behind the theoretical uncertainties due to lack of computational control 
over hadronization. Yet {\em without} hadronization bound states of quarks and antiquarks will not form; without 
the existence of kaons $K^0 - \bar K^0$ oscillations  obviously cannot occur. 
It is hadronization that provides the `cooling' of the (anti)quark degrees of freedom, which 
allows subtle quantum mechanical effects to add up coherently over macroscopic distances. 
Otherwise  
one would not have access to a super-tiny energy difference Im${\cal M}_{12} \sim 10^{-8}$ eV, 
which is very sensitive to different layers of dynamics, 
and indirect \cp~violation could not manifest itself. The same would hold for $B$ mesons and 
$B^0 - \bar B^0$ oscillations. 

\noindent 
To express it in a more down to earth way:  
\begin{itemize}
\item 
Hadronization leads to the formation of kaons and pions with masses exceeding 
greatly (current) quark masses.  
It is the {\em hadronic} phase space that suppresses the \cp~{\em conserving} rate for 
$K_L \to 3 \pi$ by a factor $\sim 500$, since the $K_L$ barely resides above the three pion threshold. 
\item 
It awards `patience'; i.e. one can `wait' for a pure $K_L$ beam to emerge after starting out with a 
beam consisting of $K^0$ and $\bar K^0$. 
\item 
It enables \cp~violation to emerge in the {\em existence} of a reaction, namely 
$K_L \to 2 \pi$ rather than an asymmetry; this greatly facilitates its observation. 
\end{itemize}
For these reasons alone we should praise hadronization as the hero in the tale of \cp~violation 
rather than the villain it is all too often portrayed. 

\begin{center}
%%%%%%%%%%%%%%
{\bf End of Interlude}
%%%%%%%%%
\end{center}

By the end of the second millenium a rich and diverse body of data on flavour dynamics had been 
accumulated, and CKM theory provided a surprisingly successful description of it. This prompted some daring spirits to perform detailed fits of the CKM triangle to infer a rather accurate prediction for 
the \cp~asymmetry in $B_d \to \psi K_S$ \cite{PARODI}: 
\beq 
{\rm sin}2\phi_1 = 0.72 \pm 0.07
\label{ACHPRED}
\eeq

%%%%%%%%%%%%%%%%
\section{The `Expected' Triumph of a Peculiar Theory}
\label{PART2}
%%%%%%%%%%%%

%%%%%%%%%%%%%%
\subsection{Establishing the CKM Ansatz as a Theory}
\label{ANSATZTOTHEORY}
%%%%%%%%%%%

The three angles $\phi_{1,2,3}$ in the CKM unitarity triangle 
(see Fig.\ref{CKMTRIANGLENOT1}  for notation) can be determined 
through \cp~asymmetries in $B_d(t) \to \psi K_S, \pi^+\pi^-$ 
and $B_d \to K^+ \pi^-$ -- in principle. In practice the angle $\phi_3$ can be extracted from 
$B^{\pm} \to D^{\rm neut}K^{\pm}$ with better theoretical control, and $B \to 3\pi , \; 4 \pi$ 
offer various experimental advantages over $B \to 2 \pi$. 
These issues will be addressed in five acts plus two interludes.

%%%%%%%%%%
\subsubsection{Act I: $B_d(t) \to \psi K_S$ and $\phi_1$ (a.k.a. $\beta$) }
\label{ACT1}
%%%%%%%%%%%%

The first published result on the \cp~asymmetry in $B_d \to \psi K_S$ was actually obtained by 
the OPAL collaboration at LEP I \cite{OPAL}, followed by the first value measured 
by CDF inside the physical range \cite{CDFPHI1} : 
\beq 
{\rm sin}2\phi_1 =0.79 \pm 0.44 
\eeq
In 2001 the two $B$ factory collaborations BABAR and BELLE 
established 
the first \cp~violation outside the $K^0 - \bar K^0$ complex : 
\beq 
{\rm sin}2\phi_1 =  
\left\{ 
\begin{array}{l} 
0.59 \pm 0.14 \pm 0.05 \; \; {\rm BABAR\; '01} \\ 
0.99 \pm 0.14 \pm 0.06 \; \; 
{\rm BELLE\;  '01}
\end{array}  
\right. 
\eeq
By 2003 the numbers from the two experiments had well converged 
allowing one to state just the world averages, which is actually a 
BABAR/BELLE average \cite{HFAG}: 
\beq 
{\rm sin}2\phi_1 =  0.675 \pm 0.026 \; \; {\rm WA \; '06}
\label{WA040506} 
\eeq
{\bf The \cp~asymmetry in $B_d \to \psi K_S$ is there, is huge and as expected even quantitatively.}  
For CKM fits based on constraints from $|V(ub)/V(cb)|$, $B^0 - \bar B^0$ oscillations and -- as 
the only \cp~sensitive observable -- $\epsilon_K$ yield \cite{UTFIT} 
\beq 
{\rm sin}2\phi_1|_{\rm CKM}  = 0.755 \pm 0.039 \; . 
\label{PHI1CKM}
\eeq
The measured value also fully agrees with the predictions from the last millenium, 
Eqs.(\ref{BEFORE},\ref{ACHPRED}). 
The CKM prediction has stayed within the $\sim 0.72 - 0.75$ interval for the last several years. 
Through 2005 it has been in impressive agreement with the data. In 2006 a hint of a deviation has emerged. It is not more than that, since it is not (yet) statistically significant. This is illustrated by Fig.\ref{CKMTRIANGLEFIT1} showing these constraints. This figure actually obscures another impressive triumph of CKM theory: 
the \cp~{\em in}sensitive observables $|V(ub)/V(cb)|$ and $\Delta M_{B_d}/\Delta M_{B_s}$ -- 
i.e. observables that do {\em not} require \cp~violation for acquiring a non-zero value -- imply 
\begin{figure}[t]
\vspace{5.0cm}
\includegraphics{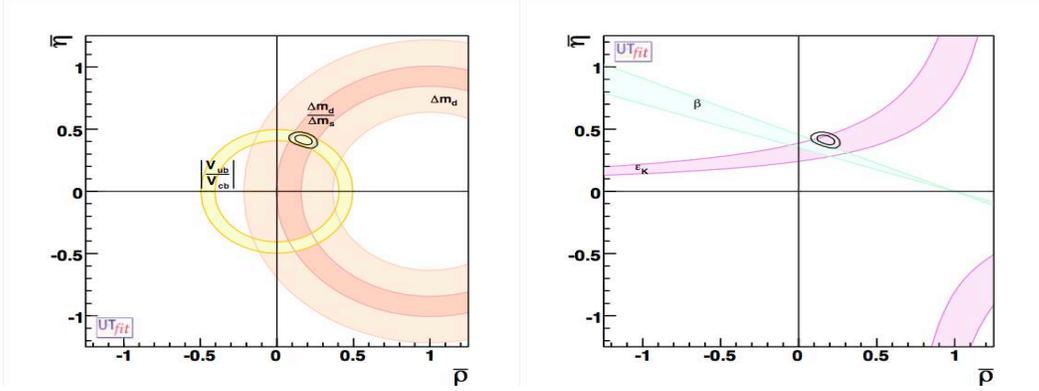}
\caption{CKM unitarity triangle from $|V(ub)/V(cb)|$ and $\Delta M_{B_d}/\Delta M_{B_s}$ on the left and  
       compared to constraints from 
$\epsilon_K$ and sin2$\phi_1/\beta$ on the right 
(courtesy of M. Pierini)
    \label{PIER} }
\end{figure}
\begin{itemize}
\item 
a non-flat CKM triangle and thus \cp~violation, see the left of Fig.~\ref{PIER} 
\item 
that is fully consistent with the observed \cp~sensitive observables $\epsilon_K$ 
and sin$2\phi_1$, see the right of Fig.~\ref{PIER}. 
\end{itemize}

%%%%%%%%%%%%%%
%%%%%%%%%%%%%%%%%%%%
\subsubsection{\cp~violation in $K$ and $B$ decays -- exactly the same, only different}
\label{AUSTRIAN}
%%%%%%%%%%%%

\begin{figure}[t]
\vspace{8.0cm}
\includegraphics{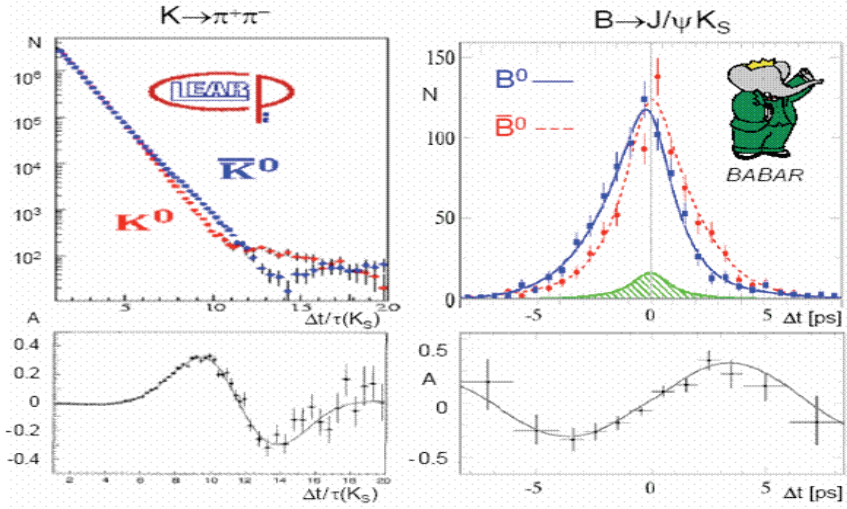}
%\centering\includegraphics[width=.9\linewidth]{CPasym_K_B.ps}
\caption{
The observed decay time distributions for $K^0$ vs. $\bar K^0$ from CPLEAR on the left 
and for $B_d$ vs.$\bar B_d$ from BABAR on the right. 
\label{KBcompfig1} }
\end{figure}

There are several similarities between $K^0 - \bar K^0$ and  
$B_d - \bar B_d$ oscillations even on the quantitative level. Their values for $x = \Delta M/\Gamma$ and 
thus for $\chi$ are very similar. It is even more intriguing that also their pattern of \cp~asymmetries in 
$K^0(t)/\bar K^0 (t) \to \pi^+\pi^-$ and $B_d(t)/\bar B_d (t) \to \psi K_S$ is very similar. Consider 
the two lower plots in Fig.\ref{KBcompfig1}, which show the asymmetry directly as a function of 
$\Delta t$: it looks intriguingly similar qualitatively and even quantitatively. The lower left plot 
shows that the difference between $K^0 \to \pi^+\pi^-$ and $\bar K^0 \to \pi^+\pi^-$ is actually 
measured in units of 10 \% for $\Delta t \sim (8 - 16) \tau_{K_S}$, which is the $K_S-K_L$ interference region.  

Clearly one can find domains in $K \to \pi^+\pi^-$ that exhibit  a truly large \cp~asymmetry. 
Nevertheless it is an empirical fact that \cp~violation in $B$ decays is much larger than in 
$K$ decays. For the mass eigenstates of neutral kaons are 
very well approximated by \cp~eigenstates, as can be read off from the  upper left plot: it shows that the 
vast majority of $K \to \pi^+\pi^-$ events follow a single exponential decay law that coincides for 
$K^0$ and $\bar K^0$ transitions. This is in marked contrast to 
the $B_d \to \psi K_S$ and $\bar B_d \to \psi K_S$ transitions, which in no domain are well approximated by a single exponential law and do not coincide at all, except for 
$\Delta t =0$, as it has to be, see Sect.\ref{EPRIMPORT1}.

%%%%%%%%%%%%%%
%%%%%%%%%%%%%

%%%%%%%%%%%%
\subsubsection{Interlude: "Praise the Gods Twice for EPR Correlations"} 
\label{EPRIMPORT1} 
%%%%%%%%%%%%

 The BABAR and BELLE analyses are based on a glorious application of quantum mechanics and in 
 particular EPR correlations\cite{EPR}.  The \cp~asymmetry in $B_d \to \psi K_S$ had been predicted to 
 exhibit a peculiar dependence on the time of decay, since it involves $B_d - \bar B_d$ oscillations 
 in an essential way: 
\beq 
{\rm rate} (B_d(t)[\bar B_d(t)] \to \psi K_S) \propto e^{-t/\tau _B} (1- [+] A {\rm sin}\Delta m_B t) \; , 
\label{ASYM}
\eeq  
At first it would seem that an asymmetry of the form given in Eq.(\ref{ASYM}) could not be measured for practical reasons. For in the reaction
\beq 
 e^+e^- \to \Upsilon (4S) \to B_d \bar B_d
\label{UPS4S}
\eeq
the point where the $B$ meson pair is produced is ill determined due to the finite size of the electron and positron beam spots: the latter amounts to about 1 mm in the longitudinal direction, while a $B$ meson typically travels only about a quarter of that distance before it decays. 
It would then seem that the length of the flight path of the $B$ mesons is poorly known and that averaging over this ignorance would greatly dilute or even eliminate the signal. 
 
It is here where the existence of a EPR correlation comes to the rescue. While the two $B$ mesons in the reaction of Eq.(\ref{UPS4S}) oscillate back and forth between a $B_d$ and $\bar B_d$, they change their flavour identity in a {\em completely correlated} way.  For the $B \bar B$ pair forms a \oc~{\em odd} state; Bose statistics then tells us that there cannot be two identical flavour hadrons in the final state: 
\beq 
 e^+e^- \to \Upsilon (4S) \to B_d \bar B_d \not \to B_d B_d, \; \bar B_d \bar B_d
 \label{NOTID}
 \eeq
 Once one of the $B$ mesons decays through a flavour specific mode, say $B_d \to l^+\nu X$ 
 [$\bar B_d \to l^- \bar \nu X$], then we know unequivocally that the other $B$ meson was a 
 $\bar B_d$ [$B_d$] at {\em that} time. The time evolution of $\bar B_d(t) [B_d(t)] \to \psi K_S$ as described by 
 Eq.(\ref{ASYM}) starts at {\em that} time as well; i.e., the relevant time parameter is the {\em interval between} 
 the two times of decay, not those times themselves. That time interval is related to -- and thus can be inferred from -- 
 the distance between the two decay vertices, which is well defined and can be measured.

%%%%%%%%%%
%\begin{figure}
%\mbox{\pdffig{file=timeasymm.pdf,width=6in}}
%\mbox{\psfig{file=timeasymm.ps,width=6in}}
%\caption{\label{obsdt}The observed decay time distributions for $B^0$ (red) and
%$\bar B^0$ (blue) decays.}
%\end{figure}
%%%%%%%%%%%
%%%%%%%%%%%%
\begin{figure}[t]
\vspace{8.0cm}
\includegraphics{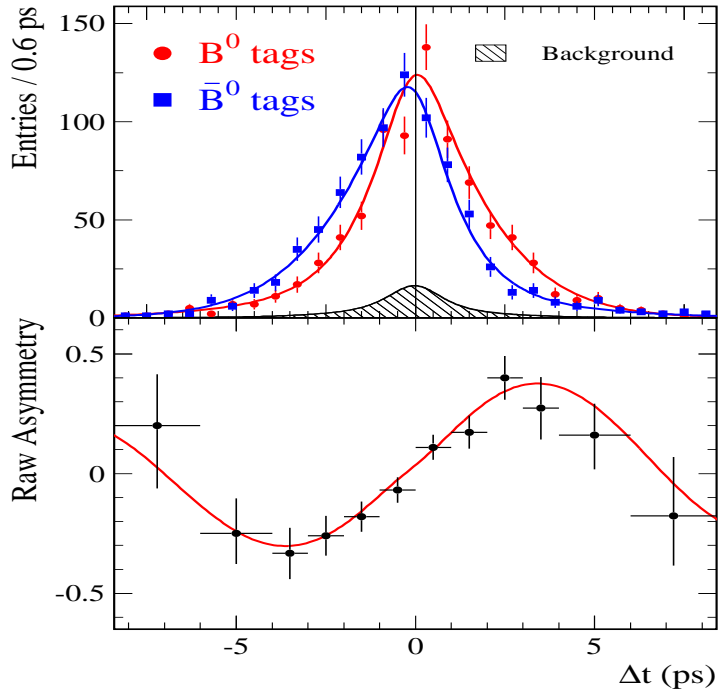}
\caption{The observed decay time distributions for $B^0$ (red) and
  $\bar B^0$ (blue) decays.}
\label{timeasymmfig} 
\end{figure}
%%%%%%%%%%%%%%%

 The great value of the EPR correlation is instrumental for another consideration as well, namely how to 
 see directly from the data that \cp~violation is matched by \ot~violation. Fig.\ref{timeasymmfig} shows two distributions, one for the 
 interval  $\Delta t$ between the times of decays $B_d \to l^{+}X$ and $\bar B_d \to \psi K_S$ and the other one for the 
\cp~conjugate process $\bar B_d \to l^{-}X$ and $B_d \to \psi K_S$. They are clearly different proving that \cp~is broken. Yet they show more: the shape of the two distributions is actually the same 
(within experimental uncertainties) the only difference being that the average of $\Delta t$ is 
{\em positive} for $(l^-X)_{\bar B} (\psi K_S)$ and 
 {\em negative} for $(l^+X)_{B} (\psi K_S)$ events. I.e., there is a (slight) preference for 
 $B_d \to \psi K_S$ 
 [$\bar B_d \to \psi K_S$] to occur {\em after} [{\em before}] and thus more [less] slowly (rather than just more rarely) than $\bar B \to l^-X$ [$B \to l^+ X$]. Invoking \cpt~invariance merely for semileptonic $B$ decays -- yet not for nonleptonic 
 transitions -- synchronizes the starting point of the $B$ and $\bar B$ decay `clocks', and the 
 EPR correlation keeps them synchronized. We thus see that \cp~and 
\ot~violation are `just' 
 different sides of the same coin. 
 As explained above, EPR correlations are essential for 
 this argument! 
 
 The reader can be forgiven for feeling that this argument is of academic interest only, since 
 \cpt~invariance of all 
 processes is based on very general arguments. Yet the main point to be noted is that EPR correlations, which 
 represent some of quantum mechanics' most puzzling features, serve as an essential precision tool, which is routinely used in these measurements. I feel it is thus inappropriate to refer to EPR correlations as a paradox.

%%%%%%%%%%
\subsubsection{Act II: $B_d(t) \to 2\pi$ and $\phi_2$ (a.k.a. $\alpha$)}
\label{ACT2}
%%%%%%%%%%%%

The situation is theoretically more complex than for $B_d (t) \to \psi K_S$ due to two 
reasons: 
\begin{itemize}
\item 
While both final states $\pi \pi$ and $\psi K_S$ are \cp~eigenstates, the former unlike the latter 
is not reached through an isoscalar transition. The two pions can form an $I=0$ or $I=2$ 
configuration (similar to $K\to 2\pi$), which in general will be affected differently by the strong interactions. 
\item 
For all practical purposes $B_d \to \psi K_S$ is described by two tree diagrams 
representing the two effective operators 
$(\bar c_L\gamma _{\mu}b_L)(\bar s_L\gamma ^{\mu}c_L)$ and 
$(\bar c_L\gamma _{\mu}\lambda _i b_L)(\bar s_L\gamma ^{\mu}\lambda _ic_L)$ with the 
$\lambda_i$ representing the $SU(3)_C$ matrices. Yet for $B\to \pi \pi$ we have 
effective operators 
$(\bar d_L\gamma _{\mu}\lambda _i b_L)(\bar q\gamma ^{\mu}\lambda _i q)$ generated by 
the Cabibbo suppressed Penguin loop diagrams in addition to the two tree 
operators $(\bar u_L\gamma _{\mu}b_L)(\bar d_L\gamma ^{\mu}u_L)$ and 
$(\bar u_L\gamma _{\mu}\lambda _i b_L)(\bar d_L\gamma ^{\mu}\lambda _iu_L)$. 
\end{itemize} 
This greater complexity manifests itself already in the phenomenological description of 
the time dependent \cp~asymmetry: 
\beq 
\frac{R_+(\Delta t) - R_-(\Delta t)}{R_+(\Delta t) + R_-(\Delta t)} = S {\rm sin}(\Delta M_d \Delta t) - 
C {\rm cos}(\Delta M_d \Delta t) \; ,  
\eeq
where $R_{+[-]}(\Delta t)$ denotes the rate for 
$B^{tag}(t)\bar B_d(t+\Delta)[\bar B^{tag}(t)B_d(t+\Delta)]$ and 
\beq 
S = \frac{2{\rm Im}\frac{q}{p}\bar \rho_{\pi^+\pi^-}}{1+\left|\frac{q}{p}\bar \rho_{\pi^+\pi^-}\right| ^2} \; , \; \; 
C = \frac{1 - \left|\frac{q}{p}\bar \rho_{\pi^+\pi^-}\right| ^2}
{1+\left|\frac{q}{p}\bar \rho_{\pi^+\pi^-}\right| ^2} \; , \; \; S^2 + C^2 \leq 1 \; . 
\eeq
As before, due to the EPR correlation between the two neutral $B$ mesons, it is the 
{\em relative} time interval $\Delta t$ between the two $B$ decays that matters, not their 
lifetime. The new feature is that one has also a cosine dependence on $\Delta t$. 

BABAR and BELLE find 
\bea 
\label{SPIPI} 
S &=&  
\left\{ 
\begin{array}{l} 
- 0.53 \pm 0.14 \pm 0.02 \; \; {\rm BABAR\; '06} \\ 
- 0.61 \pm 0.10 \pm 0.04 \; \; 
{\rm BELLE\; '06} \\
- 0.59 \pm 0.09 \hspace{1.4cm} {\rm HFAG} 
\end{array}  
\right. \\
C &=&  
\left\{ 
\begin{array}{l} 
- 0.16 \pm 0.11 \pm 0.03 \; \; {\rm BABAR\; '06} \\ 
- 0.55 \pm 0.08 \pm 0.05 \; \; 
{\rm BELLE\; '06} \\
- 0.39 \pm 0.07 \hspace{1.4cm} {\rm HFAG}
\end{array}  
\right.
\label{CPIPI}
\eea
While BABAR and BELLE agree nicely on $S$ making the HFAG average straightforward, their 
findings on $C$ indicate different messages making the HFAG average more iffy. 

$S \neq 0$ has been established and thus \cp~violation also in this channel. While 
BELLE finds $C\neq 0$ as well, BABAR's number is still 
consistent with $C =0$.  
$C\neq 0$ obviously represents {\em direct} \cp~violation. Yet it is often overlooked that 
also $S$ can reveal such \cp~violation. For if one studies $B_d$ decays into two \cp~eigenstates 
$f_a$ and $f_b$ and finds   
\beq 
S(f_a) \neq \eta(f_a) \eta(f_b) S(f_b) 
\eeq
with $\eta_i$ denoting the \cp~parities of $f_i$, then one has established 
{\em direct} \cp~violation. For the case under study that means even if $C(\pi \pi )= 0$, 
yet $S(\pi^+\pi^-) \neq - S(\psi K_S)$, one has observed unequivocally {\em direct} 
\cp~violation. One should note that such direct \cp~violation might not necessarily induce 
$C\neq 0$. For the latter requires, as explained in 
Sect.\ref{OBCP}, that two different amplitudes 
contribute coherently to $B_d \to f_b$ with non-zero relative weak as well as strong phases. 
$S(f_a) \neq \eta(f_a) \eta(f_b) S(f_b)$ on the other hand only requires that the two 
overall amplitudes for $B_d \to f_a$ and $B_d \to f_b$ possess a relative phase. This can be 
illustrated with a familiar example from CKM dynamics: If there were no Penguin operators for 
$B_d \to \pi^+\pi^-$ (or it could be ignored quantitatively), one would have 
$C(\pi^+\pi^-) = 0$, yet at the same time $S(\psi K_S) = {\rm sin}(2\phi_1)$ together with 
$S(\pi^+\pi^-)  = {\rm sin}(2\phi_2) \neq - {\rm sin}(2\phi_1)$.  I.e.,  {\em without} direct CP violation one would have to 
find $C=0$ and $S = - {\rm sin}2\phi_1$ \cite{ELEFANT}. Yet since the measured value of $S$ is within one 
sigma of - sin$2\phi_1$ this distinction is mainly of academic interest at the moment.

Once the categorical issue of whether there is {\em direct} \cp~violation has been settled, one can take up the challenge of extracting a value for 
$\phi_2$ from the data 
\footnote{The complications due to the presence of the Penguin contribution are all too often referred to as `Penguin pollution'. Personally I find it quite unfair to blame our lack of theoretical control on water fowls rather than on the guilty party, namely us. }. This can be done in a model independent way by 
analyzing $B_d(t) \to \pi^+\pi^-$, $\pi^0\pi^0$ and $B^{\pm} \to \pi^{\pm}\pi^0$ transitions 
and performing an isospin decomposition. For the Penguin contribution cannot affect 
$B_d(t) \to [\pi \pi]_{I=2}$ modes. Unfortunately there is a serious experimental bottle neck, 
namely to study $B_d (t) \to \pi^0\pi^0$ with sufficient accuracy. Therefore alternative 
decays have been suggested, in particular $B \to \rho \pi$ and $\rho \rho$. While those provide 
various advantages on the experimental side, they introduce also theoretical uncertainties, since they 
have to be extracted from $3\pi$ and $4\pi$ final states that contain also other resonances and 
configurations like the $\sigma$ state. I am most hesitant to average over the values obtained 
at present for 
$\phi_2$ from $B \to 2\pi$, $\rho \pi$ and $\rho \rho$. 

%%%%%%%%%%
\subsubsection{Act III, $1^{st}$ Version: $B_d \to K^+\pi^-$}
\label{ACT3}
%%%%%%%%%%%%
It was pointed out in a seminal paper \cite{SONI} that (rare) transitions like 
$\bar B_d \to K^- + \pi$'s have the ingredients for sizable direct \cp~asymmetries: 
\begin{itemize}
\item 
Two different amplitudes can contribute coherently, namely the highly CKM 
suppressed tree diagram with $b \to u \bar u s$ and the Penguin diagram 
with $b \to s \bar qq$. 
\item 
The tree diagram contains a large weak phase from $V(ub)$. 
\item 
The Penguin diagram with an internal charm quark loop exhibits an imaginary part, which can be 
viewed -- at least qualitatively -- as a strong phase generated by the production and subsequent 
annihilation of a $c \bar c$ pair (the diagram with an internal $u$ quark loop acts merely as a 
subtraction point allowing a proper definition of the operator). 
\item 
While the Penguin diagram with an internal top quark loop is actually not essential, the 
corresponding effective operator can 
be calculated quite reliably, since integrating out first the top quarks and then the $W$ boson leads to a truly local operator. Determining its matrix elements is however another matter. 

\end{itemize}
To translate these features into accurate numbers represents a formidable task, we have not mastered yet. In Ref. \cite{CECILIABOOK} an early and detailed effort was made to treat $\bar B_d \to K^-\pi^+$ theoretically with the following results: 
\beq 
{\rm BR}(\bar B_d \to K^- \pi ^+) \sim 10^{-5} \; , \; \; 
A_{\cp} \sim  - 0.10
\label{BKPIPRED}
\eeq 
Those numbers turn out to be rather prescient, since they are in gratifying agreement with the data 
$$
{\rm BR}(\bar B_d \to K^- \pi ^+) = (1.85 \pm 0.11) \cdot 10^{-5} 
$$
\beq 
A_{\cp} =
\left \{  
\begin{array}{l} 
 - 0.133 \pm 0.030 \pm 0.009  \; \; \; {\rm BABAR} \\
- 0.113 \pm 0.021   \; \; \; {\rm BELLE} 
\end{array}  
\right. 
\label{BKPIDATA}
\eeq 
Cynics might point out that the authors in  \cite{CECILIABOOK} did not give a specific estimate of the 
theoretical uncertainties in Eq.(\ref{BKPIPRED}). More recent authors have been more ambitious -- 
with somewhat mixed success \cite{pQCD,QCDFACT}. While the observed asymmetry in $B_d \to K\pi$ agrees 
with CKM expectations, we do not have an accurate predictions. 

%%%%%%%%%%
\subsubsection{Act III, $2^{nd}$ Version: 
$\phi_3$  from $B^+ \to D_{neut}K^+$ vs. $B^- \to D_{neut} K^-$}
\label{DNEUTK}
%%%%%%%%%

As first mentioned in 1980 \cite{CARTER}, then explained in more detail in 1985 \cite{BS85} 
and further developed in \cite{GRONWYL},  
the modes $B^{\pm} \to D_{neut}K^{\pm}$ should exhibit direct \cp~violation driven by the 
angle $\phi_3$, if the neutral $D$ mesons decay to final states that are {\em common} to 
$D^0$ and $\bar D^0$. Based on simplicity the original idea was to rely on two-body modes like 
$K_S\pi^0$, $K^+K^-$, $\pi^+\pi^-$, $K^{\pm}\pi^{\mp}$. One drawback of that method are the small 
branching ratios and low efficiencies. 

A new method was pioneered by BELLE and then implemented also by BABAR, namely to employ 
$D_{neut} \to K_S \pi^+\pi^-$ and perform a full Dalitz plot analysis. This requires a very 
considerable analysis effort -- yet once this initial investment has been made, it will pay handsome profit in the long run. For obtaining at least a decent description of the full Dalitz plot population 
provides  considerable cross checks concerning systematic uncertainties and thus a high degree of 
confidence in the results. BELLE and BABAR find:  
\beq 
\phi_3 = 
\left\{
\begin{array}{ll} 53^o  \pm18^o (stat) \pm 3^o(syst) \pm 9^o ({\rm model}) & {\rm BELLE}\\
92^o  \pm 41^o (stat) \pm 11^o(syst) \pm 12^o ({\rm model}) & {\rm BABAR}
\end{array}
\right.
\eeq 
At present these studies are severely statistics limited; one should also note that with more statistics 
one will be able to reduce in particular the model dependence. I view this method as the best one to 
extract a reliable value for $\phi_3$, where the error estimate can be defended due to the many constraints inherent in a Dalitz plot analysis. It exemplifies how the complexities of 
hadronization can be harnessed to establish confidence in the accuracy of our results. 

%%%%%%%%%%
\subsubsection{Act IV: $\phi_1$ from \cp~Violation in $B_d \to$ 3 Kaons: Snatching Victory from the Jaws of Defeat -- or Defeat from the Jaws of Victory?}
\label{KAONS}
%%%%%%%%%

Analysing \cp~violation in $B_d \to \phi K_S$ decays is a most promising way to search 
for New Physics. For the underlying quark-level transition $b \to s \bar s s$ represents a pure 
loop-effect in the SM, it is described by a {\em single} $\Delta B=1$\& $\Delta I=0$ operator (a `Penguin'), a reliable 
SM prediction exists for it \cite{GROSS} -- 
sin$2\phi_1(B_d \to \psi K_S) \simeq {\rm sin}2\phi_1(B_d \to \phi K_S)$ -- 
and the $\phi$ meson represents a {\em narrow} resonance.  

Great excitement was created when BELLE reported a large discrepancy between the predicted and 
observed \cp~asymmetry in $B_d \to \phi K_S$ in the summer of 2003:  
\beq 
{\rm sin}2\phi_1(B_d \to \phi K_S) = 
\left\{
\begin{array}{ll} - 0.96 \pm 0.5 \pm 0.10 & {\rm BELLE ('03)}\\
0.45 \pm 0.43 \pm 0.07 & {\rm BABAR ('03)}
\end{array}
\right. \;  ; 
\eeq
Based on more data taken, this discrepancy has 
shrunk considerably: the BABAR/BELLE average for 2005 yields \cite{HFAG}
\beq 
{\rm sin}2\phi_1(B_d \to \psi K_S) = 0.685 \pm 0.032
\eeq
versus  
\beq 
{\rm sin}2\phi_1(B_d \to \phi K_S) = 
\left\{
\begin{array}{ll} 0.44 \pm 0.27 \pm 0.05 & {\rm BELLE ('05)} \\
0.50 \pm 0.25 ^{+0.07}_{-0.04} & {\rm BABAR ('05)} 
\end{array}
\right. \;  ; 
\eeq
while the 2006 values read as follows: 
\beq 
{\rm sin}2\phi_1(B_d \to \psi K_S) = 0.675 \pm 0.026
\label{PHI06}
\eeq
compared to 
\beq 
{\rm sin}2\phi_1(B_d \to \phi K_S) = 
\left\{
\begin{array}{ll} 0.50 \pm 0.21 \pm 0.06 & {\rm BELLE ('06)}\\
0.12 \pm 0.31 \pm 0.10 & {\rm BABAR ('06)}\\
0.39 \pm 0.18 & {\rm HFAG ('06)}
\end{array}
\right. \;  ; 
\eeq
I summarize the situation as follows: 
\begin{itemize}
\item 
Performing dedicated \cp~studies in channels driven mainly or even predominantly by 
$b \to s q \bar q$ to search for New Physics signatures makes eminent sense since the SM 
contribution, in particular from the one-loop Penguin operator, is greatly suppressed. 
\item 
The experimental situation is far from settled, as can be seen also from how the central 
value have moved over the years. It is tantalizing to see that the $S$ contribution for the 
modes in this category -- $B_d \to \pi^0 K_S, \rho^0 K_S, \omega K_S, f_0 K_S$ -- are all low 
compared to the SM expectation Eq.(\ref{PHI06}). Yet none of them is significantly lower, and 
for none of these modes a non-zero \cp~asymmetry has been established except for 
\beq 
{\rm sin}2\phi_1(B_d \to \eta ^{\prime} K_S) = 
\left\{
\begin{array}{ll} 0.64 \pm 0.10 \pm 0.04 & {\rm BELLE ('06)}\\
0.58 \pm 0.10 \pm 0.03 & {\rm BABAR ('06)}\\
0.61 \pm 0.07 & {\rm HFAG ('06)}
\end{array}
\right. \;  ; 
\label{CPETAKS}
\eeq
\item 
Obviously there is considerable space still for significant deviations from SM predictions. 
It 
is ironic that such a smaller deviation, although not significant, is actually more believable as 
signaling an incompleteness of the SM than 
the large one originally reported by BELLE. 
While it is tempting to average over all these hadronic transitions, I would firmly resist this temptation 
for the time being, till several modes exhibit a significant asymmetry. 
\item 
One complication has to be studied, though, in particular if the observed 
value of sin$2\phi_1(B_d \to \phi K_S)$ falls below the predicted one by a moderate amount only. 
For one is actually observing $B_d \to K^+K^-K_S$. If there is a single weak phase like in the SM one finds 
\beq 
{\rm sin}2\phi_1(B_d \to \phi K_S) = - {\rm sin}2\phi_1(B_d \to `f_0(980)' K_S) \; , 
\eeq 
where $`f_0(980)'$ denotes any {\em scalar} $K^+K^-$ configuration with a mass close to that of the 
$\phi$, be it a resonance or not. A smallish pollution by such a $`f_0(980)' K_S$ -- by, say, 
10\% {\em in amplitude} --  
can thus reduce the asymmetry assigned to $B_d \to \phi K_S$ significantly -- by 20\% in this example. 
\item 
In the end it is therefore mandatory to perform a {\em full time dependent Dalitz plot analysis} 
for $B_d \to K^+K^-K_S$ and compare it with that for $B_d \to 3 K_S$ and 
$B^+ \to K^+K^-K^+, \, K^+K_SK_S$ and also with $D \to 3K$. BABAR has presented a preliminary such 
study. This is a very challenging task, but 
in my view essential. There is no `royal' way to fundamental insights. 
\footnote{The ruler of a Greek city in southern Italy once approached the resident sage with the request 
to be educated in mathematics, but in a `royal way', since he was very busy with many 
obligations. Whereupon the sage replied with admirable candor: 
"There is no royal way to mathematics."} 
\item 
An important intermediate step in this direction is given by one application of 
{\em Bianco's Razor} \cite{RIO}, namely to analyze the \cp~asymmetry in $B_d \to [K^+K^-]_MK_S$ as a 
function of the cut $M$ on the $K^+K^-$ mass. 
\end{itemize}
All of this might well lead to another triumph of the SM, when its predictions agree with accurate data in the future even for these rare transition rates dominated by loop-contributions, i.e pure quantum effects.  
It is equally possible -- personally I think it actually more likely -- that future precision data will expose 
New Physics contributions. In that sense the SM might snatch victory from the jaws of defeat -- or defeat from the jaws of victory. For us that are seeking indirect manifestations of New Physics 
the roles of victory and defeat are switched, of course.

In any case the issue has to be pursued with vigour, since these reactions provide such a natural portal to New Physics on one hand and possess such an accurate yardstick from 
$B_d \to \psi K_S$.

%%%%%%%%%%%%%%
\subsubsection{The Beginning of Act V -- \cp~Violation in Charged $B$ Decays}
\label{CPCHARGED}
%%%%%%%%%%%

So far \cp~violation has not been established yet in the decays of {\em charged} mesons, which is not 
surprising, since meson-antimeson oscillations cannot occur there and it has to be purely {\em direct} 
\cp~violation. Now BELLE \cite{CHARGEDCPV} has found strong evidence for a large \cp~asymmetry in charged $B$ decays 
with a 3.9 sigma significance, namely in $B^{\pm} \to K^{\pm}\rho^0$ observed in 
$B^{\pm} \to K^{\pm}\pi^{\pm}\pi^{\mp}$ : 
\beq 
A_{\cp} (B^{\pm} \to K^{\pm}\rho^0) = \left( 30 \pm 11 \pm 2.0 ^{+11}_{-4} \right) \% 
\label{CPCHARGED}
\eeq 
I find it a most intriguing signal since a more detailed inspection of the mass peak region shows 
a pattern as expected for a genuine effect. 
Furthermore a similar signal is seen in BABAR's data, and it would make sense to undertake a careful 
average over the two data sets. 

I view BELLE's and BABAR's analyses of the Dalitz plot for $B^{\pm} \to K^{\pm}\pi^{\pm}\pi^{\mp}$ 
as important pilot studies, from which one can infer important lessons about the strengths and pitfalls 
of such studies in general. 

%%%%%%%%%%%%%%
\subsubsection{Stop the Press: \cp~Violation in $B_d \to D^+D^-$?}
\label{STOP}
%%%%%%%%%%%%%%

Most recently BELLE has found evidence for large indirect as well as direct \cp~violation in the 
Cabibbo suppressed channel $B_d \to D^+D^-$ \cite{BELLEDD} with 4.1 $\sigma$ and 
3.2 $\sigma$ significance, respectively, the latter through $C\neq0$: 
\beq 
S(D^+D^-) = -1.13 \pm 0.37 \pm 0.09 \; , \; C(D^+D^-) = -0.91 \pm 0.23 \pm 0.06 \; \; \; 
{\rm BELLE \; '07}
\label{BELLESC}
\eeq
While the central values do not satisfy the general constraint $S^2 + C^2 \leq 1$, see 
Sect.\ref{ACT2}, this can be attributed to a likely upward statistical fluctuation. The 
$S$ term, i.e. the coefficient of the sin$\Delta M_{B_d}t$ term in the asymmetry, is just one sigma 
high compared compared to its SM prediction of $- {\rm sin}2\phi_1= - 0.675 \pm 0.026$. 
Yet the $C$ term, which unambiguously represents direct \cp~violation and is expected to be very close to zero in the SM, appears to be considerably larger. If true, it would establish the presence of New Physics. One should note though that BABAR's data indicate a somewhat different 
message \cite{BABARDD} in particular with respect to $C$: 
\beq 
S(D^+D^-) = -0.29 \pm 0.63 \pm 0.06 \; , \; C(D^+D^-) = 0.11 \pm 0.35 \pm 0.06 \; \; \; 
{\rm BABAR \; '06} \; . 
\label{BABARSC}
\eeq
 
In Ref.\cite{BELLEDD} it is suggested that New Physics could enhance the Cabibbo suppressed 
Penguin operator $b \to d \bar q q$, $q=c$ considerably so as to generate the required weak phase in the $\Delta B =1$ amplitude. It is quite conceivable that the $b \to s \bar qq$ transition 
operator is much less affected by New Physics. However one needs a very peculiar scenario. For in general 
such an operator should allow also for $q=u$, $d$, $s$ and presumably with a considerably 
higher weight. Then it would provide the dominant contribution to 
the $B_d \to \pi^+\pi^-$, $2\pi^0$ and 
$K_SK_S$ channels with a branching ratio on the about $10^{-4}$ level or probably more 
-- in clear conflict with the data. I.e., one had to postulate that the contributions 
$b \to d \bar uu$, $d \bar dd$ and $d \bar ss$ due to New Physics had to be suppressed greatly 
relative to $b \to d \bar cc$.

%%%%%%%%%%%%%
\subsection{Summary}
\label{SUMII}
%%%%%%%%%%%

As explained in Sect.\ref{PART1} while CKM forces are generated by the exchange of gauge bosons, its couplings involve elements of the CKM matrix the latter being derived from the up- and down-type 
quark {\em mass matrices}. Thus the CKM parameters are intrinsically connected with one of the central 
mysteries of the SM, namely the generation of fermion masses and family replication. 
Furthermore the hierarchy in the quark masses and the likewise hierarchical pattern of the CKM matrix 
elements strongly hints at some deeper level of dynamics about which we are quite ignorant.  
Nevertheless CKM theory with its mysterious origins has proved itself to be highly successful in describing even quantitatively a host of phenomena occurring over  a wide array of scales. It 
lead to the `Paradigm of Large \cp~Violation in $B$ Decays' as a prediction in the old-fashioned sense; 
i.e., predictions were made well before data of the required sensitivity existed. From the observation 
of a tiny and shy phenomenon -- \cp~violation in $K_L$ decays on the ${\cal O}(10^{-3})$ level -- it 
predicted without `plausible deniability' almost ubiquitous manifestations of \cp~violation about two orders of magnitude larger in $B$ decays. This big picture has been confirmed now in qualitative as well as impressively quantitative agreement with SM predictions: 
\begin{itemize}
\item 
Two \cp~insensitive observables, namely $|V(ub)/V(cb)|$ and $\Delta M_{B_d}/\Delta M_{B_s}$, imply that \cp~violation has to exist and in a way that at present is fully consistent with the measurements 
of $\epsilon$ and sin$2\phi_1$. 
\item 
Time dependent \cp~asymmetries in the range of several $\times$ 10 \% have been established in 
$B_d \to \psi K_S$, $\pi^+\pi^-$ and $\eta^{\prime} K_S$ with several others on the brink of being found. 
\item 
{\em Direct} \cp~violation of about 10\% or even larger have been discovered in $B_d \to \pi^+\pi^-$ 
and $K^- \pi^+$. 
\item 
The first significant sign of \cp~violation in a charged meson has surfaced in 
$B^{\pm} \to K^{\pm} \rho^0$. 
\item 
The optimists among us might discern the first signs of tension between data and the predictions 
of CKM theory in  $|V(ub)/V(cb)|$ \& $\Delta M_{B_d}/\Delta M_{B_s}$ vs. sin$2\phi_1$ and in the 
\cp~asymmetries in $b \to s q \bar q$ vs. $b\to c \bar c s$ driven transitions.

\end{itemize}
For all these successes it is quite inappropriate to refer anymore to CKM theory as an `ansatz' with the 
latter's patronizing flavour \footnote{The German `ansatz' refers to an educated guess.}.  Instead I would characterize these developments as "the expected triumph 
of peculiar theory". 
I will indulge myself in three more `cultural' conclusions: 
\begin{itemize}
\item 
The aforementioned `CKM Paradigm of Large \cp~Violation in $B$ Decays' is due to the confluence 
of several favourable, yet a priori less than likely factors that must be seen as gifts from nature: she had 
(i) arranged for a huge top mass, (ii) a "long" $B$ lifetime, (iii) the $\Upsilon (4S)$ resonance being above the $B \bar B$, yet below the $B\bar B^*$ thresholds and (iv) regaled us previously with charm hadrons, which prompted the development detectors with an 
effective resolution that is needed to track $B$ decays.  
\item 
`Quantum mysteries' like EPR correlations with their intrinsic non-local features were essential  
for observing \cp~violation involving $B_d- \bar B_d$ oscillations in 
$\Upsilon (4S) \to B_d \bar B_d$ and to establish that indeed there is \ot~violation commensurate 
with \cp~violation. 
\item 
While hadronization is not easily brought under quantitative theoretical control, it enhances greatly 
observable \cp~asymmetries and can provide most valuable cross checks for our interpretation of data.  
 
\end{itemize}

%%%%%%%%%%%%%%%%
\section{Probing the Flavour Paradigm of the {\em Emerging New} Standard Model}
\label{PART3}
%%%%%%%%%%%%

%%%%%%%%%
\subsection{On the Incompleteness of the SM}
\label{INCOMPLET}
%%%%%%%%

As described above the SM has scored novel -- i.e., qualitatively new -- successes in the last few years in the realm of flavour dynamics. Due to the very peculiar structure of the latter they have to be viewed as amazing. Yet even so the situation can be characterized with a slightly modified quote from Einstein: "We know a lot -- yet understand so little." 
I.e., these successes do {\em not} invalidate  the general arguments in favour of the SM being 
{\em incomplete} -- the search for New Physics is as mandatory as ever. 

You have heard about the need to search for New Physics before and what the outcome has been of such efforts so far.  And it reminds you of a quote by Samuel Beckett: 
"Ever tried? Ever failed? No matter. Try again. Fail again. Fail better." 
Only an Irishman can express profound skepticism concerning the world in such a poetic way. 
Beckett actually spent most of his life in Paris, since Parisians like to listen to someone expressing such a world view, even while they do not share it. Being in the service of Notre Dame du Lac, the home of the `Fighting Irish', I cannot just ignore such advice. 

Yet there are -- in my judgement compelling -- {\em theoretical} arguments pointing to the existence of New Physics.    
\begin{itemize}
\item 
While electric charge quantization $Q_e = 3 Q_d = - \frac{3}{2} Q_u$ 
is an essential ingredient of the SM -- it allows to vitiate the Adler-Bell-Jackiw anomaly -- it does not offer any 
understanding. It would naturally be explained through Grand Unification at very high energy scales 
implemented through, e.g., $SO(10)$ gauge dynamics. 
I call this the `guaranteed New Physics'. 
\item 
We infer from the observed width of $Z^0$ decays that there are  three (light) neutrino species. The hierarchical pattern of CKM parameters as revealed by the data is so peculiar as to suggest that some other dynamical 
layer has to underlie it. I refer to it as `strongly suspected New Physics' or {\bf ssNP}. 
We are quite in the dark about its relevant scales. 
Saying we pin our hopes for explaining the family replication on Super-String or M theory is a scholarly way of saying 
we have hardly a clue what that {\bf ssNP} is. 
\item 
What are the dynamics driving the electroweak symmetry breaking of 
$SU(2)_L\times U(1) \to U(1)_{QED}$? How can we tame  the instability of Higgs dynamics with its quadratic mass divergence? 
I find the arguments compelling that point to New Physics at the 
$\sim 1$ TeV scale -- like low-energy SUSY; therefore I call it the `confidently predicted' New Physics 
or {\bf cpNP}. 
\item 
Last and possibly least the `Strong \cp~Problem' of QCD has not been resolved. 
Similar to the other shortcomings it is a purely theoretical problem in the sense that the offending coefficient for the \op~and \cp~odd operator $\tilde G\cdot G$ can be 
fine-tuned to zero , see Sect.\ref{QCD}, -- yet in my eyes it is an intriguing problem. 

\end{itemize}
Even better, there is strong experimental evidence for New Physics: 
\begin{itemize}
\item 
The occurrence of at least two classes of neutrino oscillations has been established. 
\item 
{\em Dark Matter}: Analysis of the rotation curves of stars and galaxies reveal that there is a lot more 
`stuff' -- i.e. gravitating agents -- out there than meets the eye. About a quarter of the gravitating agents in the Universe are such dark matter, and they have to be mostly nonbaryonic. The SM has {\em no} candidate for it. 
\item 
{\em The Baryon Number of the Universe}: one finds only about one 
baryon per $10^9$ photons with the latter being mostly in the cosmic background radiation; there is 
no evidence for {\em primary} antimatter. We know standard CKM dynamics is irrelevant for the Universe's baryon number. Therefore New Physics has to exist. The success of CKM dynamics
tells us that \cp~violating phases can be large -- an insight that should be helpful for any attempt at 
explaining baryogenesis.

\item 
{\em Dark Energy}: Type 1a supernovae are considered `standard candles'; i.e. considering 
their real light output known allows to infer their distance from their apparent brightness. 
When in 1998 two teams of researchers studied them at distance scales of about 
five billion light years, they found them to be fainter as a function of their redshift than what the 
conventional picture of the Universe's decelerating expansion would yield. Unless gravitational 
forces are modified over cosmological distances, one has to conclude the Universe is filled  with 
an hitherto completely unknown agent {\em accelerating} the expansion. A tiny, yet non-zero 
cosmological constant would apparently `do the trick' -- yet it would raise more fundamental 
puzzles.

\end{itemize} 
These heavenly signals are unequivocal in pointing to New Physics, yet leave wide open 
the nature of this New Physics.

Thus we can be assured that New Physics exists `somehow' `somewhere', and quite likely 
even `nearby', namely around the TeV scale; above I have called the latter
{\bf cpNP}. The LHC program and the Linear Collider 
project are justified -- correctly -- to conduct campaigns for 
{\bf cpNP}. That is unlikely to shed light on the {\bf ssNP}, though it might. Likewise I would not 
{\em count} on 
a comprehensive and detailed program of heavy flavour studies to  
shed light on the {\bf ssNP} behind the flavour puzzle of the SM. 
Yet the argument is reasonably turned around: such a program 
will be essential to elucidate salient features of the 
{\bf cpNP} by probing the latter's flavour structure and having sensitivity to scales of order 10 TeV 
or even higher. 
One should keep in mind the following: one very popular example of 
{\bf cpNP} is supersymmetry; {\em yet it represents an organizing principle much more than even a class of theories}. I find it unlikely we can infer all required lessons by studying only flavour {\em diagonal}  
transitions. Heavy flavour decays provide a powerful and complementary probe of 
{\bf cpNP}. Their potential to reveal something about the {\bf ssNP} is a welcome extra not required 
for justifying efforts in that direction. 

Accordingly I see a dedicated heavy flavour program as an essential complement to the 
studies pursued at the high energy frontier at the TEVATRON, LHC and hopefully ILC. 
I will illustrate this assertion in the remainder of this review. 

%%%%%%%%%%
\subsection{$\Delta S \neq 0$ -- the `Established Hero'}
\label{STRANGEHERO}
%%%%%%%%%

The chapter on $\Delta S \neq 0$ transitions is a most glorious one in the history of particle physics, 
as sketched in Table \ref{tab:STRANGELESSONS}.
\begin{table}
\begin{center}
\small{\begin{tabular}{ll}
\hline
Observation      & Lesson learnt     \\
\hline
      $\tau - \theta $ Puzzle   & \op~violation \\ 
      production rate $\gg$ decay rate  & concept of families \\ 
      suppression of flavour changing neutral currents  & GIM mechanism \& existence of charm \\
    $K_L \to \pi \pi$ & \cp~violation \& existence of top \\
    \hline 
\end{tabular}}
\caption{On the History of $\Delta S \neq 0$ Studies}
\label{tab:STRANGELESSONS}
\end{center}
\end{table} 
We should note that all these features, which now are pillars of the SM, were New Physics 
{\em at that time}. Yet the discovery potential in strange decays might not have been 
exhausted.

%%%%%%%%%%%%
\subsubsection{The `Dark Horse'} 
%%%%%%%% 

The \ot~odd moment (see Sect.\ref{BASICSDISC}) 
\beq 
{\rm Pol}_{\perp}(\mu) \equiv \frac{\langle \vec s(\mu) \cdot (\vec p(\mu) \times \vec p(\pi))\rangle}
{|\vec p(\mu) \times \vec p(\pi)|} 
\eeq
measured in $K^+ \to \mu^+ \nu \pi^0$ would  
\begin{itemize}
\item 
represent genuine \ot~violation (as long as it exceeded the $10^{-5}$ level) and 
\item 
constitute prima facie evidence for \cp~violation in {\em scalar} dynamics. 
\end{itemize}
For to generate Pol$_{\perp}(\mu)$ one needs a complex phase between helicity conserving and 
violating $\Delta S =1$ amplitudes. The latter can be provided by the tree level exchange of a 
(charged) Higgs boson. 

The present upper bound has been obtained by KEK experiment E246 \cite{prdp}: 
\beq 
{\rm Pol}_{\perp}(\mu) = (-1.7 \pm 2.3 \pm 1.1)\cdot 10^{-3} \leq 5\cdot 10^{-3} \; \; {\rm 90\% \; C.L.} \; , 
\label{KEKRES}
\eeq
which is almost three orders of magnitude above the SM expectation. An experiment E06 (TREK) 
proposed for J-PARC aims at improving the sensitivity by close to two orders of magnitude down to the 
$10^{-4}$ level for Pol$_{\perp}(\mu)$. While hoping for a $10^{-3}$ signal required considerable 
optimism, the prospects for an effect $\geq 10^{-4}$ are more realistic.

%%%%%%%%%
\subsubsection{The `Second Trojan War': $K \to \pi \nu \bar \nu$ }  
\label{SECTROJAN}
%%%%%%%

According to Greek Mythology the Trojan War described in Homer's Iliad was actually 
the second war over Troja. In a similar vein I view the heroic campaign over 
$K^0 - \bar K^0$ oscillations -- $\Delta M_K$, $\epsilon_K$ and $\epsilon ^{\prime}$ -- as 
a first one to be followed by a likewise epic struggle over the two ultra-rare modes 
$K^+ \to \pi^+ \nu \bar \nu$ and 
$K_L \to \pi^0 \nu \bar \nu$. This campaign has already been opened through the observation 
of the first through three events very roughly as expected within the SM. The second one, which 
requires \cp~violation for its mere existence, so far remains unobserved at a level well above 
SM predictions. These reactions 
are like `standard candles' for the SM: their rates are functions of 
$V(td)$ with a theoretical uncertainty of about 5\% and 2\% respectively, which is mainly due to 
the uncertainty in the charm quark mass. 

While their rates could be enhanced by New Physics greatly over their SM expectation, 
I personally find that somewhat unlikely for various reasons. Therefore I suggest one should aim for 
collecting ultimately about 1000 events of these modes to extract the value of 
$V(td)$ and/or identify likely signals of New Physics.

%%%%%%%%
\subsection{The `King Kong' Scenario for New Physics Searches}
\label{KONG}
%%%%%%%

This scenario can be formulated as follows: "One is unlikely to encounter King Kong; yet 
once it happens one will have no doubt that one has come across something quite out of the ordinary!" 

What it means can be best illustrated with the historical precedent of $\Delta S \neq 0$ studies sketched 
above: the existence of New Physics can unequivocally be inferred if there is a 
{\em qualitative} conflict between data and expectation; i.e., if a theoretically `forbidden' process is 
found to proceed nevertheless -- like in $K_L \to \pi \pi$ -- or the discrepancy between expected and 
observed rates amounts to several orders of magnitude -- like in $K_L \to \mu ^+\mu^-$ or 
$\Delta M_K$. History might repeat itself in the sense that future measurements might reveal such {\em qualitative} 
conflicts, where the case for the manifestation of New Physics is easily made. This does not mean that the effects are large or straightforward to discover -- only that they are much larger than the truly minute SM effects. 

I have already mentioned one potential candidate for revealing such a qualitative conflict, namely 
the muon transverse polarization in $K_{\mu 3}$ decays. 

%%%%%%%%%%%%
\subsubsection{Electric Dipole Moments}
\label{EDMS}
%%%%%%%%%%%

The energy shift of a system placed inside a weak 
electric field can be expressed through an expansion in terms of the components of that 
field $\vec E$: 
\beq 
\Delta {\cal E} = d_i E_i + d_{ij}E_iE_j + {\cal O}(E^3) 
\eeq
The coefficients $d_i$ of the term linear in the electric field form a vector $\vec d$, 
called an electric dipole moment (EDM). For a {\em non-}degenerate system -- it does not have to be elementary -- one infers from symmetry considerations that this vector has to be proportional to 
that system's spin: 
\beq 
\vec d \propto \vec s
\eeq
Yet, since under time reversal \ot 
\beq
E_i \stackrel{\ot} \to E_i \; \; , \; \; s_i \stackrel{\ot} \to - s_i \; , 
\eeq
a non-vanishing EDM constitutes \ot~violation. 

No EDM has been observed yet; the upper bounds of the neutron and electron EDM read 
as follows \cite{PDG06}: 
\bea 
d_N &<& 5 \cdot 10^{-26} \; {\rm e\, cm} \; \; \;  {\rm [from \; ultracold\; neutrons]}\\
d_e &<& 1.5 \cdot 10^{-27} \; {\rm e\, cm} \; \; \; {\rm [from\; atomic \; EDM]}
\eea
The experimental sensitivity achieved can be illustrated as follows: (i) An 
neutron EDM of $5\cdot 10^{-26}$ e cm of an object with a radius 
$r_N \sim 10^{-13}$ cm scales to a displacement of about 7 micron, i.e. less than 
the width of human hair, for an object of the size of the earth. (ii) Expressing the uncertainty 
in the measurement of the electron's magnetic dipole moment -- 
$\delta ((g-2)/2) \sim 10^{-11}$ -- in analogy to its EDM, one finds a sensitivity level of 
$\delta(F_2(0)/2m_e) \sim 2 \cdot 10^{-22}$ e cm compared to 
$d_e< 2\cdot 10^{-26}$ e cm. 

Despite the tremendous sensitivity reached these numbers are still several orders of magnitude above what is expected in CKM theory: 
\bea 
d_N^{CKM} &\leq &  10^{-30} \; {\rm e\, cm}  \\
d_e^{CKM} &\leq &  10^{-36} \; {\rm e\, cm} \; , 
\eea
where in $d_N^{CKM}$ I have ignored any contribution from the strong \cp~problem. 
These numbers are so tiny for reasons very specific to CKM theory, namely its chirality structure 
and the pattern in the quark and lepton masses. Yet New Physics scenarios with right-handed 
currents, flavour changing neutral currents, a non-minimal Higgs sector, heavy neutrinos etc. are 
likely to generate considerably larger numbers: $10^{-28} - 10^{-26}$ e cm represents a 
very possible range there quite irrespective of whether these new forces contribute to 
$\epsilon_K$ or not. This range appears to be within reach in the foreseeable future. 
There is a vibrant multiprong program going on at several places. 
Such experiments while being of the `table top' variety require tremendous efforts,  persistence and ingenuity -- yet the insights to be gained by finding a nonzero 
EDM somewhere are tremendous.

%%%%%%%%%%%
\subsubsection{On the Brink of Major Discoveries in Charm Transitions}
\label{CHARMDEC}  
%%%%%%%%%%

The study of charm dynamics had a great past: It was instrumental in 
driving the paradigm shift from quarks as mathematical entities to physical objects and in 
providing essential support for accepting QCD as the theory of the strong interactions.  
Yet it is often viewed as one without a future. For charm's electroweak phenomenology  
is on the `dull' side: CKM parameters are known from other sources, $D^0 - \bar D^0$ oscillations 
slow, \cp~asymmetries small and loop driven decays extremely rare. 

Yet more thoughtful observers have realized that the very `dullness' of the SM phenomenology for charm 
provides us with a dual opportunity, namely to 
\begin{itemize}
\item
probe our quantitative understanding of QCD's nonperturbative dynamics thus calibrating our theoretical tools for $B$ decays and 
\item 
perform almost `zero-background' \footnote{The meaning of almost `zero-background'  has of course to be updated in the light of the 
increasing experimental sensitivity.} searches 
for New Physics. Charm transitions actually provide a unique and novel 
portal to flavour dynamics with the experimental situation 
being a priori favourable (except for the lack of Cabibbo suppression).  
While New Physics signals can exceed SM predictions on 
\cp~asymmetries by orders of magnitude, they might not be large in absolute terms, as specified later 
\cite{GROSS2}. 

\end{itemize}
The former goal provides a central motivation for the CLEO-c program at Cornell University. The latter 
perspective has received a major boost through the strong, albeit not yet conclusive evidence for 
$D^0 - \bar D^0$ oscillations presented by BABAR and BELLE in March 2007. The main three 
positive findings are: 
\begin{itemize}
\item 
Comparing the effective lifetimes for $D^0 \to K^+K^-$ and $D^0 \to K^-\pi^+$ BELLE obtains a 3.2 
$\sigma$ signal for a difference \cite{BELLEOSC1}: 
\beq 
y_{CP} = \frac{\tau (D^0 \to K^-\pi^+)}{\tau (D^0 \to K^+K^-)} - 1 = (1.31 \pm 0.32 \pm 0.25) \cdot 10^{-2}   
\label{BELLEYCP}
\eeq
In the limit of \cp~invariance, which provides a good approximation for charm decays as explained later, 
the two mass eigenstates of the $D^0 - \bar D^0$ complex are \cp~eigenstates as well. 
$D^0 \to K^+K^-$ yields the width for the \cp~even state and $D^0 \to K^-\pi^+$ the one averaged 
over the \cp~even and odd states. In this limit one has 
\beq 
y_{CP} = y_D = \frac{\Delta \Gamma_D}{2\bar \Gamma _D}
\eeq
\item 
Analyzing the time {\em dependent} Dalitz plot for $D^0(t) \to K_S\pi^+\pi^-$ BELLE finds \cite{BELLEOSC2}
\beq 
x_D \equiv \frac{\Delta M_D}{\Gamma_D} = (0.80 \pm 0.29 \pm 0.17)\cdot 10^{-2} \; , \; 
y_D = (0.33 \pm 0.24 \pm 0.15) \cdot 10^{-2} \; , 
\eeq
which amounts to a 2.4 $\sigma$ signal for $x_D \neq 0$. 
\item 
BABAR has studied the decay rate evolution for the doubly Cabibbo suppressed mode 
$D^0(t) \to K^+\pi^-$ and found \cite{BABAROSC}
\beq 
y_D^{\prime} = (0.97 \pm 0.44 \pm 0.31)\cdot 10^{-2} \; , \; 
(x_D^{\prime})^2 = (-2.2 \pm 3.0 \pm  2.1)\cdot 10^{-4} 
\eeq
representing a 3.9 $\sigma$ signal for $[y_D^{\prime}, (x_D^{\prime})^2] \neq [0,0]$ due to the 
correlations between $y_D^{\prime}$ and $ (x_D^{\prime})^2$. The observables 
$x_D^{\prime}$ and $y_D^{\prime}$ are related to $x_D$ and $y_D$ through the strong phase shift 
$\delta$ of the amplitude for $D^0 \to K^+\pi^-$ to that for $D^0 \to K^-\pi^+$: 
\beq 
x_D^{\prime} = x_D {\rm cos}\delta + y_D {\rm sin}\delta \; , \; 
y_D^{\prime} = -x_D {\rm sin}\delta + y_D {\rm cos}\delta
\label{XPRIME}
\eeq
\end{itemize} 
A `preliminary' average by the heavy Flavour Averaging Group over all relevant data yields 
\beq 
x_D = (0.85 \pm 0.32)\cdot 10^{-2} \; , \; y_D = (0.71 \pm 0.21)\cdot 10^{-2}\; , \; 
{\rm cos}\delta = 0.40 ^{+0.23}_{-0.31} 
\eeq
with 5 $\sigma$ significance for $[x_D,y_D] \neq [0,0]$ -- and the caveat that averaging over 
the existing data sets has to be taken with quite a grain of salt at present due to the 
complicated likelihood functions \cite{ASNERWS}. 

Establishing $D^0 - \bar D^0$ oscillations would provide a qualitatively new insight into 
flavour dynamics. After having discovered oscillations in {\em all three} mesons built from {\em down}-type quarks -- $K^0$, $B_d$ and $B_s$ -- it would be the first observation of oscillations with 
{\em up}-type quarks; it would also remain the only one (at least for three-family scenarios): 
while top quarks do not hadronize \cite{RAPALLO} thus removing the conditio sine qua non for 
$T^0 - \bar T^0$ oscillations, 
one cannot have $\pi^0 - \pi^0$ oscillations in the $u$ quark sector.  This means that 
analyzing flavour changing neutral currents (FlChNC) in charm transitions provides a rather unique portal to New Physics, where FlChNC could be much less suppressed for {\em up}-type than for 
{\em down-}type quarks. 

The observables $x_D$ and $y_D$ are doubly Cabibbo suppressed -- 
$\Delta M_D/\Gamma_D$, $\Delta \Gamma_D /2\Gamma_D \propto {\rm tg}^2\theta_C$ -- 
and vanish in the limit of $SU(3)_{Fl}$ symmetry. The 
history of SM predictions for them beyond these general statements is a rather checkered one with `suggested' numbers 
differing by orders of magnitude. Some of this variation is due to authors following the SM description 
for $\Delta S=2$ or $\Delta B=2$ dynamics literally by inferring ${\cal L}(\Delta C=2)$ from quark box 
diagrams treated as {\em short}-distance dynamics. This is, however, not justified. 
It is widely understood now that the usual quark box diagram is irrelevant due to its untypically severe 
GIM suppression $(m_s/m_c)^4$. Two complementary approaches have been employed 
for estimating the size of $x_D$ and $y_D$. 
(i) A systematic 
analysis based on an OPE treatment has been given in Ref.\cite{BUDOSC}. The novel feature is 
that the numerically largest contributions do not come from terms of leading order in $1/m_c$, 
but from higher-dimensional operators. For those possess a much softer 
GIM reduction of $(m_s/\mu_{had})^2$ due to nonperturbative `condensate'  terms characterized 
by a mass scale $\mu_{had}$ (even $m_s/\mu_{had}$ terms could arise in the presence of right-handed currents). The resulting expansion is expressed through powers of $1/m_c$, $m_s$ and 
$\mu_{had}$ yielding 
\beq 
\left. x_D (SM)\right|_{OPE}, \; \left. y_D (SM)\right|_{OPE} \sim {\cal O}(10^{-3}) \; . 
\eeq 
(i) The authors of 
Refs.\cite{FALK1,FALK2} find similar numbers, albeit in a quite different approach:  estimating  
$SU(3)_{Fl}$ breaking for $\Delta \Gamma_D$ from phase space differences for two-, three- and 
four-body $D$ modes they obtain $y_D(SM) \sim 0.01$ and inferring $x_D$ from $y_D$ via a 
dispersion relation they arrive at $0.001 \leq |x_D(SM)| \leq 0.01$ with $x_D$ and $y_D$ being of opposite sign. 

While one predicts similar numbers for $x_D$ and $y_D$, one should keep in mind 
that they arise in very different dynamical environments: $\Delta M_D$ is generated from 
{\em off}-shell intermediate states and thus is sensitive to New Physics, which could affect it  
considerably. $\Delta \Gamma_D$ on the other hand is shaped by 
{\em on}-shell intermediate 
states; while it is hardly sensitive to New Physics (for a dissenting opinion, see 
Ref.\cite{ALEXEI}), it involves less averaging or `smearing' than 
$\Delta M_D$ making it thus more vulnerable to violations of quark-hadron duality. 

In summary: to the best of our present knowledge even values for $x_D$ and $y_D$ as `high' as  
0.01 could be due entirely to SM dynamics of otherwise little interest. It is likewise possible that a 
large or even dominant part of $x_D \sim 0.01$ in particular is due to New Physics. While one should 
never rule out a theoretical breakthrough, I am less than confident that even the usual panacea, namely  
lattice QCD, can provide a sufficiently fine instrument in the foreseeable future. 

Yet despite this lack of an unequivocal statement from theory one wants to probe 
these oscillations as accurately as possible even in the absence of the aforementioned breakthrough, 
since they represent an intriguing quantum mechanical phenomenon and -- on the more 
practical side -- constitute an important ingredient for \cp~asymmetries arising in $D^0$ decays 
due to New Physics as explained next. 

%%%%%%%%%%%%%%%%
%\subsubsection{\cp~Violation without Oscillations -- Partial Widths, Moments etc.}
%\label{CPVCHARM}
%%%%%%%%%%%%%%%%%%

%%%%%%%%%%%%%%%%
{\bf (i) \cp~Violation without Oscillations -- Partial Widths, Moments etc.}
%%%%%%%%%%%%%%

As explained before the 
observed baryon number implies the existence of New Physics in 
\cp~violating dynamics. It would therefore be unwise not to undertake dedicated searches for \cp~asymmetries in charm decays, in particular, since those offer several pragmatic advantages. 
(i) While we do not know how to reliably compute the strong phase shifts 
required for direct \cp~violation to emerge in partial widths, we can expect them to be in general large, 
since charm decays proceed in a resonance domain. (ii) The branching ratios into relevant modes 
are relatively large. (iii) \cp~asymmetries can be linear in New Physics amplitudes thus enhancing sensitivity to the latter. (iv)  The `background' from known physics is small: within the SM the effective weak phase is highly diluted, namely $\sim {\cal O}(\lambda ^4)$. {\em Without} oscillations only 
direct \cp~violation can occur, and it can 
arise only in singly Cabibbo suppressed transitions, where one  
expects them to reach no better than the 0.1 \% level; significantly larger values would signal New Physics.  
{\em Almost any} asymmetry in Cabibbo 
allowed or doubly suppressed channels requires the intervention of New Physics, since -- in the absence of oscillations -- there is only one weak amplitude. The exception are channels containing 
a $K_S$ (or $K_L$) in the final state like $D \to K_S \pi$. There are two sources for a 
\cp~asymmetry from known dynamics: (i) There are actually two transition amplitudes involved, namely 
a Cabibbo favoured and a doubly suppressed one, $D \to \bar K^0\pi$ and $D \to K^0\pi$, respectively. 
Their relative weak CKM phase is given by $\eta A^2 \lambda ^6 \sim {\rm few} \cdot 10^{-5}$, which seems to be well beyond observability. (ii) While one has $|T(D \to \bar K^0 \pi)| = 
|T(\bar D \to K^0 \pi)|$, the well-known \cp~impurity $|p_K|\neq |q_K|$ in the $K_S$ wave function introduces a difference 
between $D^{0,+}\to K_S\pi^{0,+}$ and $\bar D^{0,-}\bar K_S \pi^{0,-}$ of 
$\frac{|q_K|^2 - |p_K|^2}{|q_K|^2 + |p_K|^2} = (3.32 \pm 0.06)\cdot 10^{-3}$ \cite{CICERONE}. 

Decays to final states of {\em more than} two pseudoscalar or one pseudoscalar and one vector meson contain 
more dynamical information than given by their  widths; their distributions as described by Dalitz plots 
or \ot{\em -odd} moments can exhibit \cp~asymmetries that can be considerably larger than those for the 
width. Final state interactions while not necessary for the emergence of such effects, can fake a signal; 
yet that can be disentangled by comparing \ot{\em -odd} moments for \cp~conjugate modes, as explained below. All \cp~asymmetries observed so far in $K_L$ and $B_d$ decays 
except one concern partial widths, i.e. 
$\Gamma (P \to f)$ vs. $\Gamma (\bar P \to \bar f)$. The one 
notable exception can teach us important lessons for future searches both in charm and $B$ decays, namely the \ot~odd moment found in $K_L\to \pi ^+ \pi ^- e^+ e^-$, as discussed 
around Eq.(\ref{PHISEHGAL}): a large asymmetry $A \simeq 14 \%$ driven by the tiny 
impurity parameter $|\epsilon_K| \sim 0.22\%$ was found in the rare mode with a 
branching ratio of about $3\cdot 10^{-7}$! I.e., one can trade the size of the branching ratio 
for that of a \cp~asymmetry. 

$D$ decays can be treated in an analogous way.  Consider the Cabibbo suppressed channel 
\footnote{This mode can exhibit direct \cp~violation even within the SM.}
\beq 
\stackrel{(-)}D \to K \bar K \pi^+\pi^-
\eeq
and define by $\phi$ now the angle between the $K \bar K$ and $\pi^+\pi^-$ planes. Then 
one has 
\bea 
\frac{d\Gamma}{d\phi}(D \to K \bar K\pi^+\pi^-) &=& \Gamma_1 {\rm cos}^2 \phi + 
\Gamma_2 {\rm sin}^2 \phi + \Gamma_3 {\rm cos} \phi {\rm sin}\phi \\
\frac{d\Gamma}{d\phi}(\bar D \to K \bar K\pi^+\pi^-) &=& \bar \Gamma_1 {\rm cos}^2 \phi + 
\bar \Gamma_2 {\rm sin}^2 \phi + \bar \Gamma_3 {\rm cos} \phi {\rm sin}\phi 
\eea
The partial width for $D[\bar D] \to K\bar K \pi^+\pi^-$ is given by 
$\Gamma_{1,2} [\bar \Gamma_{1,2}]$; $\Gamma_1 \neq \bar \Gamma_1$ or 
$\Gamma_2 \neq \bar \Gamma_2$ represents direct \cp~violation in the partial width. 
$\Gamma_3 \& \bar \Gamma_3$ constitute \ot~odd correlations. By themselves they do not necessarily 
indicate \cp~violation, since they can be induced by strong final state interactions. However 
\beq 
\Gamma_3 \neq \bar \Gamma_3 \; \; \Longrightarrow \cp~{\rm violation!}
\eeq 
It is quite possible or even likely that a difference in $\Gamma_3$ vs. $\bar \Gamma_3$ 
is significantly larger than in $\Gamma_1$ vs. $\bar \Gamma_1$ or 
$\Gamma_2$ vs. $\bar \Gamma_2$. Furthermore one can expect that differences in detection 
efficiencies can be handled by comparing $\Gamma_3$ with $\Gamma_{1,2}$ and 
$\bar \Gamma_3$ with $\bar \Gamma_{1,2}$. A pioneering search for such an effect has been 
undertaken by FOCUS \cite{PEDRINI}.

%%%%%%%%%%%%%%%%
%\subsubsection{Oscillations -- the New Portal to \cp~Violation }
%\label{CPVOSC}
%%%%%%%%%%%%%%%%%%

%%%%%%%%%%%%%%%%
{\bf (ii) Oscillations -- the New Portal to \cp~Violation} 
%%%%%%%%%%%%%%%

With oscillations on an observable level -- and it seems $x_D$, $y_D$ $\sim 0.005 - 0.01$ satisfy 
this requirement -- the possibilities for \cp~asymmetries proliferate. 

At the very least -- i.e. even if ${\cal L}(\Delta C=2)$ is generated by SM dynamics alone and thus 
does not contain any appreciable \cp~violation -- oscillations provide another stage for 
a \cp~asymmetry to surface in Cabibbo favoured channels like 
$D^0 \to K_S \rho^0$ (or $D^0 \to K_S\pi^0,\, K_S\phi$): in addition to the direct asymmetries mentioned just above -- of about 
$10^{-4}$ due to the interference between $D^0 \to \bar K^0 \rho ^0$ and 
$D^0 \to K^0 \rho ^0$ and of $(3.32 \pm 0.06)\cdot 10^{-3}$ due to $|p_K| \neq |q_K|$ -- one obtains a 
time dependent asymmetry in qualitative analogy to $B_d \to \psi K_S$ given by 
\beq 
x_D \cdot \frac{t}{\tau _D} \cdot {\rm Im}\frac{q}{p} \frac{T(\bar D^0 \to K_S\rho^0)}
{T(D^0 \to K_S\rho^0)} \simeq x_D \cdot \frac{t}{\tau _D} \cdot {\rm Im}\frac{V^*(cs)V(ud)}{V(cs)V^*(ud)} 
\simeq 2x_D \cdot \frac{t}{\tau _D} \eta (A\lambda ^2)^2 \sim x_D \cdot \frac{t}{\tau _D} \cdot 10^{-3}
\label{DTIMECPV}
\eeq
${\rm Im}\frac{q}{p} \frac{T(\bar D^0 \to K_S\phi )}{T(D^0 \to K_S\phi)}\simeq 
2\eta (A\lambda ^2)^2$ is an accurate SM prediction with{\em out} a hadronic uncertainty. Alas 
with $x_D \sim 0.01$ it amounts to a $10^{-5}$ effect and is presumably too small to be observed.

The more intriguing scenario arises, when New Physics contributes significantly to 
${\cal L}(\Delta C=2)$, which is still quite possible. I will begin by drawing on analogies with two other cases, namely the retrospective one of $K_L$ and the very topical one of $B_s$. 
(i) Let us assume that -- contrary to history -- at the time of the discovery of $K^0 - \bar K^0$ oscillations
the community had already established the SM with {\em two} families, been 
aware of the possibility of \cp~violation and the need for three families to implement the latter. 
They would then have argued that $\Delta M_K$ could be generated by long-distance 
dynamics through {\em off}-shell $K^0 \to "\pi^0, \eta , \eta ^{\prime}, 2\pi" \to \bar K^0$ etc. 
Indeed roughly  
half the observed size of $\Delta M_K$ can be produced that way. Yet they would have realized 
that long-distance dynamics can{\em not} induce \cp~violation in $K^0 \to \bar K^0$, i.e. 
$\epsilon_K \neq 0$. The latter observable is thus controlled by short-distance dynamics. 
Finding a time dependent \cp~asymmetry would then show the presence of 
physics beyond the SM then, namely the third family. (ii) $\Delta M (B_s)$ has been observed to be 
consistent with the SM prediction within mainly theoretical uncertainties; yet since those are 
still sizable, we cannot rule out that New Physics impacts $B_s - \bar B_s$ oscillations 
significantly. This issue, which is unlikely to be resolved theoretically, can be decided experimentally 
by searching for a time dependent \cp~violation in $B_s(t) \to \psi \phi$. For within the SM one predicts \cite{BS80} a very small asymmetry not exceeding 4\% in this transition since on the leading CKM level quarks of only the second and third family contribute; this will be discussed in detail in Sect.\ref{NLBS}. Yet in general one can expect New Physics contributions to $B_s - \bar B_s$ oscillations to exhibit a weak phase that is not particularly 
suppressed. Even if New Physics affects $\Delta M(B_s)$ only moderately, it could greatly enhance 
sin$2\phi (B_s \to \psi \phi)$, possibly even by an order of magnitude!

These examples can be seen as `qualitative' analogies only, not quantitatives one with 
$D^0 - \bar D^0$ oscillations being (at best) quite slow. Since $y_D$, $x_D \ll 1$, it suffices to give the 
decay rate evolution to first order in those quantities only (the general expressions can be found in 
Ref.\cite{CICERONE}): 
\bea 
\nonumber
\Gamma (D^0(t) \to K^+K^-) &\propto & e^{-\Gamma_1t}|T(D^0 \to K^+K^-)|^2 \times  \\
\nonumber 
&& 
\left[ 1 +y_D\frac{t}{\tau_D} \left( 1 - {\rm Re}\frac{q}{p}\bar \rho_{K^+K^-}\right) - 
x_D\frac{t}{\tau_D}{\rm Im}\frac{q}{p}\bar \rho_{K^+K^-}\right] \\
\nonumber 
\Gamma (\bar D^0(t) \to K^+K^-) &\propto & e^{-\Gamma_1t}|T(\bar D^0 \to K^+K^-)|^2\times 
\\ 
&& 
\left[ 1 +y_D\frac{t}{\tau_D} \left( 1 - {\rm Re}\frac{p}{q} \frac{1}{\rho_{K^+K^-}}\right) - 
x_D\frac{t}{\tau_D}{\rm Im}\frac{p}{q}\frac{1}{\rho_{K^+K^-}} \right]
\label{DKK} 
\eea
Some comments might elucidate Eqs.(\ref{DKK}): 
\begin{itemize}
\item 
\cp~invariance implies (in addition to $|T(D^0 \to K^+K^-)| = |T(\bar D^0 \to K^+K^-)|$)   
$\frac{q}{p}\bar \rho_{K^+K^-} = 1$ (and $|q|=|p|$). The transitions $D^0(t) \to K^+K^-$ and 
$\bar D^0(t) \to K^+K^-$ are then described by the same {\em single} lifetime. That is a 
consequence of the Theorem given by Eq.(\ref{EQTH}), since $K^+K^-$ is a \cp~eigenstate. 
\item 
The usual three types of \cp~violation can arise, namely the direct and indirect types -- 
$|\bar \rho_{K^+K^-}| \neq 0$ and $|q|\neq |p|$, respectively -- as well as the one involving 
the interference between the oscillation and direct decay amplitudes -- 
Im$\frac{q}{p}\bar \rho_{K^+K^-}\neq 0$ leading also to Re$\frac{q}{p}\bar \rho_{K^+K^-}\neq 1$. 
\item
Assuming for simplicity $|T(D^0 \to K^+K^-)| = |T(\bar D^0 \to K^+K^-)|$ (CKM dynamics is expected 
to induce an asymmetry not exceeding 0.1\%) and $|q/p| = 1- \epsilon_D$ one has 
$(q/p)\bar \rho_{K^+K^-} = (1-\epsilon_D) e^{i\phi_{K\bar K}}$ and thus 
\beq 
A_{\Gamma} = \frac{\Gamma (\bar D^0(t) \to K^+K^-) - \Gamma (D^0(t) \to K^+K^-)}
{\Gamma (\bar D^0(t) \to K^+K^-) + \Gamma (D^0(t) \to K^+K^-)} 
\simeq x_D\frac{t}{\tau_D} {\rm sin}\phi_{K\bar K} -  
y_D\frac{t}{\tau_D}\epsilon_D {\rm cos}\phi_{K\bar K}\; .  
\eeq
where I have assumed $|\epsilon_D| \ll 1$. 
BELLE has found \cite{BELLEOSC1}
\beq 
A_{\Gamma} = (0.01 \pm 0.30 \pm 0.15) \%
\eeq
While there is no evidence for \cp~violation in the transition, one should also note that 
the asymmetry is bounded by $x_D$. For $x_D$, $y_D \leq 0.01$, as indicated by the data, 
$A_{\Gamma}$ could hardly exceed the 1\% range. I.e., there is no real 
bound on $\phi_D$ or $\epsilon_D$ yet. The good news is that if $x_D$ 
and/or $y_D$ indeed fall into the 0.5 - 1 \% range, 
then any improvement in the experimental sensitivity for a \cp~asymmetry in 
$D^0(t) \to K^+K^-$ constrains New Physics scenarios -- or could reveal them 
\cite{GKN}! 
\end{itemize}

Another promising channel for probing \cp~symmetry is $D^0(t) \to K^+\pi^-$: since it is 
doubly Cabibbo suppressed, it should a priori exhibit a higher sensitivity to a 
New Physics amplitude.  Furthermore it cannot exhibit direct \cp~violation in the SM. 
$$
\frac{\Gamma (D^0(t) \to K^+\pi^-)}{\Gamma (D^0(t) \to K^-\pi^+)} = 
\left| \frac{T(D^0 \to K^+\pi^-)}{T(D^0 \to K^-\pi^+)}\right| ^2 \times 
$$
\beq
\left[ 1 + \left( \frac{t}{\tau_D}\right)^2\left(\frac{x_D^2 + y_D^2}{4{\rm tg}\theta_C^4}\right) 
\left|\frac{q}{p}\right|^2 
\left| \hat \rho_{K\pi}\right|^2 + \left(\frac{t}{\tau_D}\right) \left|\frac{q}{p}\right| \left| \hat \rho_{K\pi}\right|
\left( \frac{y_D^{\prime}{\rm cos}\phi_{K\pi} + 
x_D^{\prime}{\rm sin}\phi_{K\pi}}{{\rm tg}\theta_C^2}\right)
\right] 
\label{DKPI}
\eeq 
$$ \frac{\Gamma (\bar D^0(t) \to K^-\pi^+)}{\Gamma (\bar D^0(t) \to K^+\pi^-)} = 
\left| \frac{T(\bar D^0 \to K^-\pi^+)}{T(\bar D^0 \to K^+\pi^-)}\right| ^2 \times 
$$
\beq 
\left[ 1 + \left( \frac{t}{\tau_D}\right)^2\left(\frac{x_D^2 + y_D^2}{4{\rm tg}\theta_C^4}\right) \left|\frac{p}{q}\right|^2 
\left| \hat \rho_{K\pi}\right|^2 + \left(\frac{t}{\tau_D}\right) \left|\frac{p}{q}\right|  \left| \hat \rho_{K\pi}\right|
\left( \frac{y_D^{\prime}{\rm cos}\phi_{K\pi} - 
x_D^{\prime}{\rm sin}\phi_{K\pi}}{{\rm tg}\theta_C^2}\right)
\right] 
\label{BARDKPI}
\eeq
with 
\bea 
\nonumber 
\frac{q}{p} \frac{T(D^0 \to K^+\pi^-)}{T(D^0 \to K^-\pi^+)} &\equiv & 
 - \frac{1}{{\rm tg}^2\theta_C} (1-\epsilon_D) |\hat \rho _{K\pi}|e^{-i(\delta - \phi_{K\pi})} \\
 \frac{q}{p} \frac{T(\bar D^0 \to K^-\pi^+)}{T(\bar D^0 \to K^+\pi^-)} &\equiv & 
 - \frac{1}{{\rm tg}^2\theta_C} \frac{1}{1-\epsilon_D} |\hat \rho _{K\pi}|e^{-i(\delta + \phi_{K\pi})}
\eea
yielding an asymmetry 
$$  
\frac{\Gamma (\bar D^0(t) \to K^-\pi^+) - \Gamma (D^0(t) \to K^+\pi^-)}
{\Gamma (\bar D^0(t) \to K^-\pi^+) + \Gamma (D^0(t) \to K^+\pi^-)} \simeq 
$$
\beq
 \left(\frac{t}{\tau_D}\right) \left| \hat \rho_{K\pi}\right|
 \left( \frac{y_D^{\prime}{\rm cos}\phi_{K\pi}\epsilon_D - 
x_D^{\prime}{\rm sin}\phi_{K\pi}}{{\rm tg}\theta_C^2}\right) + 
 \left(\frac{t}{\tau_D}\right)^2 \left| \hat \rho_{K\pi}\right|^2 \frac{\epsilon_D(x_D^2 + y_D^2)}
 {2{\rm tg}\theta_C^4}
 \eeq
 where I have again assumed for simplicity $|\epsilon _D| \ll 1$ and {\em no direct} 
 \cp~violation. 
 
BABAR has also searched for a time dependent \cp~asymmetry in $D^0 \to K^+\pi^-$ vs. 
$\bar D^0(t) \to K^- \pi^+$, yet so far has not found any evidence for it \cite{BABAROSC} on 
the about 1 \% level. Yet again, with $x_D^{\prime}$ and $y_D^{\prime}$ capped by 
about 1\%, no nontrivial bound can be placed on the weak phase $\phi_{K\pi}$ that can be 
induced by New Physics. On the other hand any further increase in experimental sensitivity could reveal a signal. 

Oscillations, even in the extreme scenario of $x_D = 0$, $y_D \neq 0$, will induce a time 
dependance in the \ot~odd moments $\Gamma_3$ and $\bar \Gamma_3$ of 
$D \to K^+K^-\pi^+\pi^-$. 

In close qualitative analogy to $B^0$ decays can one observe \cp~violation in $D^0$ decays 
also through the {\em existence} of a transition. The reaction 
\beq 
e^+e^- \to \psi(3770) \to D^0 \bar D^0 \to f_{\pm} f^{\prime}_{\pm} 
\eeq 
with \cp $|f^{(\prime)}\rangle = \pm |f^{(\prime)}\rangle$ can occur {\em only}, if \cp~is violated, 
since \cp$|\psi (3770)\rangle = +1 \neq \cp |f_{\pm} f^{\prime}_{\pm}\rangle = (-1)^{l=1} = -1$. 
The final states $f$ and $f^{\prime}$ can be different, as long as they posses the same \cp~parity. 
More explicitly one has for $x_D \ll 1$ 
$$ 
{\rm BR}(\psi (3770) \to D^0\bar D^0 \to f_{\pm} f^{\prime}_{\pm} \simeq 
{\rm BR}(D \to f_{\pm}) {\rm BR}(D \to f_{\pm}^{\prime}) \cdot 
$$
\beq 
\left[ (2+x_D^2)\left| \frac{q}{p}  \right|^2\left|\bar \rho(f_{\pm}) -  \bar \rho(f_{\pm}^{\prime}) \right|^2 
+ x_D^2 \left| 1 - \frac{q}{p}\bar \rho(f_{\pm})\frac{q}{p}\bar \rho(f_{\pm}^{\prime})  \right|^2\right] 
\label{CPEX}
\eeq
The second contribution in the square brackets can occur only due to oscillations 
and then also for $f_{\pm}^{\prime} = f_{\pm}$; yet it is heavily suppressed by $x_D^2 \leq 10^{-4}$ 
making it practically unobservable. The first term arises even with $x_D = 0$, yet 
requires $f_{\pm}^{\prime} \neq f_{\pm}$. It is possible that 
$\left|\bar \rho(f_{\pm}) -  \bar \rho(f_{\pm}^{\prime}) \right|^2$ provides a larger signal of 
\cp~violation than either $\left|1-  |\rho(f_{\pm})|^2\right|$ or 
$\left|1-  |\rho(f_{\pm}^{\prime})|^2\right|$. 

Eq.\ref{CPEX} also holds, when the final states are not \cp~eigenstates, yet still modes common 
to $D^0$ and $\bar D^0$. Consider for example $e^+e^- \to D^0 \bar D^0 \to f_af_b$ 
with $f_a = K^+K^-$, $f_b = K^{\pm}\pi^{\mp}$. Measuring those rates will yield unique information on the strong phase shifts.

%%%%%%%%%%%%%%%%%%%
%\subsubsection{\cp~Violation in Semileptonic $D^0$ Decays}
%%%%%%%%%%%%%%

%%%%%%%%%%%%%
{\bf (iii) \cp~Violation in Semileptonic $D^0$ Decays}
%%%%%%%%%%%%%%%

$|q/p| \neq 1$ unambiguously reflects \cp~violation in $\Delta C=2$ dynamics. It can be probed most directly in semileptonic $D^0$ decays leading to `wrong sign' leptons: 
\beq 
a_{SL}(D^0) \equiv \frac{\Gamma (D^0(t) \to l^-X) - \Gamma (\bar D^0 \to l^+X)}
{\Gamma (D^0(t) \to l^-X) + \Gamma (\bar D^0 \to l^+X)} = 
\frac{|q|^4 - |p|^4}{|q|^4 + |p|^4} 
\eeq
The corresponding observable has been studied in semileptonic decays of neutral $K$ and $B$ mesons. With $a_{SL}$ being controlled by $(\Delta \Gamma/\Delta M){\rm sin}\phi_{weak}$, 
it is predicted to be small in both cases, albeit for different reasons: 
(i) While $(\Delta \Gamma_K/\Delta M_K) \sim 1$ one has sin$\phi_{weak}^K \ll 1$ leading to 
$a_{SL}^K = \delta _l  \simeq (3.32 \pm 0.06)\cdot 10^{-3}$ as observed. 
(ii) For $B^0$ on the other hand one has 
$(\Delta \Gamma_B/\Delta M_B)\ll1$ leading to $a_{SL}^B < 10^{-3}$ 
(see Sect.\ref{BSSLRAD} for details). 

For $D^0$ on the other hand both $\Delta M_D$ and $\Delta \Gamma_D$ are small, yet 
$\Delta \Gamma_D/\Delta M_D$ is not: present data indicate it is about unity or even larger; 
$a_{SL}$ is given by the smaller of $\Delta \Gamma_D/\Delta M_D$ or its inverse multiplied by 
sin$\phi_{weak}^D$, which might not be that small: i.e., while the rate for `wrong-sign' leptons is certainly small in semileptonic decays of neutral 
$D$ mesons, their \cp~asymmetry might not be at all, if New Physics intervenes to induce 
$\phi_{weak}^D$.

%%%%%%%%%%%%
\subsubsection{\cp~Violation in the Lepton Sector}
\label{CPVLEPT}
%%%%%%%%%

I find the conjecture that baryogenesis is a {\em secondary} phenomenon driven by {\em primary}  leptogenesis a most intriguing one also for philosophical reasons 
\footnote{For it would 
complete what is usually called the Copernican Revolution 
\cite{ARAB}: first our Earth was removed from the center of the Universe, then in due course our Sun, our Milky Way and local cluster; few scientists believe life exists only on our Earth. Realizing that the stuff we are mostly made out of -- protons and neutrons -- 
are just a cosmic `afterthought' fits this pattern, which culminates in the dawning realization that even {\em our} Universe is just one among innumerable others, albeit a most unusual one.}. Yet then it becomes mandatory to search for \cp~violation in the lepton sector in a dedicated fashion. 

In Sect. \ref{EDMS} I have sketched the importance of measuring {\em electric dipole moments}  as accurately as possible. The electron's EDM is a most sensitive probe of 
\cp~violation in leptodynamics. Comparing the present experimental and CKM upper bounds, 
respectively  
\beq 
d_e^{exp} \leq 1.5 \cdot 10 ^{-27} \; \; {\rm e \; cm} \; \; \; vs. \; \; \; 
d_e^{CKM} \leq 10 ^{-36} \; \; {\rm e \; cm}
\eeq
we see there is a wide window of several orders of magnitude, where New Physics could surface in an unambiguous way. This observation is reinforced by the realization that New Physics scenarios can 
naturally generate $d_e > 10^{-28}$e cm, while of only secondary significance in $\epsilon_K$, 
$\epsilon^{\prime}$ and sin$2\phi_i$. 

The importance that at least part of the HEP community attributes to finding \cp~violation in leptodynamics is best demonstrated by the efforts contemplated for observing \cp~asymmetries 
in {\em neutrino oscillations}. Clearly hadronization will be the least of the concerns, yet one has to 
disentangle genuine \cp~violation from matter enhancements, since the neutrino oscillations can be studied only in a matter, not an antimatter environment. Our colleagues involved in such endeavours 
will rue their previous complaints about hadronization and remember the wisdom of an ancient 
Greek saying: 

\begin{center}
"When the gods want to really harm you, they fulfill your wishes."   
\end{center} 

%%%%%%%%%%
\subsubsection{The Decays of $\tau$ Leptons -- the Next `Hero Candidate'}
\label{NEXTHERO}
%%%%%%%%%

Like charm hadrons the $\tau $ lepton is often viewed as  a system with a great past, but hardly a 
future. Again I think this is a very misguided view, and I will illustrate it with two examples. 

Searching for $\tau ^{\pm} \to \mu ^{\pm} \mu ^+\mu ^-$ (and its variants) -- 
processes forbidden in the SM -- is particularly intriguing, since it involves only `down-type' leptons 
of the second and third family and is thus the complete analogy of the quark lepton process 
$b \to s \bar s s$ driving $B_s \to \phi K_S$, which has recently attracted such strong attention. 
Following this analogy literally one guestimates ${\rm BR}(\tau \to 3 \mu) \sim 10^{-8}$ to be 
compared with the present bound from BELLE  
\beq 
{\rm BR}(\tau \to 3 \mu) \leq 2\cdot 10^{-7} \; . 
\eeq
It would be very interesting to know what the 
$\tau$ production rate at the hadronic colliders is and whether they could be competitive or even superior with the $B$ factories in such a search.  

In my judgment $\tau$ decays -- together with electric dipole moments for leptons and possibly $\nu$ oscillations referred to above -- provide the best stage to search for manifestations of 
\cp~breaking leptodynamics. 

The most promising channels for exhibiting \cp~asymmetries are $\tau \to \nu K \pi$, since due to 
the heaviness of the lepton and quark flavours they are most sensitive to nonminimal Higgs dynamics,  
and they can show asymmetries also in the final state distributions rather than integrated rates 
\cite{KUHN}.  

There is also a {\em unique}  opportunity in $e^+e^- \to \tau ^+ \tau ^-$: since the $\tau$ pair is produced with its spins aligned, the decay of one $\tau$ can `tag' the spin of the other $\tau$. I.e., 
one can probe {\em spin-dependent} \cp~asymmetries with {\em unpolarized} beams. This provides 
higher sensitivity and more control over systematic uncertainties. 

I feel these features are not sufficiently 
appreciated even by proponents of Super-B factories. It has been recently pointed \cite{BSTAU}  
out that based on known physics one can actually predict a 
\cp~asymmetry: 
\beq 
\frac{\Gamma(\tau^+\to K_S \pi^+ \overline \nu)-\Gamma(\tau^-\to K_S \pi^- \nu)}
{\Gamma(\tau^+\to K_S \pi^+ \overline \nu)+\Gamma(\tau^-\to K_S \pi^- \nu)}= 
(3.27 \pm 0.12)\times 10^{-3}
\label{CPKS}
\eeq
due to $K_S$'s preference for antimatter.

%%%%%%%%%%%%%%
\subsection{Future Studies of $B_{u,d}$ Decays}
\label{BFUTURE}
%%%%%%%%%%%

The successes of CKM theory to describe flavour dynamics do {\em not} tell us at all that 
New Physics does not affect $B$ decays; the message is that {\em typically} we can{\em not} count on 
a {\em numerically} massive impact there. Shifting an asymetry by, say, ten percentage points -- for example from 40 \%  to 50 \% -- might already be on the large side. Thus our aim has to be to aim for 
uncertainties that do not exceed a few percent. 

An integrated luminosity of 1 $ab^{-1}$ at the $B$ factories will fall short of such a goal for $B_d \to \pi \pi$, $B^{\pm} \to D^{neut}K^{\pm}$ and in particular also 
for the modes driven by $b \to s q \bar q$.  Even ten times that statistics  would not suffice in view of the 
`big picture', i.e. when one includes other rare transitions. Of course we are in the very fortunate 
situation that one of the LHC experiments, namely LHCb, is dedicated to undertaking precise 
measurements of the weak decays of beauty hadrons. Thus we can expect a stream of high quality data to be forthcoming over the next several years. I will briefly address one class of rare decays.

%%%%%%%%%%%%
\subsubsection{$B \to l^+l^- X$}
%%%%%%%%%%

We are just at the beginning of studying $B \to l^+l^-X$, and it has to be 
pursued in a dedicated and comprehensive manner for the following reasons: 
\begin{itemize}
\item 
With the final state being more complex than for $B \to \gamma X$, it is described by a larger number 
of observables: rates, spectra of the lepton pair masses and the lepton energies, their forward-backward asymmetries and \cp~asymmetries. 
\item 
These observables provide independent information, since there is a larger number of effective transition operators than for $B \to \gamma X$. By the same token there is a much wider window 
to find New Physics and even diagnose its salient features. 
\item 
It will take the statistics of a Super-Flavour factory to mine this wealth of information on New Physics. 
\item 
Essential insights can be gained also by analyzing the exclusive channel $B\to l^+l^-K^*$ at hadronic 
colliders like the LHC, in particular the position of the zero in the lepton forward-backward asymmetry. 
For the latter appears to be fairly insensitive to hadronization effects in this exclusive mode 
\cite{HILLER1}. It will be important to analyze quantitatively down to which level of accuracy this feature 
persists. 
\end{itemize}

%%%%%%%%%%%
\subsection{$B_s$ Decays -- an Independent Chapter in Nature's Book}
\label{BSDEC}
%%%%%%%%%%%

When the program for the $B$ factories was planned, it was thought that studying $B_s$ transitions 
will be required to construct the CKM triangle, namely to determine one of its sides and the angle 
$\phi_3$. As discussed above a powerful method has been developed to extract 
$\phi_3$ from $B^{\pm} \to D^{neut}K^{\pm}$ and a meaningful value for $|V(td)/V(ts)|$ has 
been inferred from the measured value of $\Delta M_{B_d}/\Delta M_{B_s}$.  
None of this, however, reduces the importance of a future 
comprehensive program to study $B_s$ decays -- on the contrary! With the basic CKM parameters 
fixed or to be fixed in $B_{u,d}$ decays, $B_s$ transitions can be harnessed as powerful probes 
for New Physics and its features. 

In this context it is essential to think `outside the box' -- pun intended. The point here is that several 
relations that hold in the SM (as implemented through quark box and other loop diagrams) are 
unlikely to extend beyond minimal extensions of the SM. In that sense $B_{u,d}$ and $B_s$ decays 
constitute two different and complementary chapters in Nature's book on fundamental dynamics. 

%%%%%%%%%%%%
\subsubsection{CP~Violation in Non-Leptonic $B_s$ Decays}
\label{NLBS}
%%%%%%%%%%

One class of nonleptonic $B_s$ transitions does not follow the paradigm of large \cp~violation in 
$B$ decays \cite{BS80}: 
$$ 
A_{\cp}(B_s(t) \to [\psi \phi]_{l=0}/\psi \eta ) = {\rm sin}2\phi (B_s) {\rm sin}\Delta M(B_s)t 
$$
\beq
{\rm sin}2\phi (B_s) =
{\rm Im}\left[ \frac{(V^*(tb)V(ts))^2}{|V^*(tb)V(ts)|^2}\frac{(V(cb)V^*(cs))^2}{(V(cb)V^*(cs))^2}  \right]
\simeq 2\lambda ^2 \eta   \sim 0.02  \; . 
\label{BSPSISM}
\eeq
This is easily understood: on the leading CKM level only quarks of the second and third families 
contribute to $B_s$ oscillations and $B_s \to \psi \phi$ or $\psi \eta$; therefore on that level there can be no \cp~violation making the \cp~asymmetry Cabibbo suppressed. {\em Yet New Physics of various ilks can quite conceivably generate sin$2\phi (B_s) \sim $ several $\times$ 10 \%.} 

Analyzing the decay rate evolution in proper time of 
\beq 
B_s(t) \to \phi \phi 
\label{BSPHIPHI} 
\eeq
with its direct as well as indirect \cp~violation is much more than a repetition of the 
$B_d(t) \to \phi K_S$ saga: 
\begin{itemize}
\item 
${\cal M}_{12}(B_s)$ and ${\cal M}_{12}(B_d)$ -- the off-diagonal elements in the mass matrices for 
$B_s$ and $B_d$ mesons, respectively -- provide in principle independent pieces of information 
on $\Delta B=2$ dynamics. 
\item
While the final state $\phi K_S$ is described by a single partial wave, namely $l=1$, there are 
three partial waves in $\phi \phi$, namely $l= 0,1,2$. Disentangling the three partial rates and their 
\cp~asymmetries -- or at least separating $l$ = even and odd contributions -- provides a new 
diagnostics about the underlying dynamics. 
\end{itemize}
Even in the limit of $x_s \to \infty$ \cp~violation can be searched for through the {\em existence}  
of a transition, namely $e^+ e^- \to B_s \bar B_s|_{\oc = -}\to f_{\pm}f_{\pm}^{\prime}$, 
where $f_{\pm}$ and $f_{\pm}^{\prime}$ denote \cp~eigenstates of the same \cp~parity: 
$$ 
{\rm BR}(B_s\bar B_s|_{\oc = -} \to f_{\pm} f^{\prime}_{\pm}) \simeq 
{\rm BR}(B_s \to f_{\pm}) {\rm BR}(B_s \to f_{\pm}^{\prime}) \cdot 
$$
\beq 
\left[ \left|\bar \rho(f_{\pm}) -  \bar \rho(f_{\pm}^{\prime}) \right|^2 
+  \left| 1 - \frac{q}{p}\bar \rho(f_{\pm})\frac{q}{p}\bar \rho(f_{\pm}^{\prime})  \right|^2\right] \; , 
\label{UP5S}
\eeq
where the two terms in the square brackets have the coefficients $1+\frac{1}{1+x_s^2}$ and 
$\frac{x_S^2}{1+x_s^2}$, respectively, both of which go to unity for $x_s \to \infty $.  The 
analogous expression describes also $e^+e^- \to B_d\bar B_d \to f_{\pm} f^{\prime}_{\pm}$, 
where the two terms in the square brackets carry coefficients of 1.62 and 0.38, respectively. 
Available data sets should be large enough to produce candidates.

%%%%%%%%%%%
\subsubsection{Semileptonic Modes}
\label{BSSLRAD}
%%%%%%%%%%
 
Due to the rapid $B_s$ oscillations those mesons have a practically equal probability 
to decay into `wrong' and `right' sign leptons. One can then search for an asymmetry in the 
wrong sign rate for mesons that initially were $B_s$ and $\bar B_s$: 
\beq 
a_{SL}(B_s) \equiv \frac{\Gamma (\bar B_s \to l^+X) - \Gamma (B_s \to l^-X)}
{\Gamma (\bar B_s \to l^+X) + \Gamma (B_s \to l^-X)}
\eeq
This observable is necessarily small; among other things it is proportional to 
$\frac{\Delta \Gamma _{B_s}}{\Delta M_{B_s}} \ll 1$. The CKM predictions are not very precise, yet certainly tiny \cite{LENZNEW}: 
\beq 
a_{SL}(B_s) \sim 2 \cdot 10^{-5} \; , \; a_{SL}(B_d) \sim 4 \cdot 10^{-4} \; ; 
\eeq
$a_{SL}(B_s)$ suffers a suppression quite specific to CKM dynamics; analogous to $B_s \to \psi \phi$ 
quarks of only the second and third family participate on the leading CKM level, which therefore 
cannot exhibit \cp~violation. Yet again, New Physics can enhance $a_{SL}(B_s)$, this time  
by two orders of magnitude up to the 1\% level.

%%%%%%%%
\subsection{Instead of a Summary: On the Future HEP Landscape -- a Call to Well-Reasoned 
Action}
\label{SUM5}
%%%%%%%% 

The situation of the SM, as it enters the third millenium, can be characterized through 
several statements: 
\begin{enumerate}
\item 
There is a new dimension due to the findings on $B$ decays:  one has established the first 
\cp~asymmetries outside the $K^0 - \bar K^0$ complex in four $B_d$ modes-- as predicted 
qualitativly as well as quantitatively by CKM dynamics: $B_d (t) \to \psi K_S$, 
$B_d(t) \to \pi^+\pi^-$, $B_d \to K^+\pi^-$ and $B_d(t) \to \eta^{\prime}K_S$. 
Taken together with the other established signals -- $K^0(t) \to 2\pi$ and 
$|\eta_{+-}| \neq |\eta_{00}|$ -- we see that in all these cases except for $B_d \to K^+\pi^-$ 
the intervention of meson-antimeson oscillations was instrumental in \cp~violation becoming 
observable. This is why I write $B_d[K^0] (t) \to f$. For practical reasons this holds even for $|\eta_{+-}| \neq |\eta_{00}|$. 

For the first time strong evidence has emerged for \cp~violation in the decays of a charged state, namely 
in $B^{\pm} \to K^{\pm} \rho^0$. 

The SM's success here can be stated more succinctly as follows: 
\begin{itemize}
\item 
From a tiny signal of $|\eta_{+-}| \simeq 0.0023$ one successfully 
infers \cp~asymetries in $B$ decays two orders 
of magnitude larger, namely sin$2\phi_1 \simeq 0.7$ in $B_d(t) \to \psi K_S$. 
\item 
From the measured values of two \cp~insensitive quantities -- $|V(ub)/V(cb)|$ in semileptonic $B$ 
decays and $|V(td)/V(ts)|$ in $B^0 - \bar B^0$ oscillations -- one deduces the existence of 
\cp~violation in $K_L \to 2\pi$ and $B_d (t) \to \psi K_S$ even in quantitative agreement with the data. 
\end{itemize}

We know now that CKM dynamics provides at least the lion's share in the observed \cp~asymmetries. 
The CKM description thus has become a {\em tested} theory. Rather then searching for 
{\em alternatives} to CKM dynamics we hunt for {\em corrections} to it. We have already learnt one lesson of a general lesson: \cp~violation has been `de-mystified; i.e., weak phases expressing  
\cp~violation can be as large as 90$^o$. 
\item 
None of these novel successes of the SM invalidate the theoretical arguments for it being incomplete. 
There is also clean evidence of mostly heavenly origin for New Physics, namely 
(i) neutrino oscillations, (ii) dark matter, (iii) presumably dark energy, 
(iv) probably the baryon number of our Universe and 
(v) possibly the Strong \cp~Problem. 

\item 
Flavour dynamics has become even more intriguing due to the emergence of neutrino 
oscillations. We do not understand the structure of the CKM matrix in any profound way -- and 
neither the PMNS matrix, its leptonic counterpart. Presumably we do understand why they look different, since only neutrinos can possess Majorana masses, which can give rise to the 
`see-saw' mechanism. 

Sometimes it is thought that the existence of two puzzles makes their resolution harder. I feel 
the opposite way: having a larger set of observables allows us to direct more questions to Nature, 
if we are sufficiently persistent, and learn from her answers. 
\footnote{Allow me a historical analogy: in the 1950's it was once suggested to a French politician 
that the 
French government's lack of enthusiasm for German re-unification showed that the French had not learnt to overcome their dislike of Germany. He replied with aplomb: "On the contrary, Monsieur!  
We truly love Germany and are therefore overjoyed that there are two Germanies we can love. Why would we change that?"}
\item 
The next `Grand Challenge' after studying the dynamics behind the electroweak phase transition is 
to find \cp~violation in the lepton sector -- anywhere. 
\item 
While the quantization of electric charge is an essential ingredient of the SM,  it  does not offer any understanding of it. It would naturally be explained through 
Grand Unification at very high energy scales. I refer to it as the `guaranteed New Physics', 
see Sect.\ref{INCOMPLET}. 
\item 
The SM's success in describing flavour transitions is not matched by a deeper understanding of the flavour structure, namely the patterns in the fermion masses and CKM parameters. 
For those do not appear to be of an accidental nature. I have referred to the dynamics generating the flavour structure as the 
`strongly suggested' New Physics ({\bf ssNP}), see Sect.\ref{INCOMPLET}. 
\item 
Discovering the {\bf cpNP} that drives the electroweak phase transition 
has been the justification for the LHC program, which will come online soon. Personally I am very partisan to the idea that the {\bf cpNP} will be of the 
SUSY type. Yet SUSY is an organizing principle rather than a class of theories, let alone a theory. 
We are actually quite ignorant about how to implement the one empirical feature of SUSY that has been established beyond any doubt, namely that it is broken. 
\item 
The LHC is very  likely to uncover the {\bf cpNP}, and I have not given up hope that the 
TEVATRON will catch the first glimpses of it. Yet the LHC and a forteriori the TEVATRON  are primarily 
discovery machines. The ILC project is motivated as a more surgical probe to map out the salient features of that {\bf cpNP}. 
\item 
This {\bf cpNP} is unlikely to shed light on the {\bf ssNP} behind the flavour puzzle of the SM, although one should not rule out such a most fortunate development. On the other hand New Physics even at the 
$\sim $ 10 - 100 TeV scale could well affect flavour transitions significantly through virtual effects. A comprehensive 
and dedicated program of heavy flavour studies might actually elucidate salient features of the 
{\bf cpNP} that could not be probed in any other way. Such a program is thus 
complementary to the one pursued at the TEVATRON, the LHC and hopefully at the ILC and -- 
I firmly believe  -- actually necessary rather than a luxury to identify the {\bf cpNP}. 

To put it in more general terms: Heavy flavour studies are of fundamental importance, 
many of its lessons cannot be obtained any other way and they cannot become obsolete.  
I.e., no matter what studies of high $p_{\perp}$ physics at the LHC and ILC will or 
will not show -- comprehensive and detailed studies of flavour dynamics will remain crucial 
in our efforts to reveal Nature's Grand Design. 
\item 
Yet a note of caution has to be expressed as well. Crucial manifestations of New Physics in flavour dynamics are likely to be subtle. Thus we have to succeed in acquiring data as well as interpreting them  
with {\em high precision}. Obviously this represents a stiff challenge -- however one that I believe we can meet, if we prepare ourselves properly.

\end{enumerate}
One of three possible scenarios will emerge in the next several years. 
\begin{enumerate} 
\item 
{\em The optimal scenario}: New Physics has been observed in "high $p_{\perp}$ physics", i.e. through the production of new quanta at the TEVATRON and/or LHC. Then it is {\em imperative} to study the impact of such New Physics on flavour dynamics; even if it should turn out to have none, this is an important piece of information, no matter how frustrating it would be to my 
experimental colleagues. Knowing the typical mass scale of that New Physics from collider data will be of great help to estimate its impact on heavy flavour transitions.  

\item 
{\em The intriguing scenario}: Deviations from the SM have been established in heavy flavour decays -- like the $B \to \phi K_S$ \cp~asymmetry or an excess in $\Gamma (K\to \pi \nu \bar \nu )$ -- without a clear signal for New Physics in high $p_{\perp}$ physics. A variant of this scenario has already emerged through the observations of neutrino 
oscillations. 

\item 
{\em The frustrating scenario}: No deviation from SM predictions have been identified. 

\end{enumerate}
I am optimistic it will be the `optimal' scenario, quite possibly with some elements of the 'intriguing' one. Of course one cannot rule out the `frustrating' scenario; yet we should not treat it as a case for defeatism: a possible failure to identify New Physics in future experiments at the hadronic colliders (or the $B$ factories) does not -- in my judgment -- invalidate the persuasiveness of the theoretical arguments and experimental evidence pointing to the incompleteness of the SM. 
It `merely' means we have to increase the sensitivity of our probes. 
{\bf I firmly believe a 
Super-flavour factory with a luminosity of order $10^{36}$ cm$^{-2}$ s$^{-1}$ or more for the 
study of beauty, charm and $\tau$ decays has to be an integral part of our future efforts towards deciphering Nature's basic code.}  
For a handful of even perfectly measured transitions will not be sufficient for the task at hand -- a 
{\em comprehensive} body of {\em accurate} data will be essential. {\bf Likewise we need a new round 
of experiments that can measure the rates for $K \to \pi \nu \bar \nu$ {\em accurately} 
with sample sizes $\sim {\cal O}(10^3)$ and mount another serious effort to probe the 
muon transverse polarization in $K_{\mu3}$ decays.}

\vspace{0.5cm}

{\bf Acknowledgments:} This work was supported by the NSF under the grant number PHY-0355098.

\vspace{4mm}

%%%%%%%%%%%%%%%%

\end{document}